%% file: wcpaper.tex
\documentclass[a4paper,11pt]{article}
\pdfoutput=1 

\input defines.tex

\usepackage{graphicx, subfigure}
\usepackage{longtable}
\usepackage{multirow}
\usepackage{pdflscape}
\usepackage{rotating}

\usepackage{jheppub} 

\usepackage[T1]{fontenc} 

\usepackage{atlasphysics}
\usepackage{epstopdf}
\usepackage{amsmath}

\usepackage{preprintcover}  
\PreprintCoverPaperTitle{\boldmath Measurement of the production of a $W$ boson in association with a charm quark in $pp$ collisions at $\sqrt{s}=7$\,TeV{} with the ATLAS detector}  
\PreprintIdNumber{CERN-PH-EP-2014-007}  
\PreprintCoverAbstract{The production of a $W$ boson in association with a single charm quark is studied using 4.6~fb$^{-1}$ of 
$pp$ collision data at $\sqrt{s}=7$\,\TeV\ collected with the ATLAS detector at the Large Hadron Collider.  
In events in which a $W$ boson decays to an electron or muon,
the charm quark is tagged either by its semileptonic decay to a muon or by the presence of a charmed meson.
The integrated and differential cross sections as a function of the pseudorapidity of the lepton from the $W$-boson decay are measured.
Results are compared to the predictions of next-to-leading-order QCD calculations obtained from various parton
distribution function parameterisations. The ratio of the strange-to-down sea-quark distributions is determined to be $0.96^{+0.26}_{-0.30}$ at $Q^2=1.9\,\GeV^2$, which supports the hypothesis of an SU(3)-symmetric composition of the light-quark sea. 
Additionally, the cross-section ratio $\sigma(W^+ + \overline c)/\sigma(W^- + c)$ is compared to the predictions obtained using parton distribution function parameterisations with different assumptions about the $s$--$\overline s$ quark asymmetry.}  
\PreprintJournalName{JHEP}  

\title{\boldmath Measurement of the production of a $W$ boson in association with a charm quark in $pp$ collisions at $\sqrt{s}=7$\,TeV{} with the ATLAS detector}

\author{The ATLAS Collaboration}

\abstract{The production of a $W$ boson in association with a single charm quark is studied using 4.6~fb$^{-1}$ of 
$pp$ collision data at $\sqrt{s}=7$\,\TeV\ collected with the ATLAS detector at the Large Hadron Collider.  
In events in which a $W$ boson decays to an electron or muon,
the charm quark is tagged either by its semileptonic decay to a muon or by the presence of a charmed meson.
The integrated and differential cross sections as a function of the pseudorapidity of the lepton from the $W$-boson decay are measured.
Results are compared to the predictions of next-to-leading-order QCD calculations obtained from various parton
distribution function parameterisations. The ratio of the strange-to-down sea-quark distributions is determined to be $0.96^{+0.26}_{-0.30}$ at $Q^2=1.9\,\GeV^2$, which supports the hypothesis of an SU(3)-symmetric composition of the light-quark sea. 
Additionally, the cross-section ratio $\sigma(W^+ + \overline c)/\sigma(W^- + c)$ is compared to the predictions obtained using parton distribution function parameterisations with different assumptions about the $s$--$\overline s$ quark asymmetry.}

\begin{document} 
\maketitle
\flushbottom

\input introduction.tex
\input detector.tex

\input dataandmc.tex

\input wselection.tex

\input wd.tex

\input wc.tex

\input unfolding.tex
\input systematics.tex
\input results.tex

\input additionalresults_WD.tex

\input additionalresults_Wc.tex

\input conclusion.tex

\input Acknowledgements.tex

\clearpage

\bibliographystyle{atlasBibStyleWithTitle}
\bibliography{references}


\onecolumn 
\clearpage
\input atlas_authlist.tex

\end{document}

%% file: defines.tex
\def\mtw{\ensuremath{m^W_{\mathrm{T}}}} 

\newcommand{\ptnu}{\ensuremath{p_{\mathrm{T}}^{\nu}}\ }
\newcommand{\ptmu}{\ensuremath{p_{\mathrm{T}}^{\mu}}\ }
\newcommand{\pte}{\ensuremath{p_{\mathrm{T}}^{e}}\ }
\newcommand{\ptmum}{\ensuremath{p_{\mathrm{T}}^{\mu}}}
\newcommand{\ptem}{\ensuremath{p_{\mathrm{T}}^{e}}}
\newcommand{\ptell}{\ensuremath{p_{\mathrm{T}}^{\ell}}\ }
\newcommand{\etaell}{\ensuremath{\eta^{\ell}}\ }

\newcommand{\ptD}{\ensuremath{p_{\mathrm{T}}^{D^{(*)}}}}
\newcommand{\etaD}{\ensuremath{\eta^{D^{(*)}}}}
\newcommand{\ptDNoSp}{\ensuremath{p_{\mathrm{T}}^{D^{(*)}}}}

\newcommand{\DpDec}  {\ensuremath{D^{+}\rightarrow K^{-}\pi^{+}\pi^{+}}\ } 
\newcommand{\DpDecNoSp}  {\ensuremath{D^{+}\rightarrow K^{-}\pi^{+}\pi^{+}}} 
\newcommand{\DstarDec}  {\ensuremath{D^{*+}\rightarrow D^{0}\pi^{+} }\ } 
\newcommand{\DsDec}  {\ensuremath{D^{*+}\rightarrow D^{0}\pi^{+} \rightarrow (K^{-}\pi^{+})\pi^{+}}\ } 
\newcommand{\DsDecNoSp}  {\ensuremath{D^{*+}\rightarrow D^{0}\pi^{+} \rightarrow (K^{-}\pi^{+})\pi^{+}}} 
\newcommand{\DsatDec}  {\ensuremath{D^{*+}\rightarrow D^{0}\pi^{+} \rightarrow (K^{-}\pi^{+}\pi^{0})\pi^{+}}\ } 
 
\newcommand{\DkpppDec}  {\ensuremath{D^{*+}\rightarrow D^{0}\pi^{+} \rightarrow (K^{-}\pi^{+}\pi^{-}\pi^{+})\pi^{+}}\ }
\newcommand{\DkpppDecNoSp}  {\ensuremath{D^{*+}\rightarrow D^{0}\pi^{+} \rightarrow (K^{-}\pi^{+}\pi^{-}\pi^{+})\pi^{+}}} 
\newcommand{\DzsDec}  {\ensuremath{D^{0} \rightarrow K^{-}\pi^{+}}\ } 
\newcommand{\DzsDecNoSp}  {\ensuremath{D^{0} \rightarrow K^{-}\pi^{+}}}
\newcommand{\DzsatDec}  {\ensuremath{D^{0}\rightarrow K^{-}\pi^{+}\pi^{0}}\ } 
\newcommand{\DzsatDecNoSp}  {\ensuremath{D^{0}\rightarrow K^{-}\pi^{+}\pi^{0}}} 
\newcommand{\DzkpppDec}  {\ensuremath{D^{0} \rightarrow K^{-}\pi^{+}\pi^{-}\pi^{+}}\ } 
\newcommand{\DzkpppDecNoSp}  {\ensuremath{D^{0} \rightarrow K^{-}\pi^{+}\pi^{-}\pi^{+}}} 
\newcommand{\DzkporpppDec}  {\ensuremath{D^{0} \rightarrow K^{-}\pi^{+}(\pi^{-}\pi^{+})}} 
 
\newcommand{\DpDecboth}  {\ensuremath{D^{\pm}\rightarrow K^{\mp}\pi^{\pm}\pi^{\pm}}} 
\newcommand{\DsDecboth}  {\ensuremath{D^{*\pm}\rightarrow D^{0}\pi^{\pm} \rightarrow (K^{\mp}\pi^{\pm})\pi^{\pm}}\ } 
\newcommand{\DsDecbothNoSp}  {\ensuremath{D^{*\pm}\rightarrow D^{0}\pi^{\pm} \rightarrow (K^{\mp}\pi^{\pm})\pi^{\pm}}} 
\newcommand{\DsatDecboth}  {\ensuremath{D^{*\pm}\rightarrow D^{0}\pi^{\pm} \rightarrow (K^{\mp}\pi^{\pm}\pi^{0})\pi^{\pm}}\ } 
\newcommand{\DkpppDecboth}  {\ensuremath{D^{*\pm}\rightarrow D^{0}\pi^{\pm} \rightarrow (K^{\mp}\pi^{\pm}\pi^{\mp}\pi^{\pm})\pi^{\pm}}\ }

\newcommand{\Wca}  {\ensuremath{ W+c}}
\newcommand{\Wpca}  {\ensuremath{ W^++\overline{c}}}
\newcommand{\Wmca}  {\ensuremath{ W^-+c}}
\newcommand{\Wce}  {\ensuremath{ Wc\mbox{-}\rm{jet}}}

\newcommand{\WDe}  {\ensuremath{ WD^{(*)}}}
\newcommand{\WpDe}  {\ensuremath{ W^+D^{(*)-}}}
\newcommand{\WmDe}  {\ensuremath{ W^-D^{(*)+}}}
\newcommand{\De}  {\ensuremath{ D^{(*)}}\,}
\newcommand{\DeNoSp}  {\ensuremath{ D^{(*)}}}
\newcommand{\WDse}  {\ensuremath{ WD^{*}}}
\newcommand{\WDpe}  {\ensuremath{ WD}}

\newcommand{\WpDs}  {\ensuremath{ W^{+}D^{*-}}\ }
\newcommand{\WpDp}  {\ensuremath{ W^{+}D^{-}}\ }
\newcommand{\WmDs}  {\ensuremath{ W^{-}D^{*+}}\ }
\newcommand{\WmDp}  {\ensuremath{ W^{-}D^{+}}\ }
\newcommand{\xsecratio}  {\ensuremath{ \sigma_{\rm fid}^{\rm OS-SS}(WD^{(*)})/\sigma_{\rm fid}(W)}\ } 
\newcommand{\xsecratioNoSp}  {\ensuremath{ \sigma_{\rm fid}^{\rm OS-SS}(WD^{(*)})/\sigma_{\rm fid}(W)}}

\newcommand{\asymratio}  {\ensuremath{ \sigma_{\rm fid}^{\rm OS-SS}(W^{+}D^{(*)-})/\sigma_{\rm fid}^{\rm OS-SS}(W^{-}D^{(*)+})}\ }
\newcommand{\asymratioNoSp}  {\ensuremath{ \sigma_{\rm fid}^{\rm OS-SS}(W^{+}D^{(*)-})/\sigma_{\rm fid}^{\rm OS-SS}(W^{-}D^{(*)+})}}

\newcommand{\asymratioDp}  {\ensuremath{ \sigma_{\rm fid}^{\rm OS-SS}(W^{+}D^{-})/\sigma_{\rm fid}^{\rm OS-SS}(W^{-}D^{+})}\ }

\newcommand{\asymratioDs}  {\ensuremath{ \sigma_{\rm fid}^{\rm OS-SS}(W^{+}D^{*-})/\sigma_{\rm fid}^{\rm OS-SS}(W^{-}D^{*+})}\ } 
\newcommand{\asymratioc}  {\ensuremath{ \sigma_{\rm fid}^{\rm OS-SS}(W^{+}\overline{c}\mbox{-}{\rm jet})/\sigma_{\rm fid}^{\rm OS-SS}(W^{-}c\mbox{-}{\rm jet})}\ } 
\newcommand{\asymratiocNoSp}  {\ensuremath{ \sigma_{\rm fid}^{\rm OS-SS}(W^{+}\overline{c}\mbox{-}{\rm jet})/\sigma_{\rm fid}^{\rm OS-SS}(W^{-}c\mbox{-}{\rm jet})}} 

\newcommand{\xsecWD}  {\ensuremath{ \sigma_{\rm fid}^{\rm OS-SS}(WD^{(*)})}\ }

\newcommand{\xsecWDNoSp}  {\ensuremath{ \sigma_{\rm fid}^{\rm OS-SS}(WD^{(*)})}}
\newcommand{\xsecWc}  {\ensuremath{ \sigma_{\rm fid}^{\rm OS-SS}(Wc\mbox{-}\rm{jet})}\ }
\newcommand{\xsecWcNoSp}  {\ensuremath{ \sigma_{\rm fid}^{\rm OS-SS}(Wc\mbox{-}\rm{jet})}}
\newcommand{\xsecplusWc}  {\ensuremath{ \sigma_{\rm fid}^{\rm OS-SS}(W^+\overline{c}\mbox{-}\rm{jet})}}
\newcommand{\xsecminusWc}  {\ensuremath{ \sigma_{\rm fid}^{\rm OS-SS}(W^-c\mbox{-}\rm{jet})}}
\newcommand{\xsecplusDp}  {\ensuremath{ \sigma_{\rm fid}^{\rm OS-SS}(W^{+}D^{-})}} 
\newcommand{\xsecminusDp}  {\ensuremath{ \sigma_{\rm fid}^{\rm OS-SS}(W^{-}D^{+})}} 
\newcommand{\xsecplusDs}  {\ensuremath{ \sigma_{\rm fid}^{\rm OS-SS}(W^{+}D^{*-})}}
\newcommand{\xsecminusDs}  {\ensuremath{ \sigma_{\rm fid}^{\rm OS-SS}(W^{-}D^{*+})}}

\newcommand{\RC}{\ensuremath{R_c^\pm}}

\def\mm{\ifmmode  {\mathrm{\ mm}}\else
                   \textrm{~mm}\fi}%

\newcommand{\Dp}  {\ensuremath{D^{+}}\ } 
\newcommand{\Dpm}  {\ensuremath{D}} 
\newcommand{\Ds}  {\ensuremath{D^{*}}} 
\newcommand{\Dz}  {\ensuremath{D^{0}}\ } 

\newcommand{\OSSS} {\textsc{OS--SS}}
\newcommand{\OS} {\textsc{OS}}
\newcommand{\SSS} {\textsc{SS}}

\def\PythiaSix{\textsc{Pythia}}
\def\Pythia{\textsc{Pythia}}
\def\Herwig{\textsc{Herwig}}
\def\Herwigpp{{Herwig\raise0.2ex\hbox{\scriptsize\kern-0.5pt+\kern-0.5pt+}}}
\def\PowHeg{\textsc{Powheg}}
\def\MCatNLO{\textsc{MC@NLO}}
\def\Jimmy{\textsc{Jimmy}}
\def\Alpgen{\textsc{Alpgen}}
\def\aMCatNLO{aMC@NLO}
\def\EvtGen{EvtGen}
\def\CT{\textsc{CT10}}
\def\MSTW{\textsc{MSTW2008}}
\def\HERA{\textsc{HERA\-PDF1.5}}
\def\epWZ{ATLAS-epWZ12}
\def\NNPDF{\textsc{NNPDF2.3}}
\def\NNPDFcoll{\textsc{NNPDF2.3coll}}
\def\Tauola{\textsc{Tauola}}
\def\Photos{\textsc{Photos}}

%% file: introduction.tex
\section{Introduction}
\label{s:introduction}

The production of a $W$ boson in association with a single charm quark in proton--proton collisions is described 
at leading order (LO) in perturbative quantum chromodynamics (QCD) 
by the scattering of a gluon and a down-type 
quark ($d$, $s$, $b$). The relative contribution from each of the three families in the initial state is 
determined by the parton distribution functions (PDF) of the proton 
and by the quark-mixing matrix elements  $V_{cd}$, $V_{cs}$ and $V_{cb}$. In proton--proton collisions at a centre-of-mass energy of $\sqrt{s}=7$~\TeV, 
$gs\rightarrow W^- c$ and $g\bar{s}\rightarrow W^+\bar{c}$ production channels are dominant, while the reaction 
initiated by a $d$-quark contributes about 10\%~\cite{Stirling:2012vh}, 
being suppressed by the quark-mixing matrix element 
$V_{cd}$. The contribution of processes that include $b$-quarks is negligible.
The next-to-leading-order (NLO)
QCD terms~\cite{wznlo} are dominated by one-loop corrections 
to the subprocess $gs\rightarrow Wc$ and the tree-level $2\rightarrow 3$ processes $gg\rightarrow sWc$ 
and $qs\rightarrow qWc$. 
Processes with charm quarks in the initial state are not considered for this analysis as explained in section~\ref{subs:theory}. 
Since the $gs\rightarrow Wc$ process and its higher-order corrections are dominant, the $pp\rightarrow WcX$ production is directly sensitive to the $s$-quark distribution function in the proton at momentum-transfer values on the order of the $W$-boson mass ($m_W$). 

The $s$-quark PDF has been determined by neutrino--nucleon deep inelastic scattering (DIS) 
experiments~\cite{Mason:2007zz,nutev} 
at momentum transfer squared $Q^2\sim 10\GeV^2$ and momentum
fraction $x\sim0.1$.  However, the interpretation of these data is sensitive to the modelling
of $c$-quark fragmentation and nuclear corrections; some 
analyses~\cite{Martin:2009iq,Alekhin:2009ni,Ball:2008by}
indicate that the $s$-quark sea is suppressed relative to the $d$-quark sea 
at all values of $x$ while 
others~\cite{Lai:2010vv} suggest that SU(3) symmetry is restored as $x$ decreases.
A recent joint analysis of inclusive $W$ and $Z$ production
data from ATLAS at $Q^2\sim m_W^2$ and DIS data from HERA has bolstered
the case for an SU(3)-symmetric sea at $x\sim 0.01$~\cite{Aad:2012sb}.  The main result of that analysis was obtained 
under the assumption
that the $s$- and $\overline{s}$-quark distributions are equal.  However, fits to
the neutrino DIS data from NuTeV~\cite{nutev,Martin:2009iq,Ball:2009mk} prefer a small asymmetry between the $s$ and $\overline{s}$ sea. 

The possibility of using \Wca\ events as probes of the strange-quark distribution function has been 
discussed for some time~\cite{baur,giele}. While the cross section for this process was
measured with a precision of 20--30\%\ at the
Tevatron~\cite{cdfwc1,cdfwc2,d0wc}, the large production rates available at the Large Hadron Collider (LHC) provide the first
opportunity for a measurement with sufficient precision to constrain the $s$-quark PDF at $x\sim0.01$.
A measurement of the \Wca\ production cross-section at the LHC was performed recently by CMS~\cite{Chatrchyan:2013uja} and exploited to constrain the $s$-quark PDF in ref.~\cite{Chatrchyan:2013mza}.

This paper presents a measurement of the 
production of a $W$ boson in association with a single charm quark using 4.6~fb$^{-1}$ of 
$pp$ collision data at $\sqrt{s}=7$~\TeV\, collected with the ATLAS detector at the LHC~in 2011. 
In events where a $W$ boson decays to an electron or muon,
the charm quark is tagged either by the presence of a jet of particles containing its
semileptonic decay to a muon (hereafter referred to as a soft muon) or by the presence of a charmed hadron 
reconstructed in the decay modes \DpDec and \DstarDec with $D^{0} \rightarrow K^{-}\pi^{+}$, \DzsatDec or \DzkpppDecNoSp, 
and their charge conjugates.  

The relative sign of the charges of the $W$ boson and the soft muon, or the \De meson\footnote{Throughout this paper, \De refers to $D^{*\pm}$ and $D^\pm$}, is exploited to reduce the
 backgrounds substantially. In \Wca\ production, the final-state $W$ boson is always
 accompanied by a charm quark with charge of opposite sign,
 that is \Wpca\ or \Wmca. The soft muon and the
 \De meson have the same-sign charge as the charm quark and thus a
 charge opposite to the $W$ boson and its corresponding decay
 lepton. Requiring the $W$ boson and the soft
 muon or \De meson to be of opposite charge therefore selects the
 $W+c$ signal with very high purity. Most backgrounds are evenly distributed between
 events with opposite-sign (OS) and same-sign (SS) charge. Therefore,
 an important strategy used in this analysis is to determine the $W+c$ yields by measuring the difference between the number of opposite-sign and same-sign charge events (\OSSS).  Since the kinematics of pair-produced charm and anti-charm quarks are the same, the pair-produced quarks do not contribute to distributions formed from \OSSS\ events.

The integrated and differential cross sections
as a function of the pseudorapidity of the lepton from the $W$-boson decay are measured
for the fiducial region defined by lepton transverse momentum 
$\ptell>20$\,\GeV\ and  pseudorapidity $|\eta^{\ell}|<2.5$,\footnote{ATLAS uses a right-handed coordinate system with its origin
at the nominal interaction point (IP) in the centre of the 
detector and the $z$-axis coinciding with the axis of the beam
pipe. The $x$-axis points from the IP to the centre of the LHC
ring, and the $y$-axis points upward. Cylindrical coordinates
($r$,$\phi$) are used in the transverse plane, $\phi$ being the azimuthal
angle around the beam pipe. The pseudorapidity is defined in
terms of the polar angle $\theta$ as $\eta=-\ln \tan(\theta/2)$.} neutrino transverse momentum $\ptnu > 25$\,\GeV\ and $W$-boson transverse mass
$\mtw > 40$\,\GeV.\footnote{\mtw\ is defined as $\mtw=\sqrt{2\pt^\ell\met(1-\rm{cos}\Delta\phi)}$ where $\Delta\phi$ is the azimuthal separation between the directions of the lepton and the missing momentum in the transverse plane.}   

The two tagging methods are sensitive to different charm-quark kinematic regions and have different dominant systematic uncertainties.
In the analysis referred to as the \Wce\ analysis, the selection requires that a soft muon is associated with a jet reconstructed in the calorimeter with a minimum transverse momentum $\pt>25$\,\GeV\ and pseudorapidity $|\eta|<2.5$.
The cross section is evaluated for the production of a $W$ boson in association with a particle-level jet containing a weakly decaying charmed hadron with $\pt>5$\,\GeV\ and within a cone of $\Delta R=\sqrt{(\Delta \phi)^2 + (\Delta \eta)^2}=0.3$ from the jet axis as described in section \ref{s:phasespace}.
In addition, the cross section is reported in a fiducial region where the above mentioned charmed hadron is required to decay semileptonically to a muon with $\pt^\mu>4$\,\GeV~and pseudorapidity $|\eta^\mu|<2.5$ with $\Delta R<0.5$ from the jet axis.

The analysis referred to as the \WDe{} analysis does not require a reconstructed calorimeter jet and is thus sensitive to charmed hadrons from $c$-quarks at lower transverse momenta. Differential cross sections are measured separately as a function of the 
\DeNoSp-meson transverse momentum \ptD\ and 
of the pseudorapidity of the lepton from the $W$-boson decay $|\eta^{\ell}|$.
The data are then integrated and the fiducial cross section is measured
for \De mesons with $\ptD>8$\,\GeV~and 
$|\etaD|<2.2$.\\
The measurements are performed separately for events with a positively and a negatively charged $W$ boson,
and the ratio 
\begin{displaymath}
\RC \equiv \sigma(W^{+}+\overline{c})/\sigma(W^{-}+c)
\end{displaymath}
is also measured.
All measurements are
compared to predictions of NLO QCD calculations obtained with various PDF sets
and the sensitivity to the choice of PDFs is presented.  \\

%% file: detector.tex
\section{The ATLAS detector}
\label{s:detector}
The ATLAS detector is described in detail in 
ref.~\cite{Aad:2008zzm}. The inner detector (ID) is used to measure the momenta and trajectories of charged particles. The ID has full coverage in the azimuthal angle 
$\phi$ and over the pseudorapidity range
$|\eta|<2.5$. It consists of three subsystems: a silicon-pixel detector, a silicon-strip tracker
(SCT) and a transition-radiation straw-tube tracker (TRT).  These
detectors are located
inside a solenoid  that provides a 2~T axial field.  
The ID barrel (end-caps)
consists of three ($2\times3$) pixel layers, four ($2\times9$) double-layers of single-sided
silicon strips with a 40 mrad stereo angle, and 73 ($2\times160$) layers of
TRT straws. These detectors have position resolutions of typically 10, 17
and 130 $\mu$m in the azimuthal ($r\phi$) coordinate and, 
in the case of the pixel and SCT, 115
and 580 $\mu$m, respectively, for the second measured coordinate ($z$). A track from a particle
traversing the barrel detector has typically 11 silicon hits (3 pixel
clusters and 8 strip clusters), and more than 30 straw hits.

The electromagnetic calorimeters use liquid argon as the
active detector medium. They consist of accordion-shaped 
electrodes and lead absorbers, and are divided into one barrel ($|\eta|<1.475$) and two end-cap components ($1.375<|\eta|<3.2$). The technology used for the hadronic calorimeters varies with $\eta$.
In the barrel region ($|\eta|<1.7$), the detector is made of 
scintillator tiles with steel absorbers.  In the end-cap region 
($1.5<|\eta|<3.2$), the detector uses liquid argon and copper.
A forward calorimeter consisting of liquid argon and tungsten/copper absorbers 
has both electromagnetic and hadronic sections, and extends the coverage to $|\eta| < 4.9$.

The muon spectrometer (MS) contains one barrel and two end-cap air-core toroid magnets, 
each consisting of eight superconducting coils arranged
symmetrically in azimuth, and surrounding the calorimeter. 
Three layers of precision tracking stations, consisting of drift tubes
and cathode strip chambers, provide precise muon momentum measurements over the range
$|\eta|<2.7$. Resistive plate and thin-gap chambers provide muon 
triggering capability over the range $|\eta|<2.4$.

The ATLAS trigger consists of three levels of event selection: 
a first level implemented 
using custom-made electronics, which selects events at a design rate of at most 75\,kHz, 
and two successive software-based higher levels using fast online algorithms for 
the second level and reconstruction software which is close to the offline algorithms for the third level.

%% file: dataandmc.tex
\section{Data and Monte Carlo samples}
\label{s:dataandmc}

This analysis is based on data collected with the \mbox{ATLAS} detector in the year 2011 during periods with stable $pp$ collisions at $\sqrt{s}=7\,\TeV$ in which all relevant detector components are fully operational. The resulting data sample corresponds to an integrated luminosity of $4.6~\rm{fb^{-1}}$ with an uncertainty of 1.8\%~\cite{r:lumi}.

MC samples are used to compute efficiencies, to model kinematic distributions of signal and 
background processes and to interpret the results. The signal is defined to be the production of a $W$ boson in association with a
single charm quark.  Background processes include the production of  $W$+light jets (i.e. light-quark and gluon jets, hereafter referred to as $W$+light), $W+c\overline{c}$ and $W+b\overline{b}$, while the contribution from $W+b$ production is negligible. $Z$+jets (including $\gamma^{*}$+jets), top-quark pairs, single top quarks, dibosons and multijet events also contribute to the background.

The \WDe\ signal events are generated with \PythiaSix\ 6.423~\cite{Sjostrand:2006za} where the \EvtGen~\cite{Lange:2001uf} program is used to model the charm decays. The CTEQ6L1 PDF~\cite{Pumplin:2002vw} is used for all LO MC generators.

$W$ bosons produced in association with $c$-jets, $b$-jets and light jets are generated separately using \Alpgen\ 2.13~\cite{Mangano:2002ea} interfaced to \Herwig\ 6.520~\cite{Corcella:2002jc} for the parton shower and hadronisation, as well as \Jimmy\ 4.31~\cite{Butterworth:1996zw} for the underlying event. Exclusive samples with zero to four additional partons and an inclusive sample with five or more additional partons are used.  Overlaps between different \Alpgen\ samples with heavy-flavour quarks originating from the matrix element and from the parton shower are removed. In addition, the MLM~\cite{Mangano:2001xp} matching scheme is applied to remove 
overlaps between events with the same parton multiplicity generated by the matrix element or
the parton shower.

A dedicated sample generated with \Alpgen\ and \PythiaSix\ for 
the parton shower and hadronisation is used for the $W$ boson plus $c$-jet signal process. 
In this sample, the fragmentation fractions are 
reweighted to those derived from the combination of measurements 
in $e^+e^-$ and $ep$ collisions \cite{r:ffrac}, 
the momentum fraction of $c$-hadrons is reweighted to that given by \Herwigpp\ 2.6.3~\cite{Bahr:2008pv}, 
the semileptonic branching ratios of $c$-hadrons are 
rescaled to the world average values \cite{PDG} 
and the distribution of the momentum of outgoing muons in the $c$-hadron rest frame is reweighted to that provided by \EvtGen. 

Inclusive $W$ production is generated using 
the \PowHeg\ r1556~\cite{Nason:2004rx, Frixione:2007vw, Alioli:2010xd, Alioli:2008gx} generator 
interfaced to \PythiaSix\ for parton shower, hadronisation and 
underlying-event modelling. For systematic studies, samples generated using \PowHeg\ 
or \MCatNLO\ 4.01~\cite{Frixione:2002ik}, where the parton shower and hadronisation are modelled 
by \Herwig{} and the underlying event 
by \Jimmy, are used. The \CT~\cite{Lai:2010vv} PDF is used for the 
NLO matrix-element calculations, while showering is performed with the CTEQ6L1
PDF. 

Background from $Z$+jets events is generated with \Alpgen\ interfaced 
to \Herwig\ and \Jimmy{} using the same configuration as for $W$+jets events. 
For the diboson backgrounds ($WW$, $WZ$ and $ZZ$), MC samples generated with \Herwig\ are used. 
The $t\overline{t}$ background is obtained from the \MCatNLO\ generator with \Herwig\ used for the parton shower and hadronisation, 
while single-top production is based on the \textsc{Acer} 3.7~\cite{Kersevan:2004yg} MC generator (interfaced to \PythiaSix) 
in the $t$-channel, and \MCatNLO\ in the $s$-channel and for associated production with a $W$ boson.
When \PythiaSix{} or \Herwig{} is used, \Tauola~\cite{Jadach:1993hs} and \Photos~\cite{Golonka:2005pn} are employed to model the decay of $\tau$-leptons and the radiation of photons, respectively.

The background processes are normalised to NNLO predictions for inclusive $W$, $Z$ and 
$t\overline{t}$ production~\cite{Anastasiou:2003ds, Cacciari:2011hy} and to NLO predictions for the 
other processes~\cite{Campbell:2011bn, Campbell:2004ch}. 
The properties of the multijet background events are determined using data-driven techniques. 

Multiple $pp$ collisions per bunch crossing (pileup) are modelled by 
overlaying minimum-bias events generated using \PythiaSix\ with the hard process. 

The MC events are passed through a detailed simulation of the ATLAS detector response~\cite{Aad:2010ah} based on \textsc{GEANT4}~\cite{Agostinelli:2002hh}.

%% file: wselection.tex
\section{Object reconstruction and $\boldsymbol{W}$-boson selection}
\label{s:wselection}
$W$ bosons are reconstructed in their leptonic decay channels, 
$W\rightarrow e\nu$ and $W\rightarrow\mu\nu$.  While the
\WDe\  and \Wce\ analyses share a common selection strategy for the $W$ bosons, 
the selection requirements are optimised separately for the two analyses to
better suppress the respective backgrounds.
The measured cross sections for the two analyses
are extrapolated to a common fiducial region of $W$-boson kinematics (see section~\ref{s:unfolding}).

Data used for this analysis are triggered either by a single-muon
trigger with a requirement on the pseudorapidity of $|\eta^\mu|<2.4$
and on the transverse momentum of $\ptmu>18$\,\GeV,
or by a single-electron trigger with pseudorapidity coverage of $|\eta^e|<2.47$
and a threshold for the
transverse momentum $\pt^e$ of 20\,\GeV\ or 22\,\GeV, depending on the data-taking period.

Events are required to have at least one vertex. The vertex with the highest sum of the squared transverse
momenta of the associated tracks is selected as the primary vertex. Jets are reconstructed with the FastJet package~\cite{bib:antikt3} which uses the infrared- and collinear-safe anti-$k_t$ algorithm~\cite{bib:antikt2} with radius parameter $R=0.4$.
The input from the calorimeter is based on three-dimensional topological clusters~\cite{bib:topo} and
jet energies are calibrated using the EM+JES scheme~\cite{jetcalib1}.
The presence of neutrinos is inferred from the missing
transverse momentum. The magnitude (\met) and azimuthal direction are measured from the vector sum of the transverse
momenta of calibrated physics objects~\cite{r:MetPerf}. Low-\pt\
tracks are used to recover soft particles which are not measured in the
calorimeters~\cite{Aad:2012re}. 

Exactly one lepton fulfilling the isolation requirements discussed below is allowed in each
event. Events with additional isolated electrons or muons are vetoed to
suppress background from $Z$ and $t\bar{t}$ events. The selection
applied to veto leptons is looser than the one used
for signal leptons to ensure higher background rejection.  
Trigger and reconstruction scale factors
are applied to the MC simulation so that the simulation efficiencies match those measured in data.

\subsection[$W\rightarrow e\nu$]{$\boldsymbol{W\rightarrow e\nu}$}

Electrons with transverse momentum $\pte>25$\,\GeV\ and in the pseudorapidity range $|\eta^e|<2.47$, excluding the
calorimeter transition region $1.37<|\eta|<1.52$, are selected. 
Electrons are required to
satisfy the ``tight'' identification criteria described in ref.~\cite{Aad:2014fxa}. 

In the \WDe\ analysis, stringent requirements on the electron isolation and track impact parameter are applied in order to reject electrons from multijet background events. The \WDe\ analysis uses track- and calorimeter-based isolation requirements: the track-based isolation criterion requires the sum of transverse momenta, $\Sigma_{\Delta R < 0.4} \pt^{\rm track}$, of tracks with $\pt^{\rm track}>1$\,\GeV\ within a cone of radius $\Delta R=0.4$ around the electron's direction to be less than 10\% of the electron's transverse momentum. The track associated with the electron is excluded from the calculation of the isolation requirement. The calorimeter-based isolation requirement is analogously defined as the sum of transverse energies in the calorimeter cells (including electromagnetic and hadronic sections, and excluding contributions from the electron itself), $\Sigma_{\Delta R < 0.4} E_{\rm T}^{\rm cells}$, and is required to be less than 40\% of \mbox{\ptem}. To improve the measurement of the electron track parameters, tracks associated 
with the electron are refitted by the Gaussian Sum Filter described in ref.~\cite{r:gsf} which takes into account bremsstrahlung effects. The transverse impact parameter significance is required to be \mbox{$|d_0|/\sigma(d_0)<3$}, where $\sigma(d_0)$ is the uncertainty on the measured $d_0$, and the longitudinal impact parameter is required to be $|z_0|<1$\,mm.

The \Wce\ analysis uses only a calorimeter-based isolation
requirement,  $\Sigma_{\Delta R < 0.3} E_{\rm T}^{\rm cells}<3$\,\GeV. The electron must
be separated by $\Delta R>0.5$ from any jet.
In both analyses, a minimum \met\ of 25\,\GeV\ and a minimum \mtw\, of 40\,\GeV\ are
required.

\subsection[$W\rightarrow \mu\nu$]{$\boldsymbol{W\rightarrow \mu\nu}$}
Muon candidates are formed from associated tracks in the ID
and the MS that are combined using a $\chi^2$-matching
procedure~\cite{atlasmuons,atlasmuonreco}. 
Muons are required to have
$\ptmum > 20$\,\GeV{} and a pseudorapidity range $|\eta^\mu|<2.4$.  
The set of ID hit requirements described in \cite{ATLAS-CONF-2011-063} together with an additional condition of at least one hit in the first pixel layer is applied to select high-quality tracks.

The \WDe\ analysis uses a selection on the longitudinal impact parameter 
$|z_0|<1$\,mm to remove muons from cosmic-ray background. In order to suppress background from heavy-flavour decays, the muon is required to have a transverse impact parameter satisfying $|d_{0}|/\sigma(d_0)<2.5$ and track- and calorimeter-based isolation requirements are applied: $\Sigma_{\Delta R < 0.4} \pt^{\rm track}<0.1\cdot\ptmu$ and $\Sigma_{\Delta R < 0.4} E_{\rm T}^{\rm cells}<0.2\cdot\ptmu$, where the contribution from the muon itself is excluded.

The \Wce\ analysis also uses a combination of track- and calorimeter-based isolation: \mbox{$\Sigma_{\Delta
  R < 0.3} \pt^{\rm track}$} is required to be less than 2.5\,\GeV\ and
$\Sigma_{\Delta R < 0.2} E_{\rm T}^{\rm cells}$ to be less than 4\,\GeV. Additionally, muons must be separated by $\Delta R>0.4$ from any jet with
$\pt>25$\,\GeV.

The \WDe\ analysis requires a minimum \met\ of 25\,\GeV\ and \mtw\ is required to be greater
than 40\,\GeV. A lower \met\ threshold of 20\,\GeV\ is applied in the \Wce\ analysis. However, a more stringent \mtw\ requirement of 60\,\GeV\ is imposed to improve the suppression of the multijet background.

%% file: wd.tex
\section{Event yields for $\boldsymbol{\WDe}$ final states}
\label{s:analysis}
\label{s:wd}

The measurement of the cross section for the production of a single charmed meson in association
with a $W$ boson proceeds as follows.  First, leptonic $W$-boson decays are selected as described in section~\ref{s:wselection}.  The inclusive $W$
yield is then determined for each lepton species and each charge separately.  Second, in events satisfying the $W$ selection criteria, \De mesons are reconstructed for the decay modes \DpDec and \DstarDec with
\DzsDecNoSp, \DzsatDecNoSp, and \DzkpppDecNoSp. 
Third, distributions of the mass (for the \Dpm) or the mass difference $\Delta m =m(D^*)-m(D^0)$ (for the \Ds) are formed in the OS and SS samples.
Fourth, the charge correlation between the lepton and the \De meson is exploited to extract the 
single-charm component by forming the \OSSS{} mass or mass difference distributions.
Fifth, the \WDe\ yield is extracted for each decay mode in bins of \ptDNoSp\ or $|\eta^\ell|$ by fitting the mass or mass difference distribution in \OSSS\ events after combining the electron and muon channels. The raw yields 
are then corrected for detector acceptance, reconstruction
efficiency and migration due to finite \ptD\ resolution as discussed in section~\ref{s:unfolding}.
The corrected results are presented as ratios of the \WDe\ production cross section to the inclusive $W$ production cross section, that is \xsecratioNoSp.  
In addition, the cross-section ratio $\RC$ is calculated.   

Finally, the inclusive $W\rightarrow e \nu$ and $W\rightarrow \mu \nu$ cross sections, which are measured with different fiducial requirements, are corrected to a common kinematic region and combined. The inclusive $W$ cross section is used to obtain fully corrected measurements of \xsecWDNoSp.

\subsection [Determination of inclusive $W$ yields]{Determination of inclusive $\boldsymbol{W}$ yields}
\label{s:wdyield}
The inclusive $W\rightarrow\mu\nu$ and  $W\rightarrow e\nu$ yields are determined from the number of data events passing the event selection after subtracting the expected background contributions. The backgrounds from real leptons are estimated from simulation. Among these sources of backgrounds, the contributions from the $W\rightarrow\tau\nu$ and $Z\rightarrow\ell\ell$ channels are dominant and amount to 1.8(2.8)\% and 1.2(3.6)\% of the total yield in the electron(muon) channel, respectively. Contributions from $t\overline{t}$, single-top and diboson channels are small, at the level of 0.4\%. 

The background from multijet events is important due to the large production cross section. In the electron channel, multijet events pass the electron selection due to misidentified hadrons, converted photons and semileptonic heavy-flavour decays. In the muon channel, muons from heavy-flavour decays as well as decays in flight of pions and kaons are the dominant sources. Since predictions for the normalisation and the composition of these backgrounds suffer from large uncertainties, data-driven techniques are applied. To minimize systematic uncertainties arising from correlations between \met{} and \mtw{} in the multijet background, the signal yields are derived from binned maximum-likelihood fits~\cite{r:barlow} of templates to the \met\ distributions in a expanded sample obtained by removing the requirements on \met\ and \mtw. These fits discriminate the electroweak and top-quark processes from the multijet background. $W$ yields in the signal region are determined from the fitted yield at $\met>25$\,\GeV\ after removing electroweak and top-quark background events and applying a small correction for the efficiency of the \mtw{} requirement\footnote{The efficiency of the $\mtw>40$\,\GeV\ requirement in signal events is determined from simulation and amounts to 98\%.}. 

\begin{figure*}
\begin{center}
\includegraphics[width=0.42\textwidth]{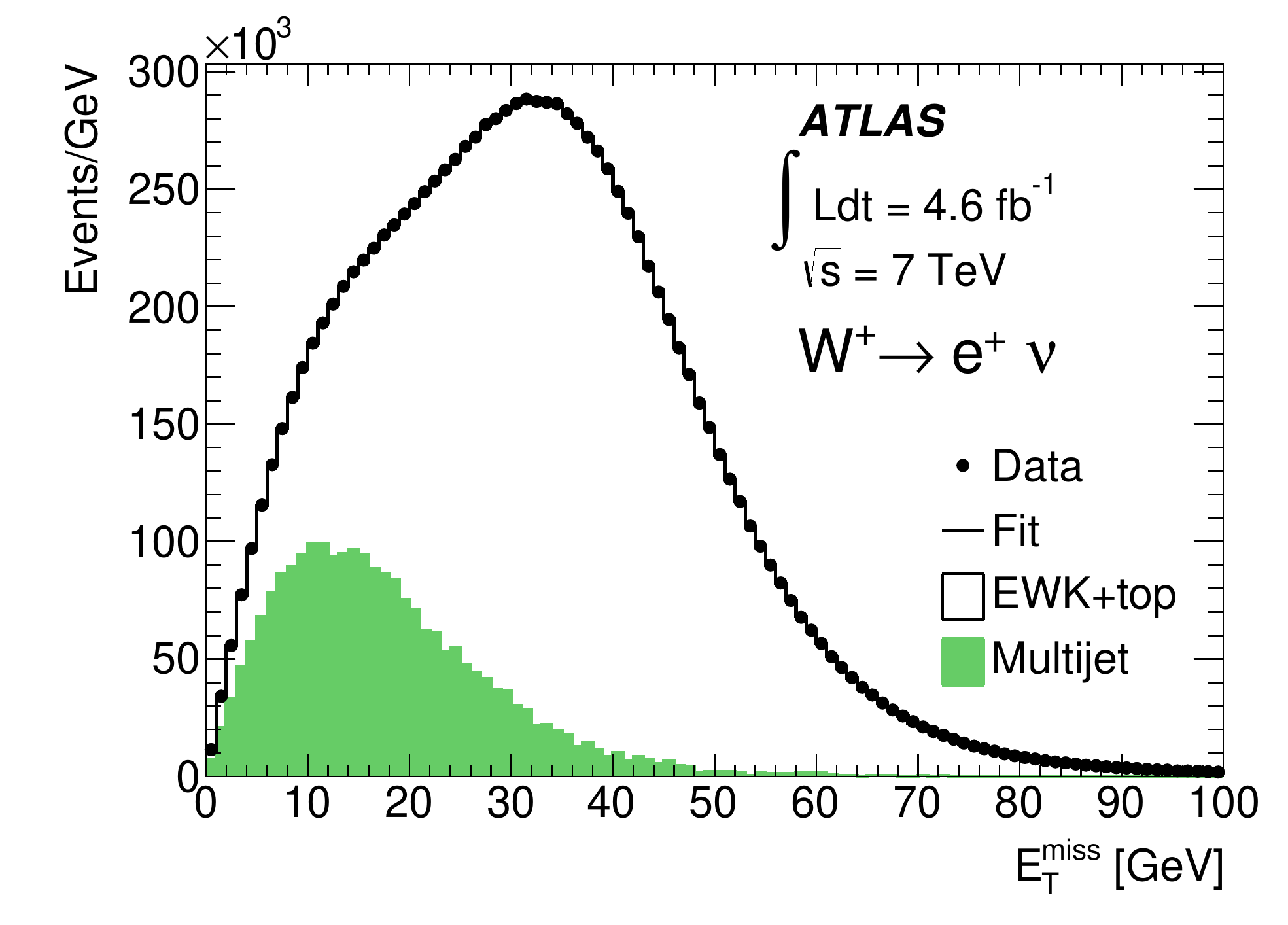}  
\includegraphics[width=0.42\textwidth]{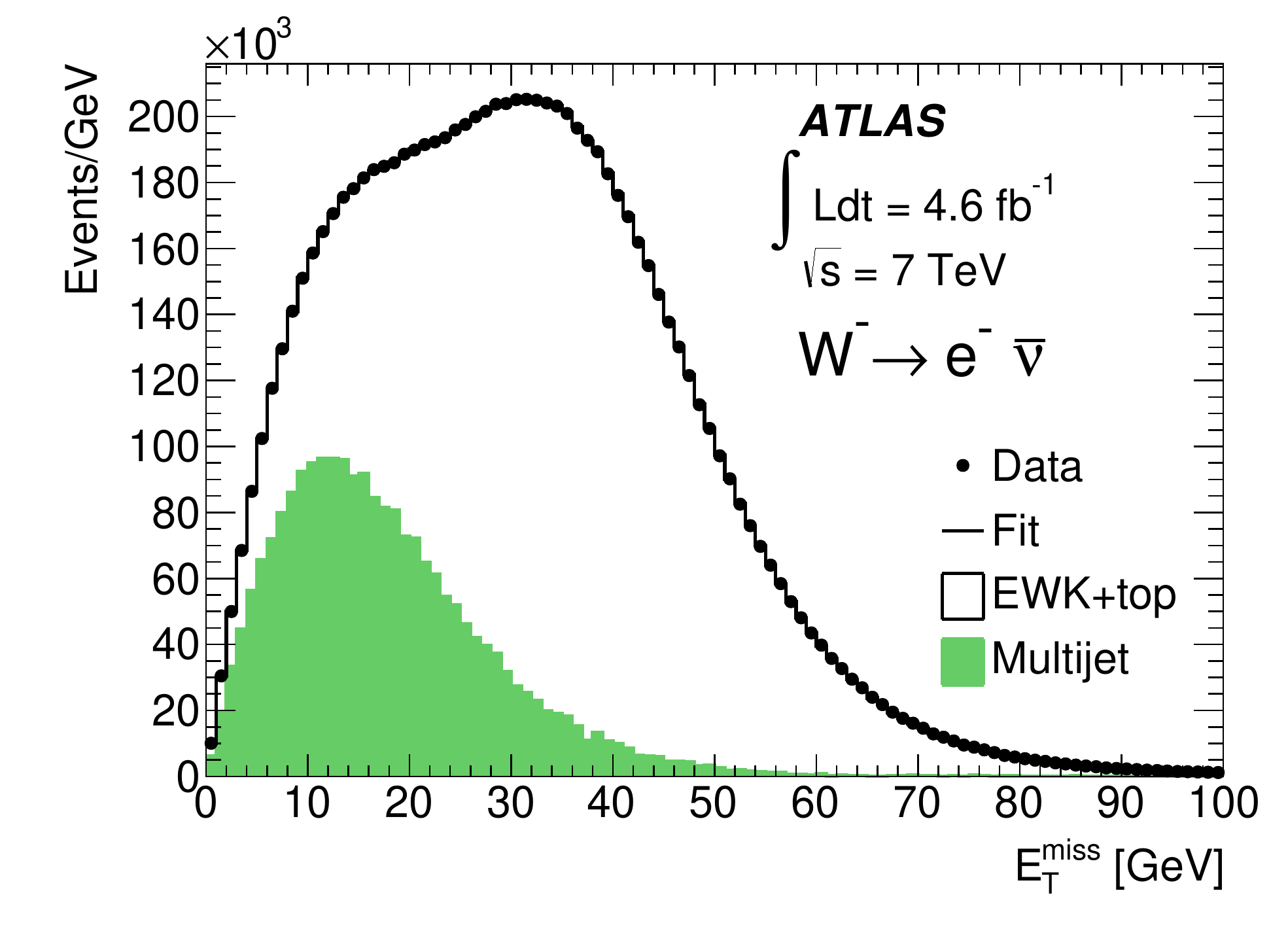}\\    
\includegraphics[width=0.42\textwidth]{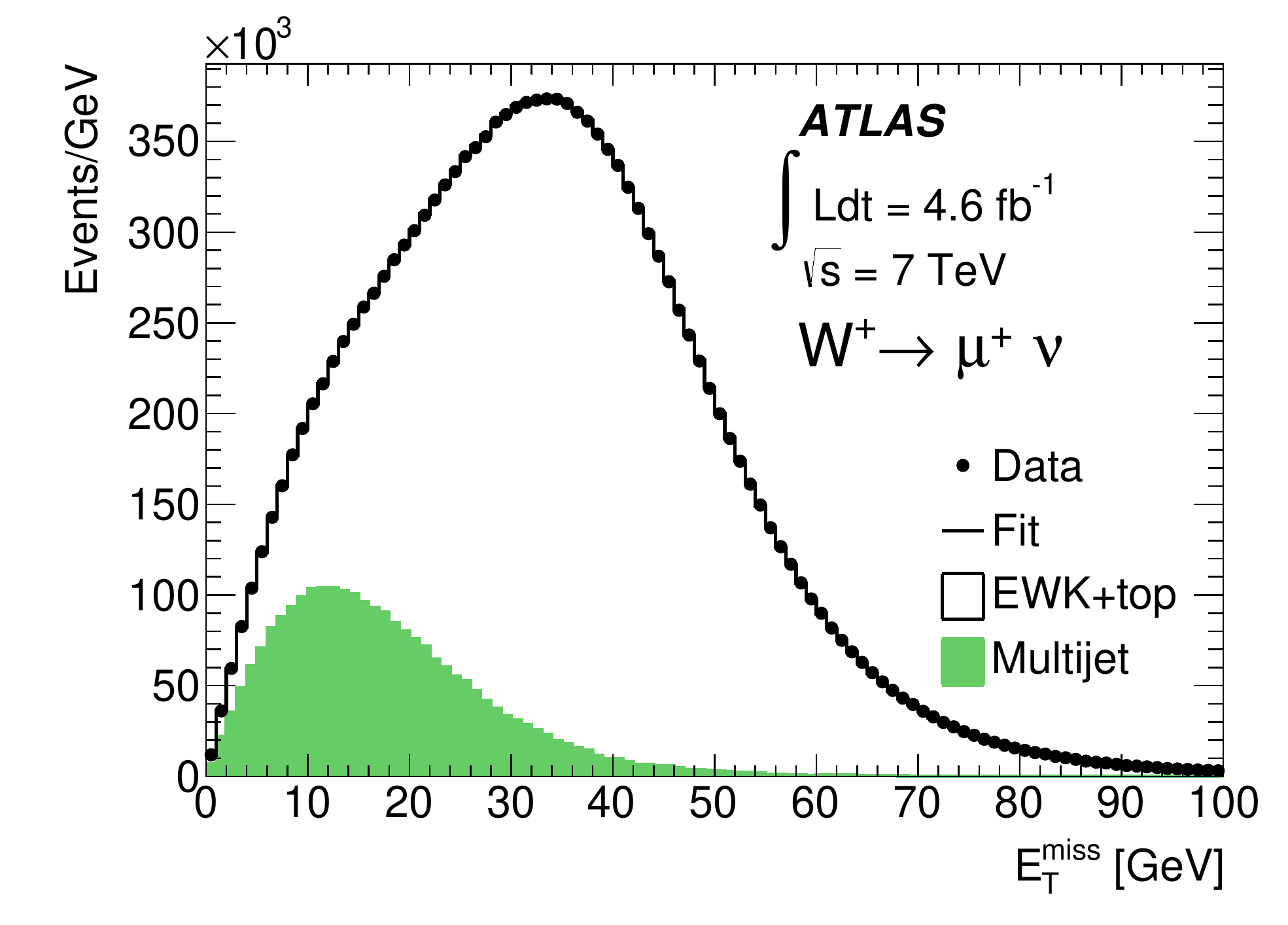}  
\includegraphics[width=0.42\textwidth]{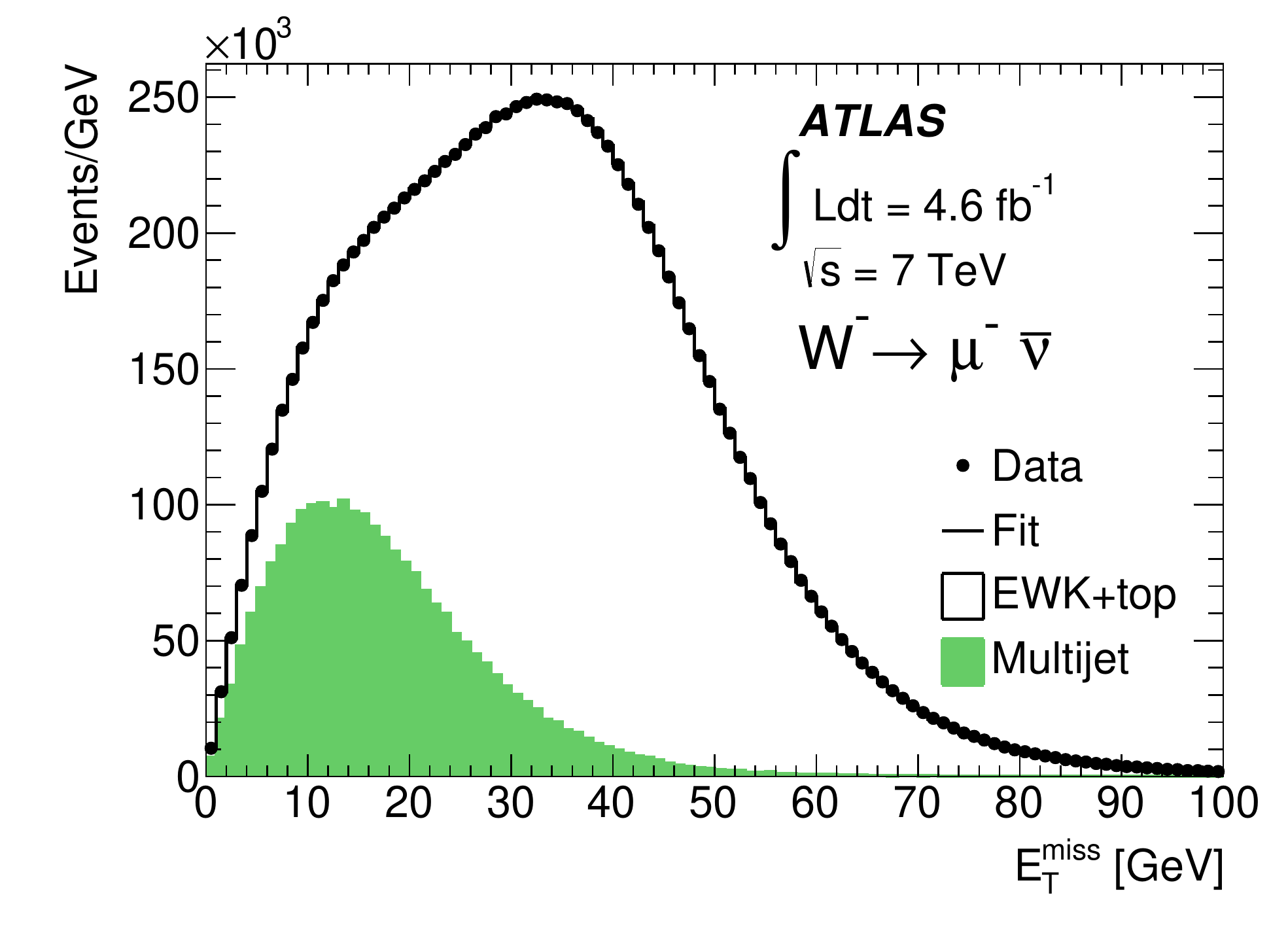}
 \caption{Result of the binned maximum-likelihood fit to the \met\ distribution for $W^+\rightarrow e^+\nu$ (top left),  $W^-\rightarrow e^-\overline{\nu}$ (top right),  $W^+\rightarrow\mu^+\nu$ (bottom left) and $W^-\rightarrow\mu^-\overline{\nu}$ (bottom right). The data are shown by filled markers and the fit result by the solid line. The multijet template, normalised according to the fit result, is shown by the filled histogram. The shape of the distribution for the electroweak and top-quark processes is obtained from simulation. Electroweak processes include $W$, $Z$ and diboson processes.}
\label{fig:wincl}
\end{center}
\end{figure*}

The templates for electroweak and top-quark events are taken from simulation and the relative normalisations of the individual processes are fixed according to the corresponding cross sections. The multijet templates are derived from data in a control region defined like the signal region except that the lepton transverse impact parameter requirement is reversed, the lepton transverse momentum requirement is lowered, and no requirement is imposed on \met{} or \mtw{}. This selection defines a region orthogonal to the signal region and enriches the heavy-flavour content in the sample. The contamination from electroweak and top-quark events in the control region is estimated from MC simulation and subtracted. The resulting fits, presented in figure~\ref{fig:wincl}, show that the template predictions provide a good description of the data. The stability of the fit result is verified by repeating the fits with varied bin width and fit ranges.

The size of the multijet background in the signal region is 3.1(4.2)\% for $W^+(W^-)$ in the electron channel and 2.0(3.0)\% for $W^+(W^-)$ in the muon channel. The reconstructed inclusive $W$ yields after background subtraction are $6.9\cdot10^6$ for $W^+\rightarrow e^+\nu$, $4.7\cdot10^6$ for $W^-\rightarrow e^-\overline{\nu}$, $9.3\cdot10^6$ for $W^+\rightarrow\mu^+\nu$ and $6.0\cdot10^6$ for $W^-\rightarrow\mu^-\overline{\nu}$. The maximum-likelihood fits are also performed in four bins of $|\eta^\ell|$.

Systematic effects arising from the modelling of the electroweak and top-quark event template shapes are investigated by studying the dependence of the fit result on the choice of generator and PDF set. The uncertainty in the multijet template shapes is determined using alternative control region definitions in data and in simulation. Systematic uncertainties on the extrapolation to the signal region are assessed by comparing the efficiencies of the \mtw\ requirement produced by different generators and different hadronisation and underlying event models. The systematic uncertainty due to the MC-based subtraction of the electroweak and top-quark backgrounds is taken into account by propagating the corresponding cross-section uncertainties. These studies result in a relative systematic uncertainty with respect to the number of $W$ candidates of 2.6(0.8)\% for the electron(muon) channel and 1.3\% for the combined measurement.

\subsection[$D$-meson selection]{$\boldsymbol{D}$-meson selection}

Charmed hadrons from events passing the $W$ selection are reconstructed in the decay modes \DpDec and \DstarDec with \DzsDecNoSp, \DzsatDec or \DzkpppDec and their charge conjugates. The selection is optimised to maximise the signal-to-background ratio using simulated events. \De candidates are reconstructed from good-quality tracks within $|\eta|<2.5$.
Each track is required to have at least six hits in the SCT and to have a hit in the innermost layer of the pixel detector unless it passes through a region where no hit is expected.
Only tracks with no more than one missing pixel or SCT hit are used. A requirement of $|z_{0}|\sin\theta<15$\,mm is applied to suppress background from pileup events. No requirement on the track impact parameter in the transverse plane is applied to avoid any bias in the selection of tracks coming from a displaced $D$-meson decay vertex.

The $D^0(D^+)$ candidates are reconstructed from tracks with $\pt > 500\;(800)$\,\MeV. The track associated with the lepton from the decay of the $W$ boson is excluded. Sets of tracks with the appropriate charge combination in a cone of $\Delta R=0.6$ are fit to a common vertex hypothesis in three dimensions. All possible pion or kaon mass assignments are considered for the tracks. The vertex fit is required to have $\chi^2<5$. Vertex candidates are retained if their distance to the primary vertex in the transverse plane, $L_{xy}$, is greater than 0 (1)\,mm for \mbox{\Dz($D^+$)} candidates. The transverse impact parameter of the candidate's flight path with respect to the primary vertex is required to fulfil the requirement $|d_0|<1$\,mm. A relaxed requirement of $|d_0|<10$\,mm is applied in the \DzsatDec mode, since, in this channel, the impact parameter distribution is distorted due to the $\pi^0$ meson, which is not reconstructed.
Combinatorial background in the \Dp reconstruction is reduced by requiring $\cos\theta^{\ast}(K)>-0.8$ where $\theta^{\ast}(K)$ is the angle between the kaon in the $K\pi\pi$ rest frame and the $K\pi\pi$ line of flight in the laboratory frame.
\Dp candidates with $m(K\pi\pi)-m(K\pi)<180$\,\MeV\ are rejected\footnote{Here the mass difference is calculated for both pion candidates and the \Dp candidate is rejected if either combination fulfils the requirement.} to suppress background from \Ds\ decays. Background from $D_s^+\rightarrow\phi\pi^+\rightarrow(K^+K^-)\pi^+$ is significantly reduced by removing any \Dp candidate containing a pair of oppositely charged particles with an invariant mass within 8\,\MeV\ of the nominal $\phi$ mass~\cite{PDG} when the kaon mass hypothesis is assumed for both tracks. 

For the \Ds\ channel, suitable ranges are selected in the $m(D^0)$ invariant mass spectrum: 
$m(K\pi(\pi\pi))$ is required to be within 40\,\MeV\ of the nominal \Dz mass~\cite{PDG} for the \DzkporpppDec~channels, 
while the \DzsatDec decay is reconstructed from the satellite peak~\cite{r:goldhaber} at 1.5\,\GeV$<m(K\pi)<1.7$\,\GeV.
An additional track ($\pi_{\rm{slow}}^+$) with $\pt>400$\,\MeV, $\left |d_0\right |<1$\,mm and charge opposite to the kaon is used to form \DstarDec candidates. In order to reduce combinatorial background, an isolation requirement of $\pt(D^0)/\Sigma\pt>0.5$ is imposed where the sum runs over all tracks with $\pt>500$\,\MeV\ in a cone of radius $\Delta R=0.4$ around the $D^0$ candidate. Duplicate removal is performed in the \DzkpppDec channel: if two or more \Ds\ candidates share the same $\pi_{\rm{slow}}^+$ track as well as two tracks from the decay of the associated $D^0$, only the candidate with the mass closest to the nominal \Dz mass is retained. Finally, kinematic requirements of $\ptD>8$\,\GeV\ and $|\etaD|<2.2$ are applied.

If several \De candidates are found in an event, all of them are used  for the inclusive cross-section measurement. They enter the OS or SS samples, depending on the charge of the charmed meson and of the $W$-boson candidate. The fraction of events with multiple candidates is about 10\% for \DpDecNoSp, \DzsDec and \DzsatDec and about 30\% for \DzkpppDecNoSp.

\begin{figure}
\begin{center}
\includegraphics[width=0.46\textwidth]{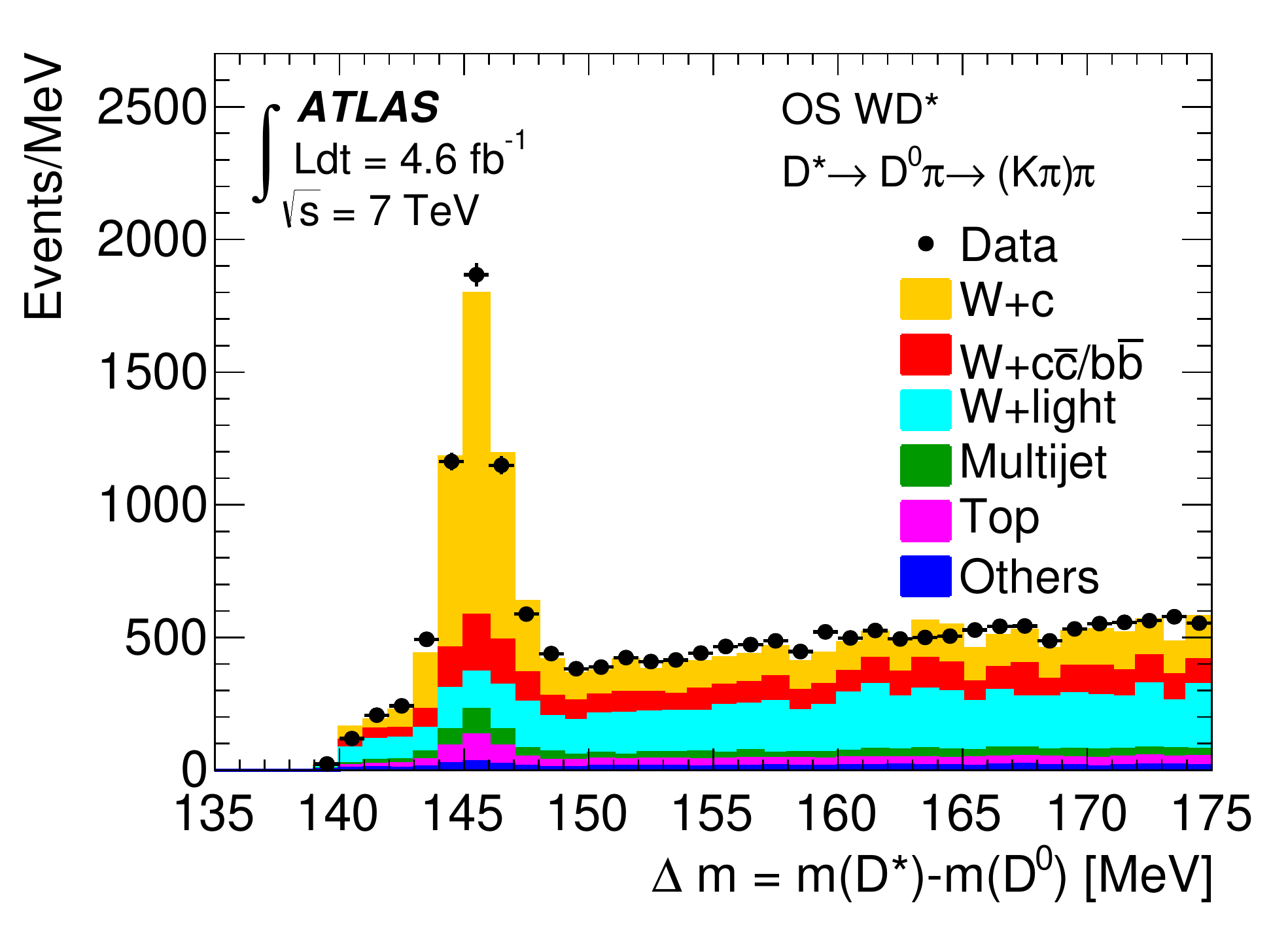}  
\includegraphics[width=0.46\textwidth]{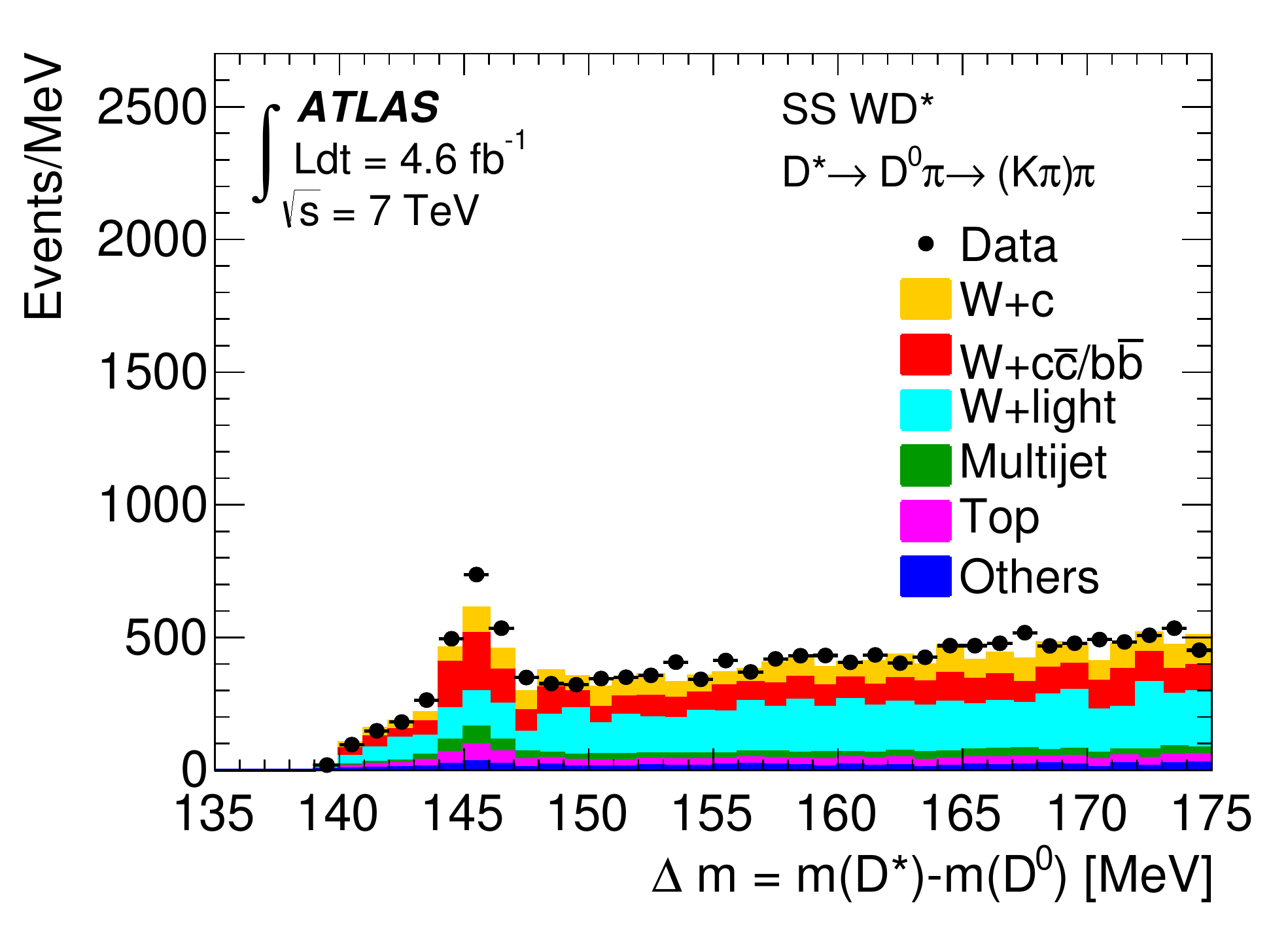} 
 \caption{Reconstructed $\Delta m = m(D^{*})-m(D^0)$ distributions in OS (left) and SS (right) events passing the selection for \DsDecbothNoSp. The data distributions are shown by the filled markers. The coloured stacked histograms represent the predictions for the signal and background processes. The predictions are obtained from simulation, except for the multijet background, which is estimated from the data control region with semileptonic $b$-quark decays. The difference in normalisation between data and predictions is consistent with the uncertainty on the background normalisations.}
\label{fig:osss}
\end{center}
\end{figure}

For illustration, the $\Delta m$ distribution for \DsDec events is shown in figure~\ref{fig:osss} for OS and SS events passing the \WDe\ selection in data and is compared to simulation. The simulation provides a good description of the data.
The difference in normalisation is less than the uncertainty on the normalisations of the simulated backgrounds.
The dominant background is from $W$+light events. All other background sources are small and further reduced after the \OSSS\ subtraction.

\subsection{Determination of \OSSS\ yields}
\label{s:qcdbkginWc}
In order to suppress background contributions, the charge correlation of the \WDe\ signal process is 
exploited by measuring the number of \OSSS\ events ($N^\OSSS$). 
The remaining background after the \OSSS\ subtraction is predominantly $W$+light events 
in which the OS/SS asymmetry is due to the 
correlation of the charge of the $W$ boson and the
associated quark, and to the charge conservation among the
fragmentation products of the quark.
A smaller contribution is from top-quark events and semileptonic decays in $c\overline{c}$/$b\overline{b}$ events with a $D$ meson from the charm hadronisation. The fraction of \WDe\ events with $W\rightarrow\tau\nu$ is determined from simulation to be 2.5\% and is subtracted from the number of signal events. All other backgrounds are negligible after the \OSSS\ subtraction.
The \WDe\ yields in \OSSS\ events are obtained from fits to the mass (for the \Dpm\ meson) or $\Delta m$ (for the \Ds\ meson) distributions.
In the fit, template histograms from a data control region are used to model the signal shape and functional forms are used for the combinatorial background. The data control regions and the fitting methodology are discussed in the following sections. 

In contrast to the combinatorial background, top-quark and heavy-flavour production include real $D$ mesons that produce a peak in the signal region of the mass distribution and requires special treatment. The background from top-quark events is estimated from MC simulation. The background from heavy-flavour production in \OSSS\ events depends on the relative contributions from $c\overline{c}$ and $b\overline{b}$ events and is estimated using a data-driven technique. The peaking backgrounds are subtracted from the fitted \WDe\ yields as outlined below. 

\begin{figure*}
\begin{center}
\includegraphics[width=0.4\textwidth]{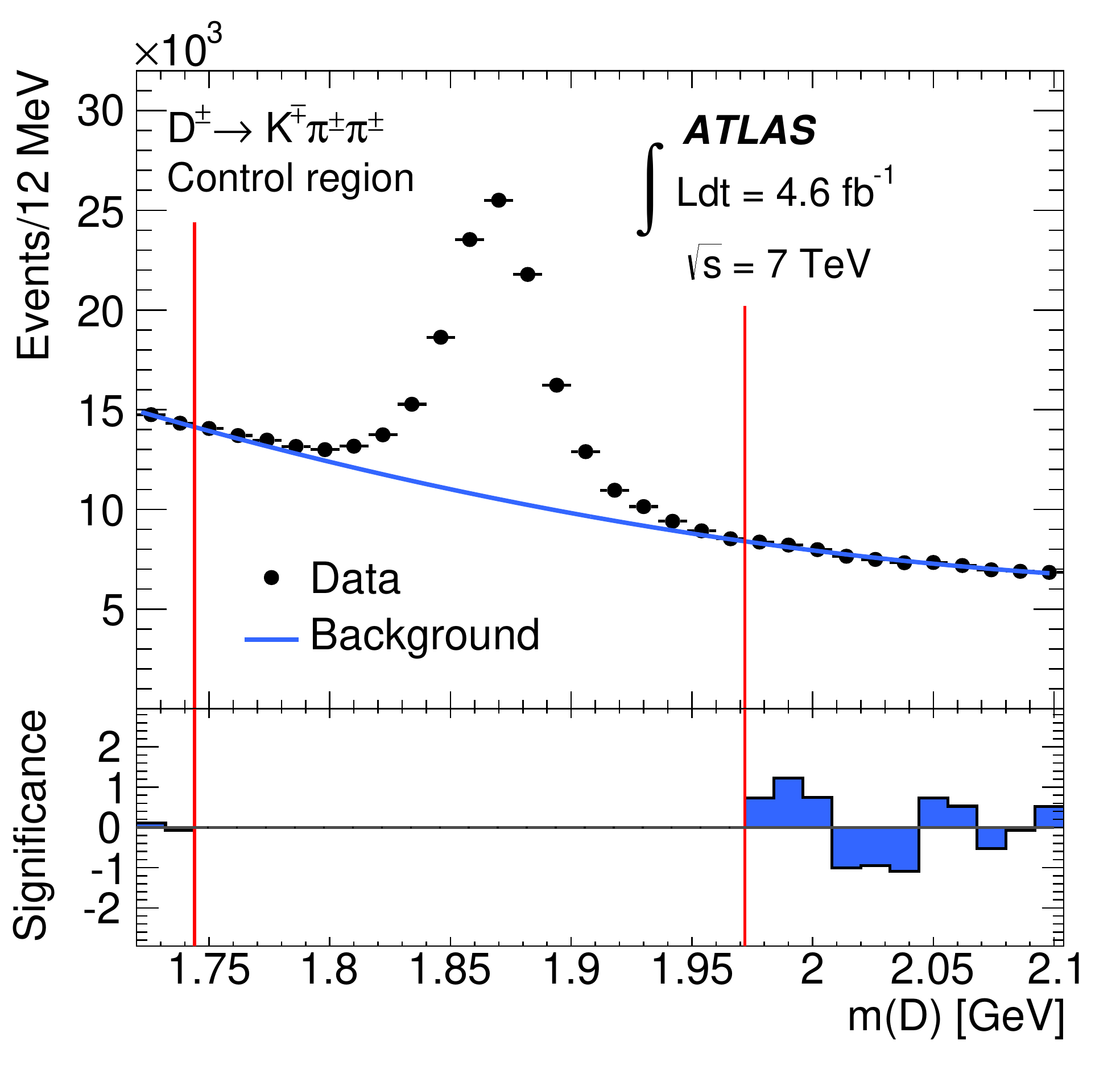}  
\includegraphics[width=0.4\textwidth]{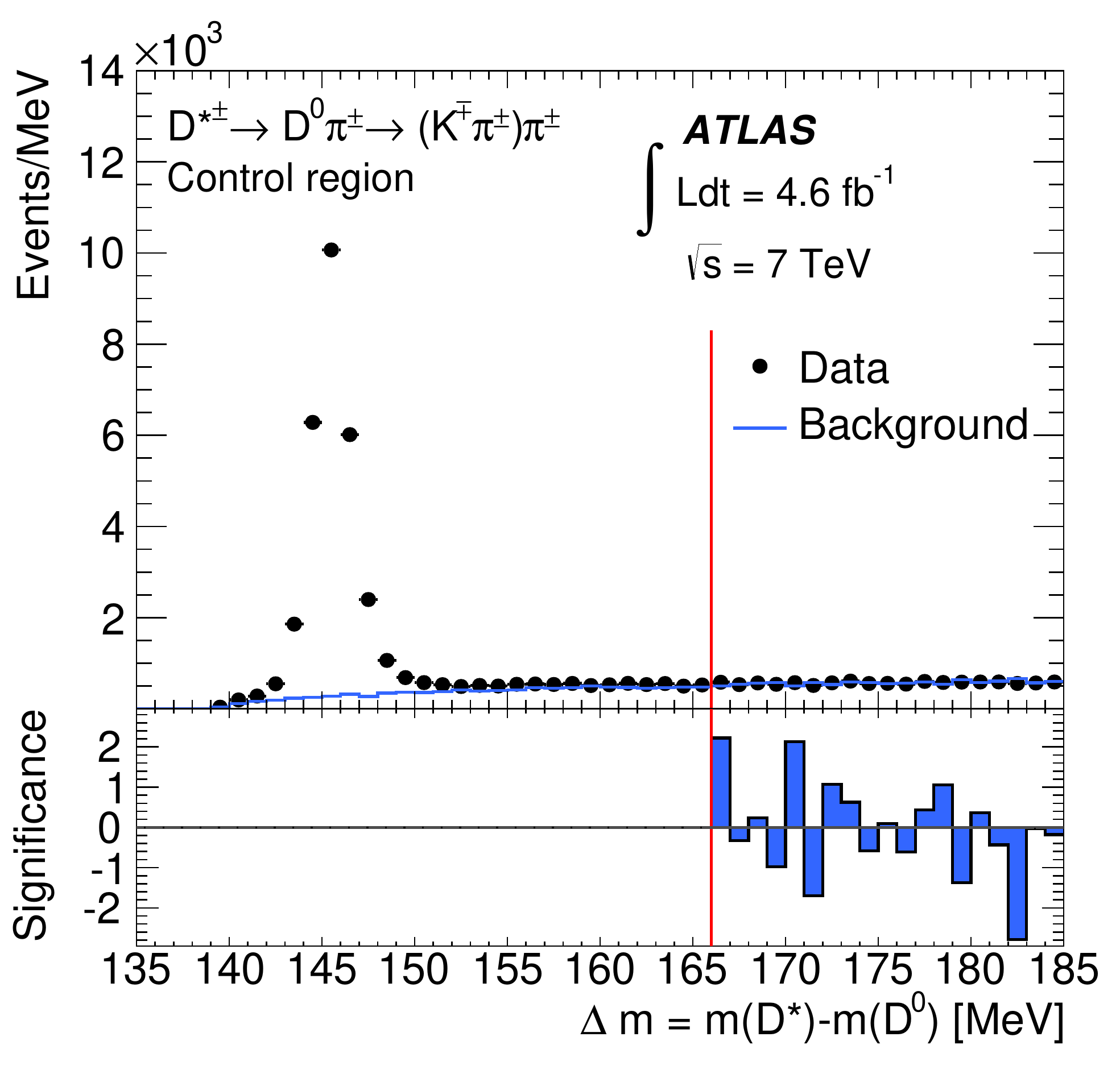} \\  
\includegraphics[width=0.4\textwidth]{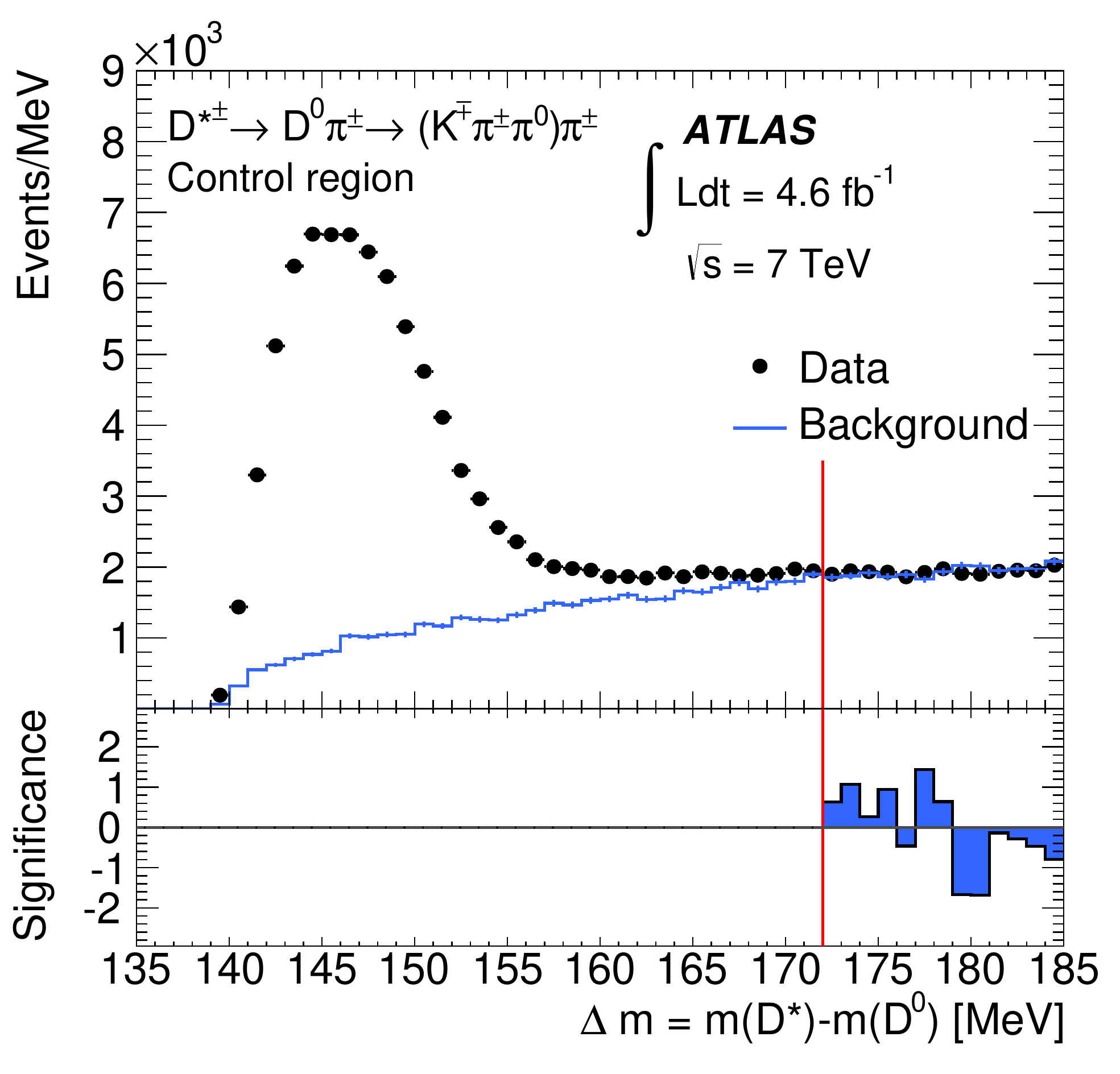}   
\includegraphics[width=0.4\textwidth]{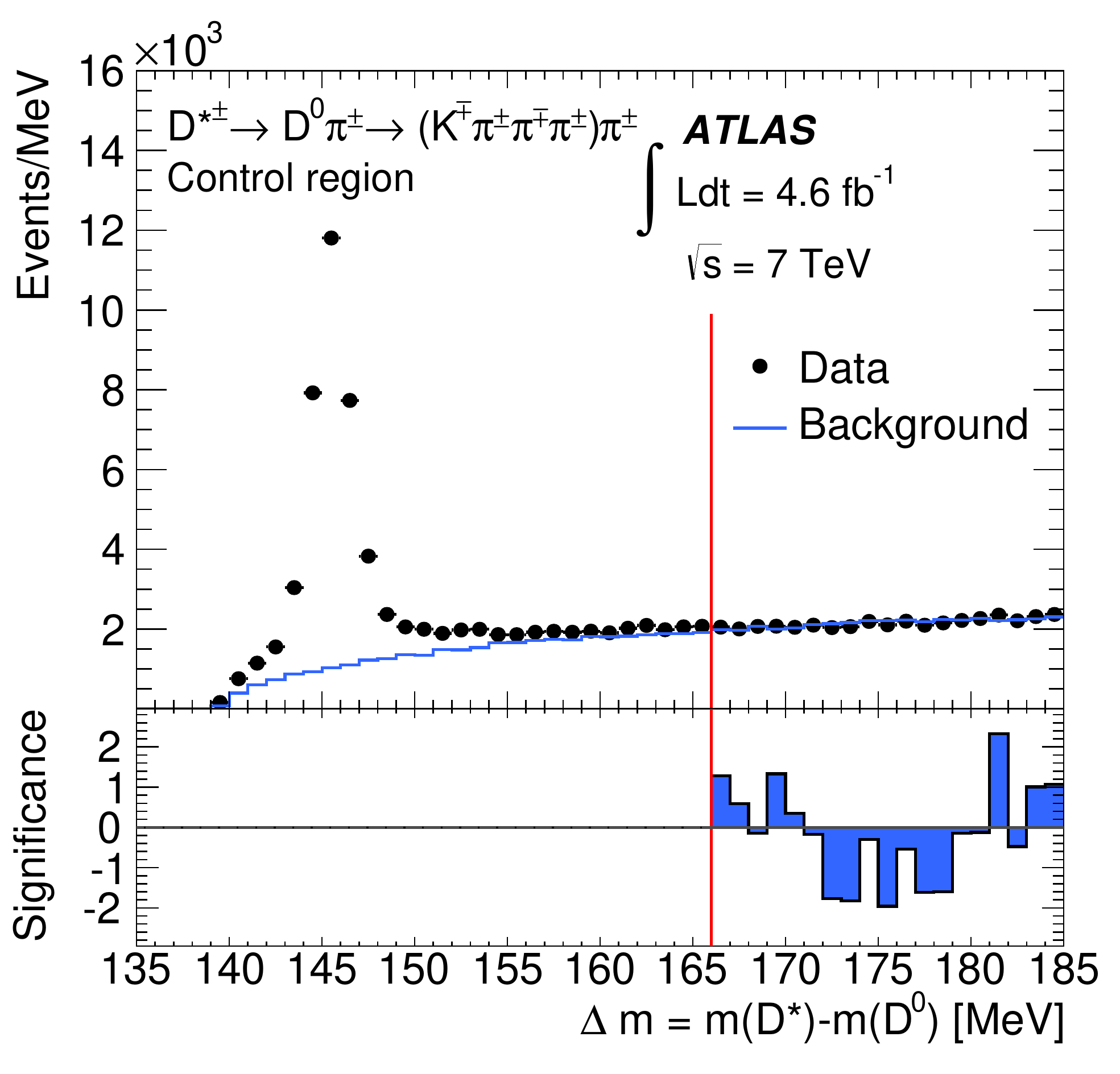}
 \caption{Reconstructed $m(D)$ and $\Delta m = m(D^{*})-m(D^0)$ distributions in the sample with semileptonic $b$-quark decays for \DpDecboth (top left), \DsDecboth (top right), \DsatDecboth (bottom left) and \DkpppDecboth (bottom right). The data distributions are shown by the filled markers. The solid line represents the shape of the combinatorial background that is subtracted in order to obtain the signal templates for the fits to the $WD^{(*)}$ signal region. The combinatorial background is normalised in the sidebands of the $m(D)$ and $\Delta m$ distributions indicated by the vertical lines. The lower panels show the statistical significance of the deviations of the fit from the data in the sideband region.}
\label{fig:semilep}
\end{center}
\end{figure*}

\subsubsection[Fits to $\Delta m$ and $m(D)$ distributions]{Fits to $\boldsymbol{\Delta m}$ and $\boldsymbol{m(D)}$ distributions}
\paragraph{Signal templates}
The signal templates for the $m(D)$ and $\Delta m$ distributions are obtained from a data control region dominated by events with semileptonic $b$-quark decays. Due to the abundant production of $b\overline{b}$ pairs, this control region provides a large sample of $D$-meson decays with similar decay kinematics to the signal events and is thus well suited for this purpose. Semileptonic $b$-quark decays are characterised by non-isolated leptons with lower \pt\ and lower \met\ than in signal events. Furthermore, the decay products of $b$-quarks tend to be collimated within a jet and the average spatial separation between the lepton and the $D$ meson from the $b$-quark decay is thus smaller than in \WDe\ events. The control sample with semileptonic $b$-quark decays is selected in the muon channel. The muon selection follows the signal selection 
but with the isolation and impact parameter requirements removed 
and with a lower muon momentum requirement of
$\ptmu>18$\,\GeV. In addition, the \met\ and \mtw\ requirements are inverted to make the control region orthogonal to the signal region. Other selection criteria are identical to those applied to the \WDe\ signal sample. The muon and the reconstructed \De candidate are required to be within $\Delta R(\mu,D^{(*)})<1$ and to have an invariant mass consistent with a $B$-meson decay, $m(\mu D^{(*)})<5$\,\GeV. Finally, the muon and the \De are required to have opposite charge. 

The mass distributions of reconstructed $D$ candidates in the control region as well as the $\Delta m$ distributions for \Ds\ candidates are shown in figure~\ref{fig:semilep}. The template histograms used for the fit in the \WDe\ signal region are obtained from these distributions by subtracting the combinatorial background. The continuously falling spectrum of the combinatorial background in the $D$ channel is described by a second-order polynomial with parameters extracted from a fit to the mass sidebands.
In the \Ds\ channel, events with a wrong-sign soft-pion track are used to model the shape of the combinatorial background in the $\Delta m$ distribution.
In this configuration, the soft pion has the same charge as the muon and the kaon. 
The normalisation of the combinatorial background is obtained in the region of high $\Delta m$ where the signal contribution is negligible. Appropriate template predictions are obtained for each bin in \ptD\ and for each charge. For the fits in bins of $|\eta^\ell|$, the signal templates are obtained from the inclusive sample. The procedure used to derive the \Ds\ templates from data in events with semileptonic $b$-quark decays removes combinatorial background but can potentially retain events from other $D$-meson decay modes which can produce broad reflections at low $\Delta m$. Since these reflections cannot be discriminated efficiently from the reconstructed signal, they are treated as signal events in the fits. The effect of these reflections on the fitted yields is later included 
as part of the calculation of the reconstruction efficiency.

\paragraph{Shape of combinatorial background}
\label{combbg}
The shape of the combinatorial background in the signal region due to $W$+light production is determined in a control region defined by $L_{xy}<-1\;(0)$\,mm for $D$(\Ds) decays. A functional form is used to fit the shape of the background in the $W$+light control region: a second-order polynomial in the $m(D)$ distribution and a logarithmic function for the background in the $\Delta m$ distribution.
These functions are found to provide a good description of the combinatorial background in all \ptD\ bins; their parameters are determined for each decay mode from a fit 
in the control region and are fixed in the fits to the signal region. 

\begin{figure*}
\begin{center}
\includegraphics[width=0.38\textwidth]{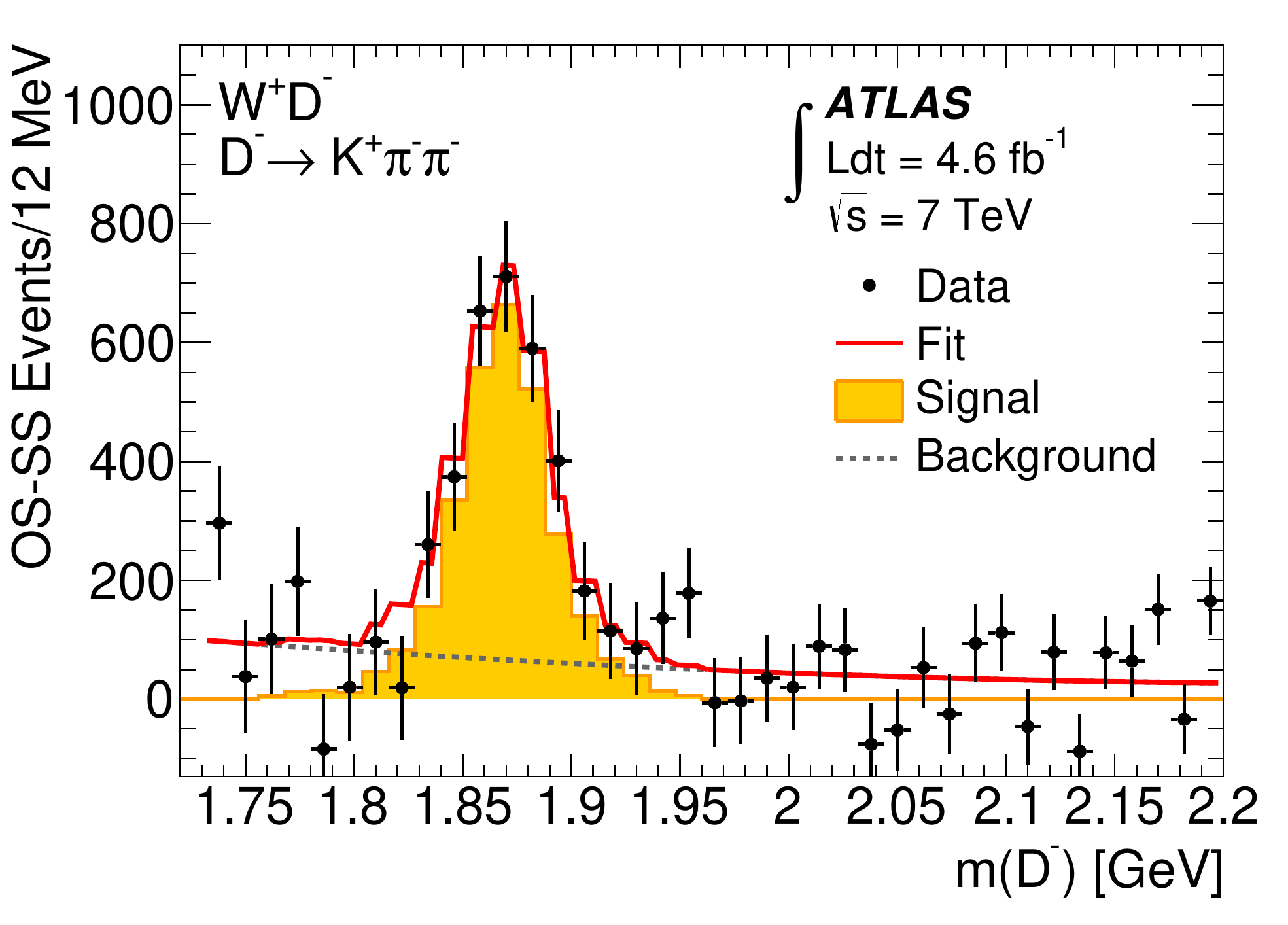}  
\includegraphics[width=0.38\textwidth]{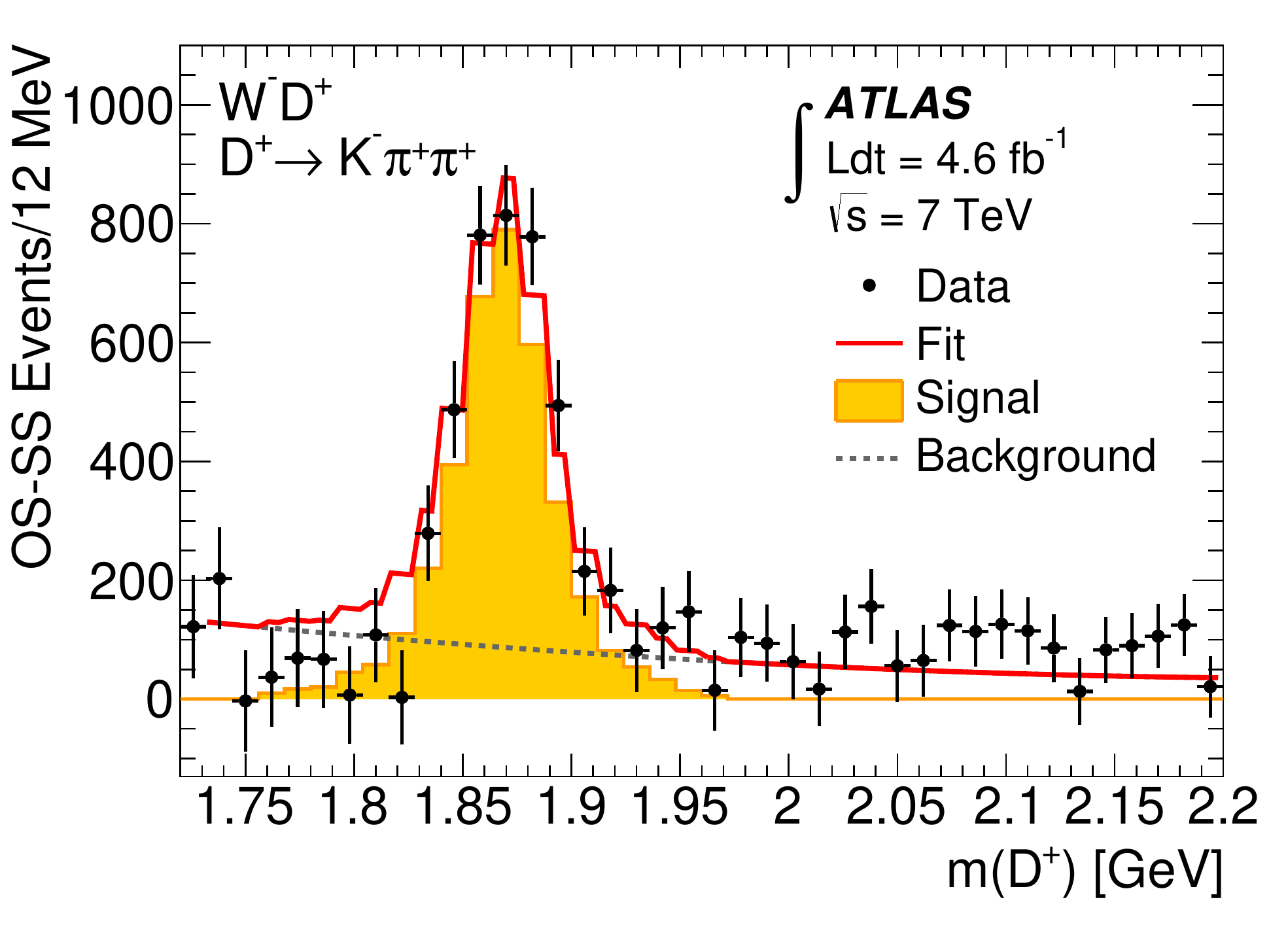}  \\
\includegraphics[width=0.38\textwidth]{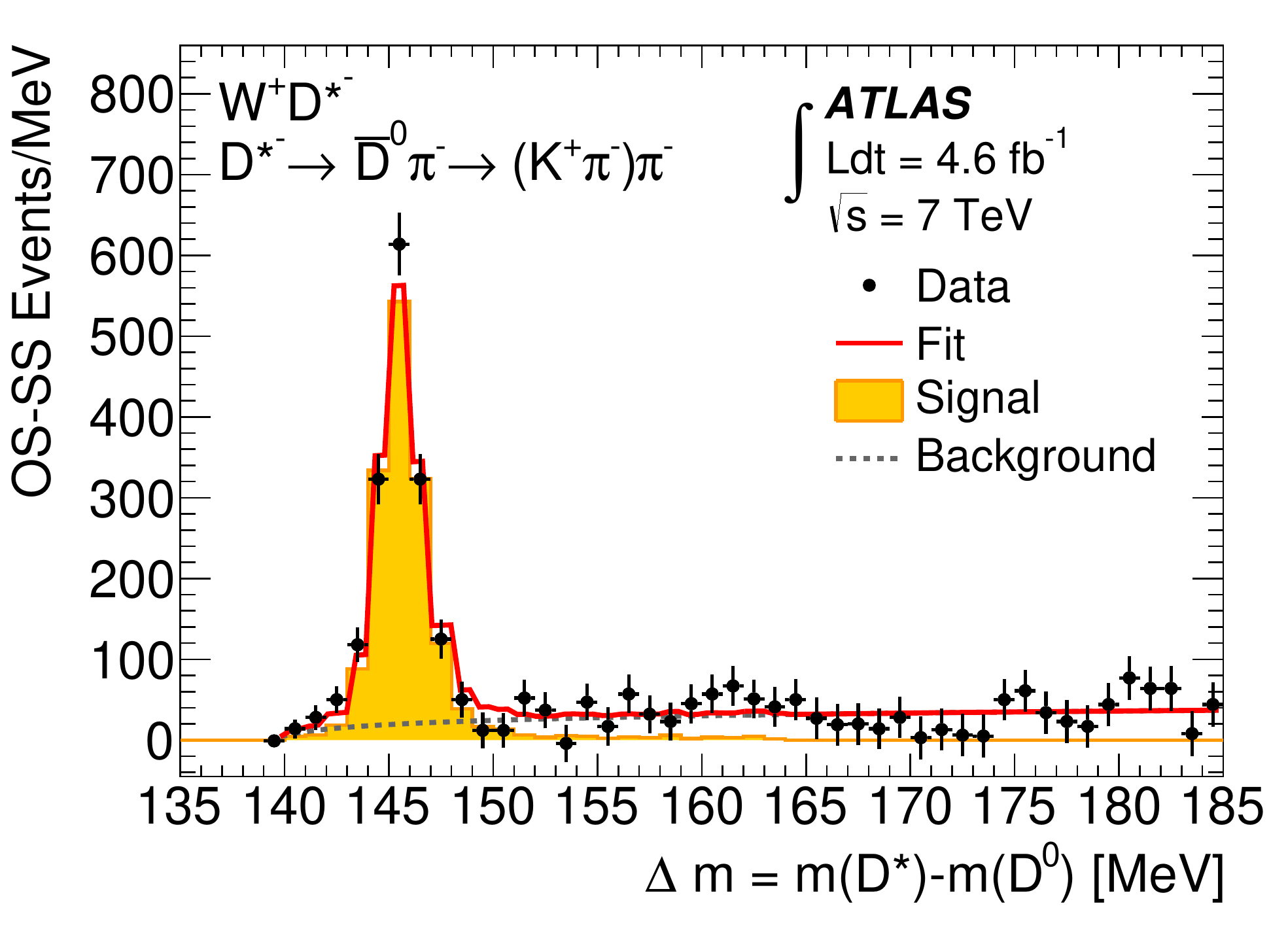} 
\includegraphics[width=0.38\textwidth]{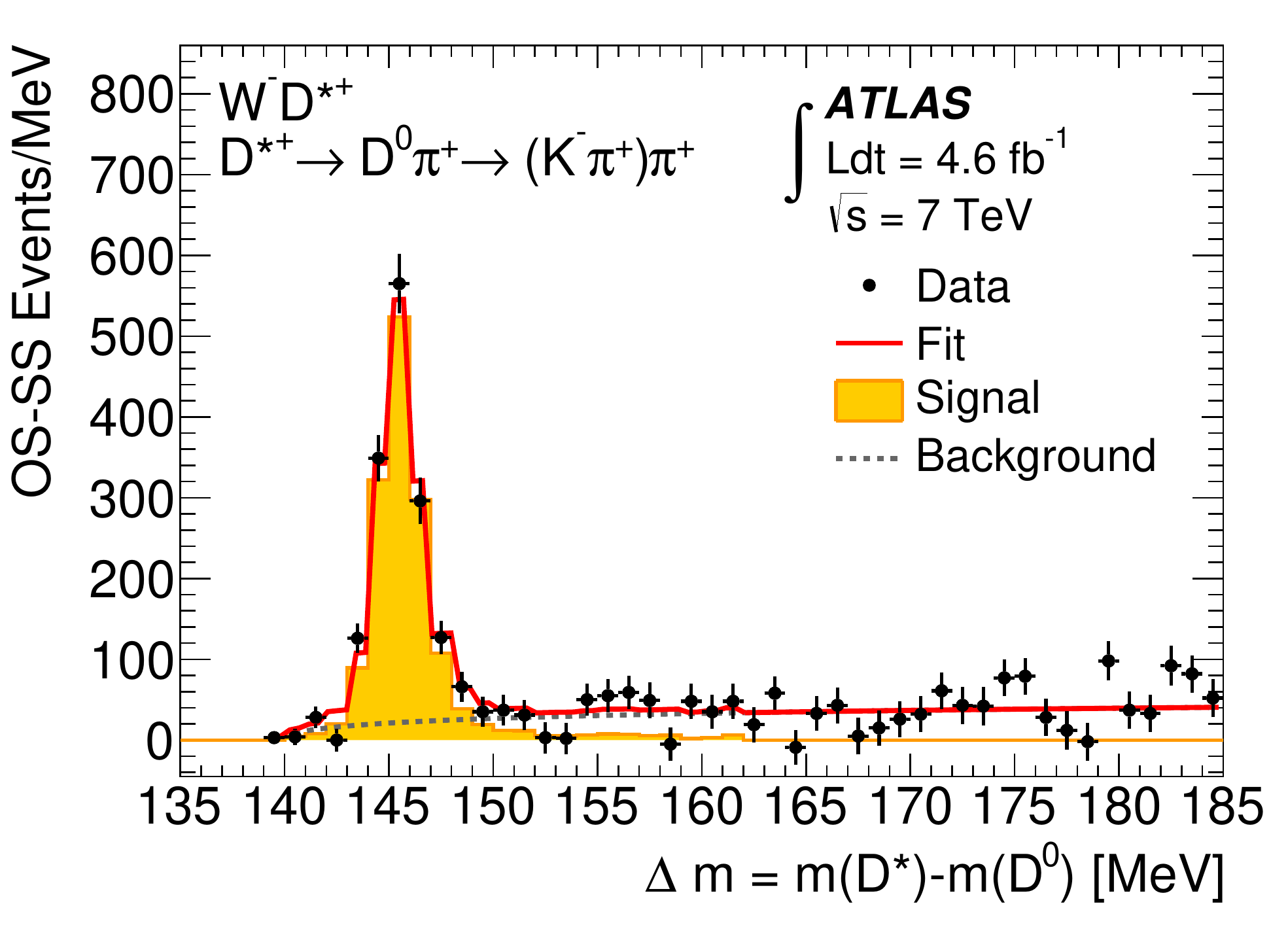} \\
\includegraphics[width=0.38\textwidth]{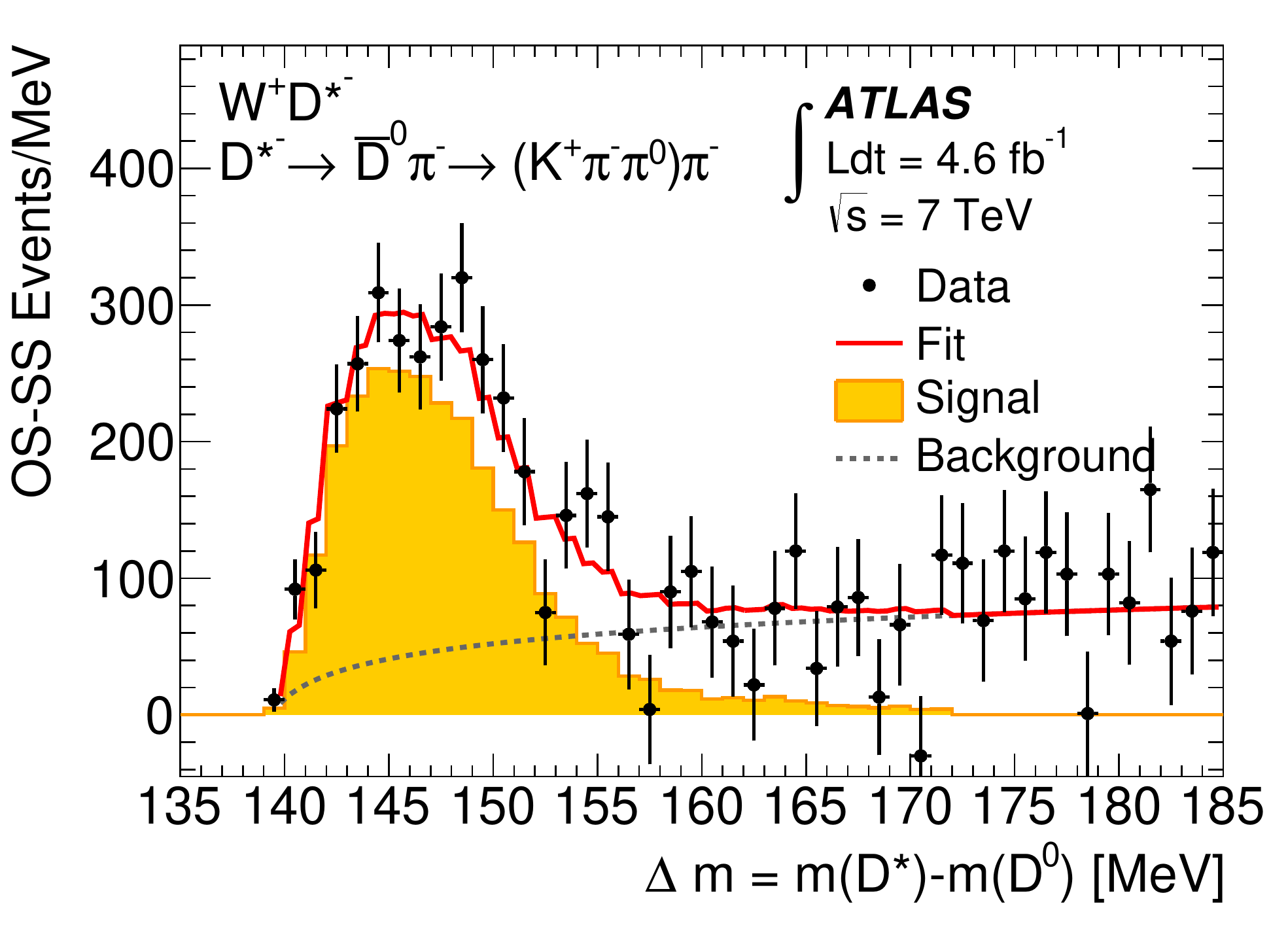}  
\includegraphics[width=0.38\textwidth]{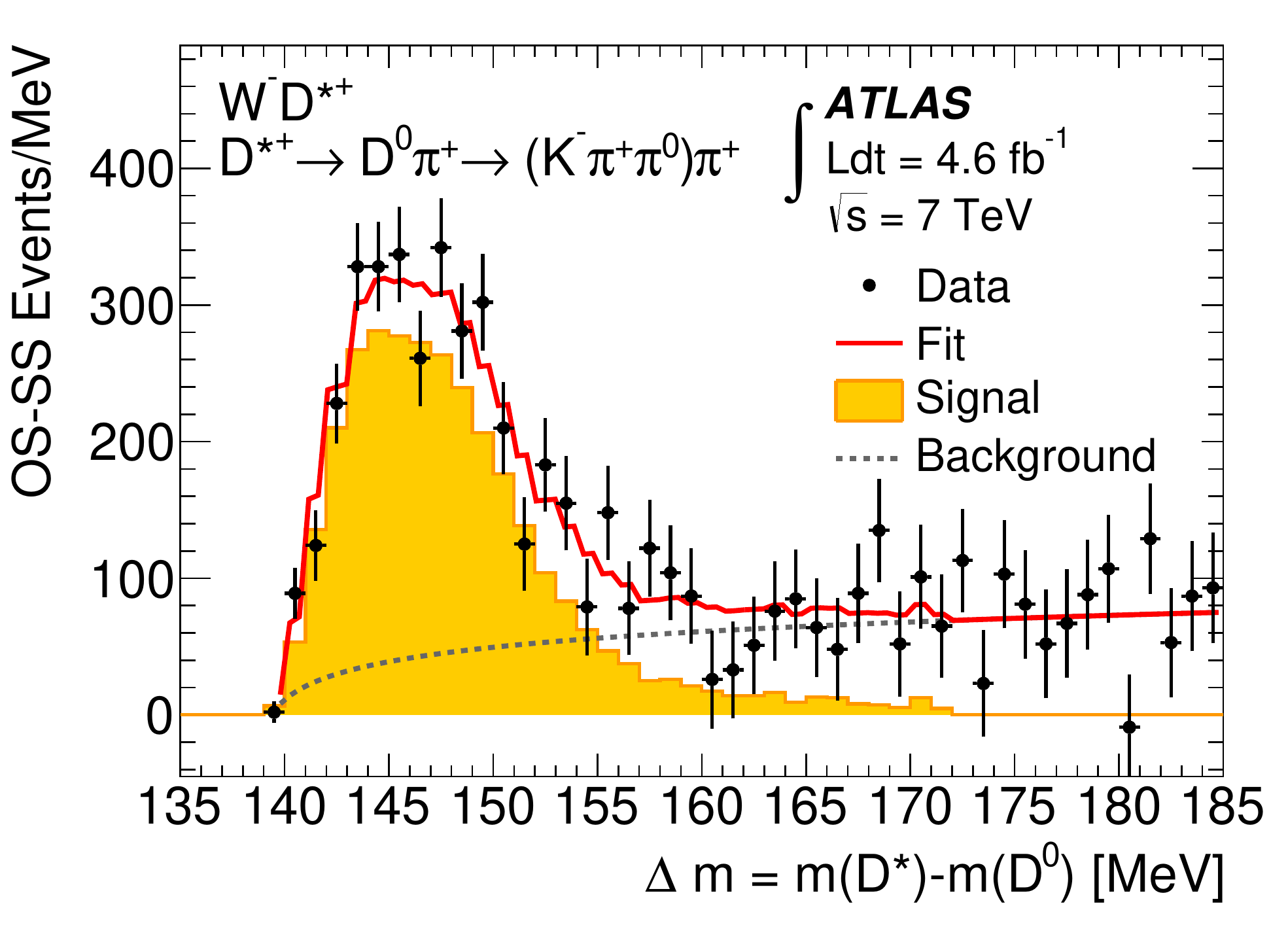}  \\
\includegraphics[width=0.38\textwidth]{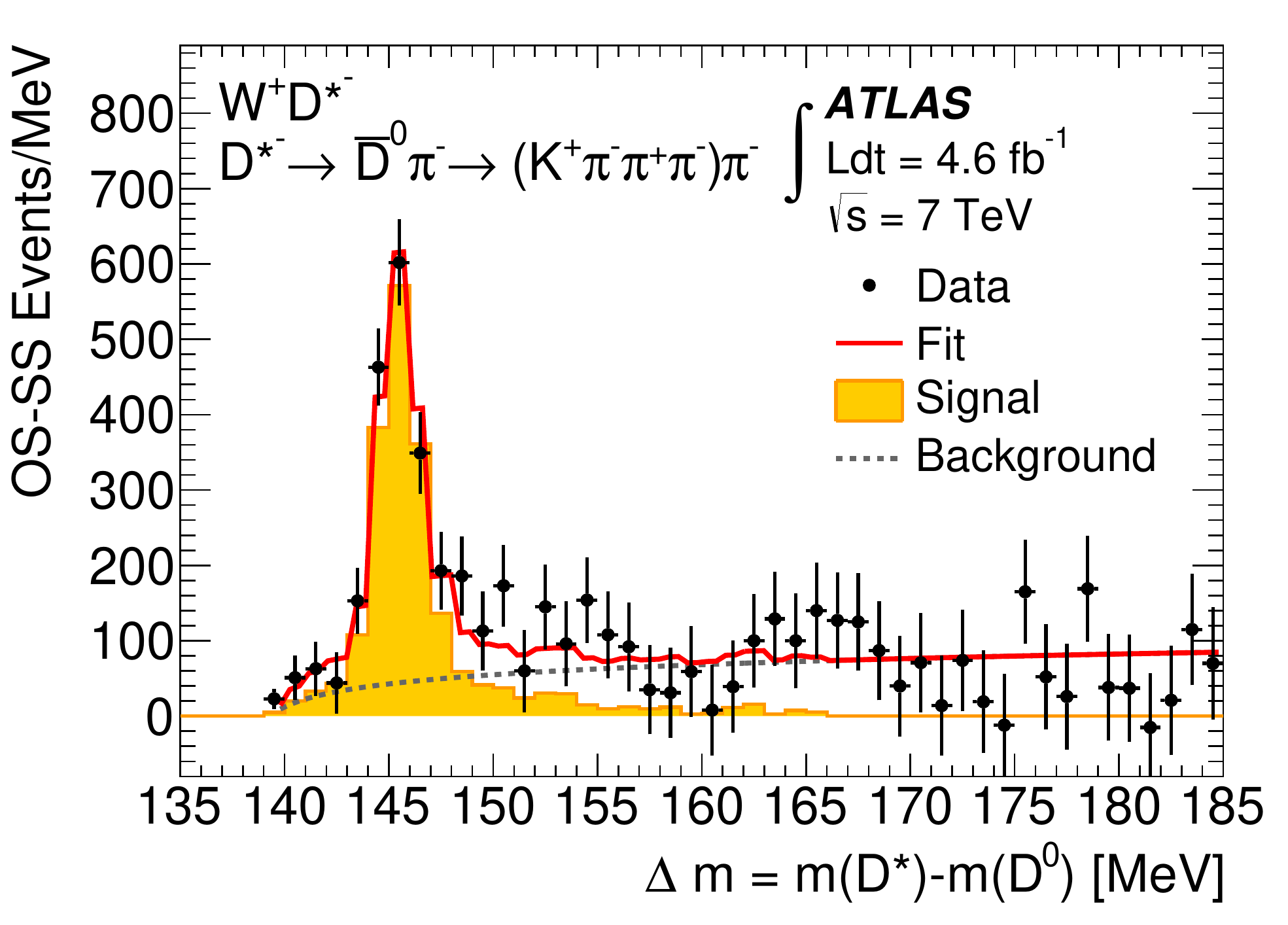} 
\includegraphics[width=0.38\textwidth]{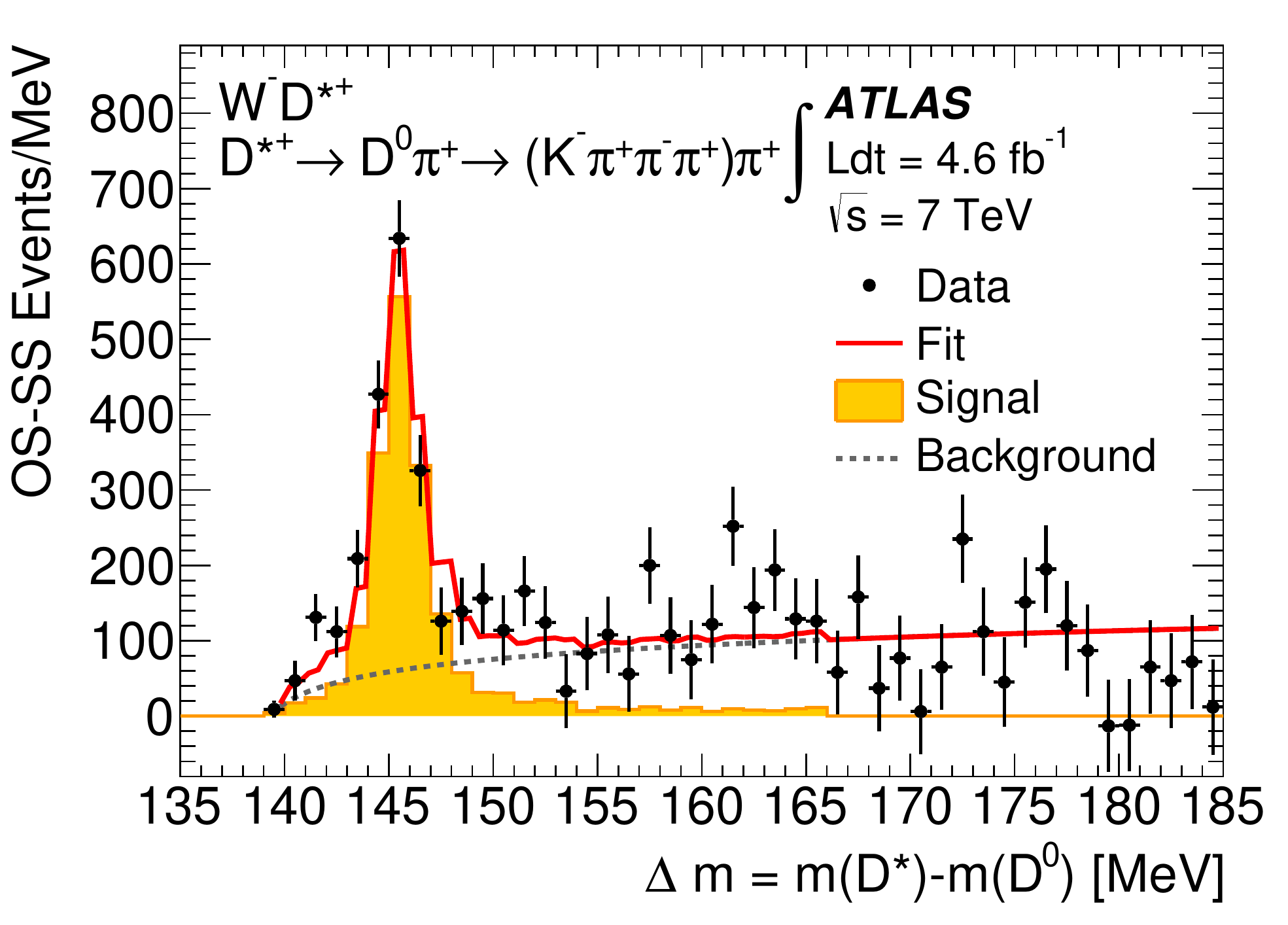} 
 \caption{Results of the fits to the distributions of $m(D)$ and $\Delta m = m(D^{*})-m(D^0)$ in \OSSS\ $W^\pm D^{(*)\mp}$ events. The fit results are shown for $W^+D^{(*)-}$ (left) and $W^-D^{(*)+}$ (right) in the inclusive sample defined by $\ptD>8$\,\GeV\ and $|\etaD|<2.2$: \DpDecboth (top row), \DsDecboth (second row), \DsatDecboth (third row) and \DkpppDecboth (bottom row). The data distributions are shown by the filled markers, where the error bars show the statistical uncertainty. The fit result is shown by the solid line. The filled histogram represents the signal template normalised according to the fit result, while the contribution of the combinatorial background is shown by the dotted line.}
\label{fig:fitsignal}
\end{center}
\end{figure*}

\paragraph{Fit results}
In the fits to the \WDe\ signal region, the normalisations of the signal template and the background function are the only free parameters. The results of the fits to $m(D)$ and $\Delta m$ in the signal region are shown in figure~\ref{fig:fitsignal}. The electron and muon channels are combined to decrease the statistical uncertainty. In addition to the inclusive samples, the fits are performed in four bins of \ptD\ and four bins of $|\eta^\ell|$. Good-quality fits are obtained in all kinematic regions. 

\subsubsection{Removal of heavy-flavour backgrounds}
\label{wdqcdsub}
The \WDe\ yields are obtained by subtracting the expected contribution from heavy-flavour pair production from the fitted yields. $D$ mesons from heavy-flavour pair-produced events give rise to a peak in the mass spectra for both the OS and SS events (see figure~\ref{fig:osss}). 
Given the large uncertainties in predicting
this background from simulation, a data-driven technique is used to estimate it.
The number of background events in the signal region is obtained from the ABCD method~\cite{CMS:2011aa} applied to background-dominated control regions defined by reversing the lepton isolation requirement or the missing transverse momentum and \mtw\ requirements.
 In these control regions, the number of events is taken to be the number of events in the $m(D)$ or $\Delta m$ peaks, since the combinatorial background contribution is already taken into account in the fitting procedure described in the preceding sections. The number of events in the peak is determined by a fit to the $m(D)$ or $\Delta m$ distributions in the control regions. The peaking background in the \OSSS\ \WDe\ signal, which is dominated by heavy-flavour pair production, is found to be less than 1\% in all bins of \ptD\ and $|\eta^\ell|$. Systematic uncertainties on the amount of background from heavy-flavour production arise due to the extrapolation uncertainty from the background control region to the signal region. For the integrated measurement, these uncertainties are small for all decay modes, at a level of 0.3--0.5\%. The systematic uncertainties range from 0.3\%\ to 2.1\%\ depending on the bin of \ptD\ and the \De decay mode.

\subsection{Resulting \OSSS\ yields}
After subtracting the contribution of heavy-flavour events from the fitted yields, the remaining background is predominantly due to top-quark events and is estimated using MC simulation. The uncertainty on the predicted number of top-quark events due to the uncertainty on the cross section and the fiducial acceptance is 10\% in the low \ptD\ bins, while it is increased to 20\% in the two highest \ptD\ bins to cover the uncertainty due to the limited number of MC events.

The yields are determined in each \ptD\ and $|\eta^\ell|$ bin, separately for \WpDe\ and \WmDe\ and are quoted for the four \De decay modes in a range of 140\,\MeV\,$<\Delta m<155$\,\MeV\ for \DzsDec and \DzkpppDec, 140\,\MeV\,$<\Delta m<170$\,\MeV\ for \DzsatDec and 1.8\,\GeV\,$<m(D)<1.94$\,\GeV. The measured yields in the integrated data sample after combining the electron and muon channels are 7990$\pm$370 for the $D$ mode and 3590$\pm$130, 8850$\pm$290 and 5890$\pm$270 for the $D^*$ modes ($K\pi$, $K\pi\pi^0$, $K3\pi$). The expected numbers of background events are summarised in table~\ref{tab:wd_yields}.

\begin{table}[htb]
\begin{center}
\begin{tabular}{| l |c |c |c |c |}
\hline
  $N^\OSSS$($W\rightarrow\ell\nu$)  	&  $D^\pm$ 				& $D^{*\pm}$    	& $D^{*\pm}$ 	&  $D^{*\pm}$ \\
  	&     & $K\pi$    	& $K\pi\pi^0$ 	&  $K3\pi$ \\
  \hline
   $W$+light               &  1680$\pm$440   		&  620$\pm$120 	&  3440$\pm$240	& 2240$\pm$210 \\
   Top                         &    228$\pm$23    		&  86$\pm$9   		&  290$\pm$29 	& 123$\pm$12   \\
  Multijet                 &      35$\pm$18 	  	&  22$\pm$11 		&  50$\pm$25 		&  80$\pm$40   \\
  $W\rightarrow\tau\nu$          & 148$\pm$7 	  		&  68$\pm$4 		&  114$\pm$7		&  89$\pm$6  \\
  \hline
  Total background                 &  2090$\pm$440  	  	&  800$\pm$130	&  3890$\pm$240 	& 2540$\pm$210\\ 
\hline
\end{tabular}
\caption{Estimated background in \OSSS\ events in the four decay modes of the \WDe\ analysis. The electron and the muon channels are combined. The uncertainties include statistical and systematic contributions.}
\label{tab:wd_yields} 
\end{center}
\end{table}

%% file: wc.tex
\section{Event yields for $\boldsymbol{\Wce}$ final states}
\label{s:wc}
In a complementary approach, inclusive charm-quark production is studied by exploiting the semileptonic decays of charm quarks into muons. In this approach, a charm quark is identified by reconstructing the jet of particles produced by its hadronisation and finding an associated soft muon from its semileptonic decay.
The single-charm yield for each $W$-boson charge is determined from the \OSSS\ yields. 
The analysis is performed on separate samples of events with exactly one and exactly two reconstructed jets 
as well as on the combined sample of events with one or two jets. 
The electron and muon decay channels of the $W$ boson are analysed separately and subsequently combined. 

\subsection{Charm-jet selection}
In addition to the event selection described in section \ref{s:wselection}, 
events are required to have either one or two jets with $\pt>25$\,\GeV\ and $|\eta|<2.5$. 
In order to remove jets reconstructed from energy deposits from particles produced in pileup events,
the $\pt$ sum of tracks inside the jet and
associated with the primary vertex divided by the $\pt$ sum of 
all tracks inside the jet is required to be larger than 0.75. 

One and only one jet is 
required to contain a soft muon with $\pt>4$\,\GeV\ and $|\eta|<2.5$. A good match between the ID and MS tracks of the soft muon is required.
The same set of ID hit requirements \cite{ATLAS-CONF-2011-063} that is used for muon candidates from $W$-boson decays is applied to soft muons in addition to two impact parameter requirements: $|d_{0}|<3$\,\mm{} and $|z_{0}\cdot\sin\theta|<3$\,\mm{}.
Exactly one muon is required to be associated with the jet within a cone of radius $\Delta R=0.5$; the small fraction of events with jets containing more than one muon is discarded. 
The soft-muon tagging (SMT) efficiency and mistagging rate are measured in data \cite{softmu}. 
The overall $c$-tagging efficiency is about 4\%, due mainly to the low branching ratio of charmed hadrons to muons (approximately 10\%). 
The light-quark mistagging efficiency is around 0.2\% depending on the jet kinematics. 
Scale factors are applied to correct the MC simulation efficiencies to those measured in data. 
Efficiency scale factors are applied to $b$- and  $c$-jets.
Scale factors for the mistagging rates are applied to light jets.
Two additional requirements, with minor impact on the signal, are applied in the muon channel 
to suppress the $Z$+jets and the $\Upsilon$ backgrounds. 
First, the $c$-jet candidate is required to have either a track multiplicity of at least three or an electromagnetic-to-total energy fraction of less than 0.8.
Second, the event is discarded if the invariant mass of the soft muon and the muon from the decay of the $W$-boson candidate is close to either the $Z$-boson mass (i.e. 80--100\,\GeV) or the $\Upsilon$ mass (i.e. 8--11\,\GeV).

\subsection{Determination of \OSSS\ yields}
Since most backgrounds are nearly OS/SS symmetric, the number of \OSSS\ events is a good estimator of the signal yield. Nonetheless, residual asymmetries in the backgrounds necessitate an additional subtraction. The signal yields are determined from:

\begin{equation}
N^\OSSS_{\Wce}=N^\OSSS_{\rm{data}}-\sum_{\rm{bkg}}A_{\rm{bkg}}\cdot N^{\OS+\SSS}_{\rm{bkg}},
\label{equ:bkg_subtraction}
\end{equation}
where $N^{\OS+\SSS}_{\rm{bkg}}$ is the sum of the number of background events in the OS ($N_{\rm{bkg}}^{\OS}$) and SS ($N_{\rm{bkg}}^{\SSS}$) samples and the asymmetry $A_{\rm{bkg}}$ is defined as
\begin{equation}
A_{\rm{bkg}}=N_{\rm{bkg}}^{\OSSS}/N_{\rm{bkg}}^{\OS+\SSS}.
\end{equation}
\label{equ:Asym}

Backgrounds to the \Wce\ candidate sample include the production of $W$+light, $W$
plus heavy-flavour quark pairs ($c\bar{c}$, $b\bar{b}$), multijet events, $Z$ and, to a
lesser extent, single and pair-produced top quarks, and dibosons.

The background from $W$+light events and multijet events is estimated with data-driven methods. 
$Z$ events, in which one of the muons from the $Z$ decay radiates a photon that is mistakenly reconstructed as a jet, are a significant background source in the muon channel and are thus determined using a data-driven method. Smaller backgrounds from top-quark and diboson production, and the
$Z$+jets background in the electron channel, are estimated from MC simulations. Backgrounds from $W+b\bar{b}$, $W+c\bar{c}$ are negligible since they are OS/SS symmetric.

\subsubsection{Backgrounds and yield in the electron channel}
\label{sec:wc_ele_yields}

\begin{figure}[t]
\begin{center}
\includegraphics[width=0.42\linewidth]{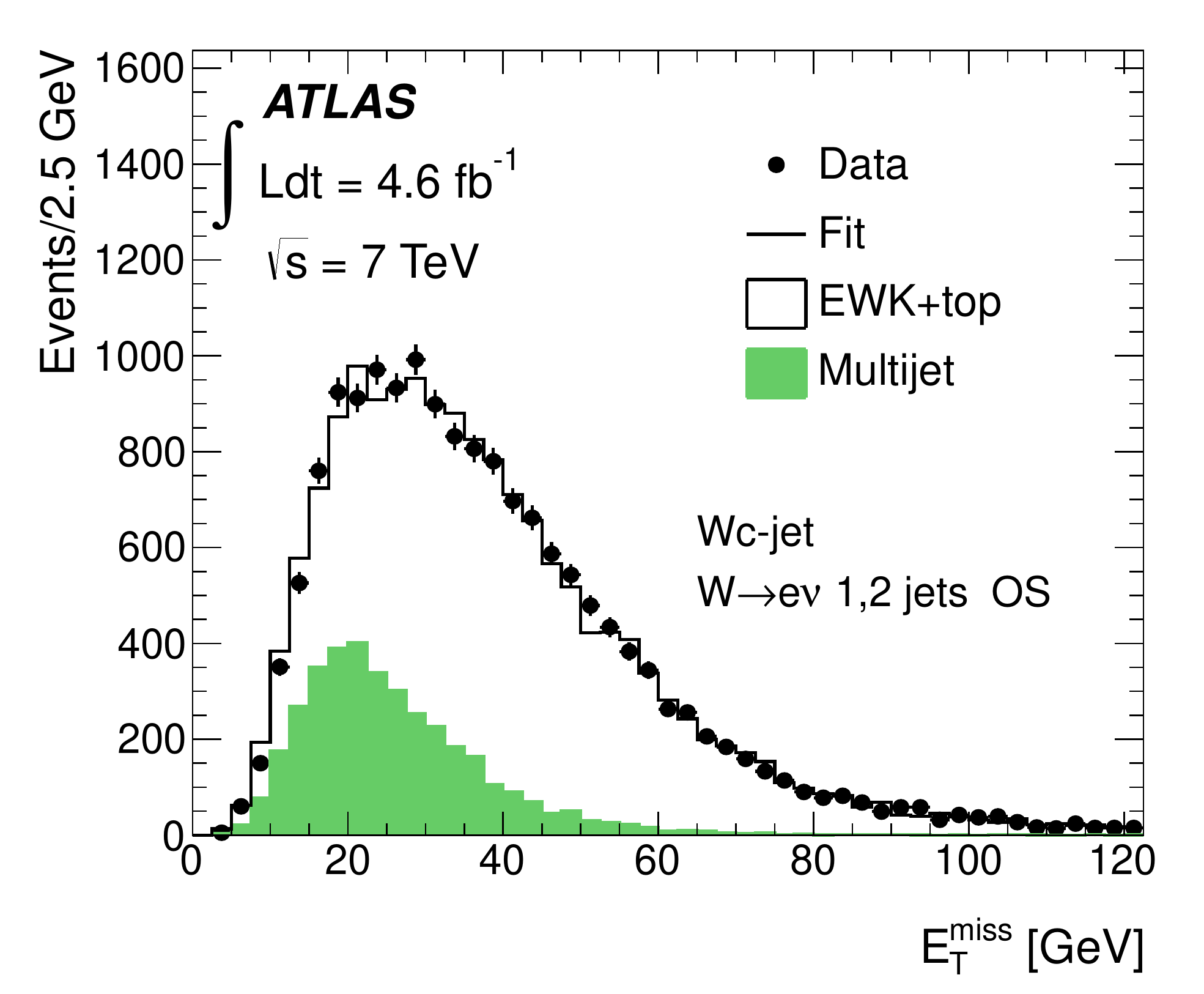} 
\includegraphics[width=0.42\linewidth]{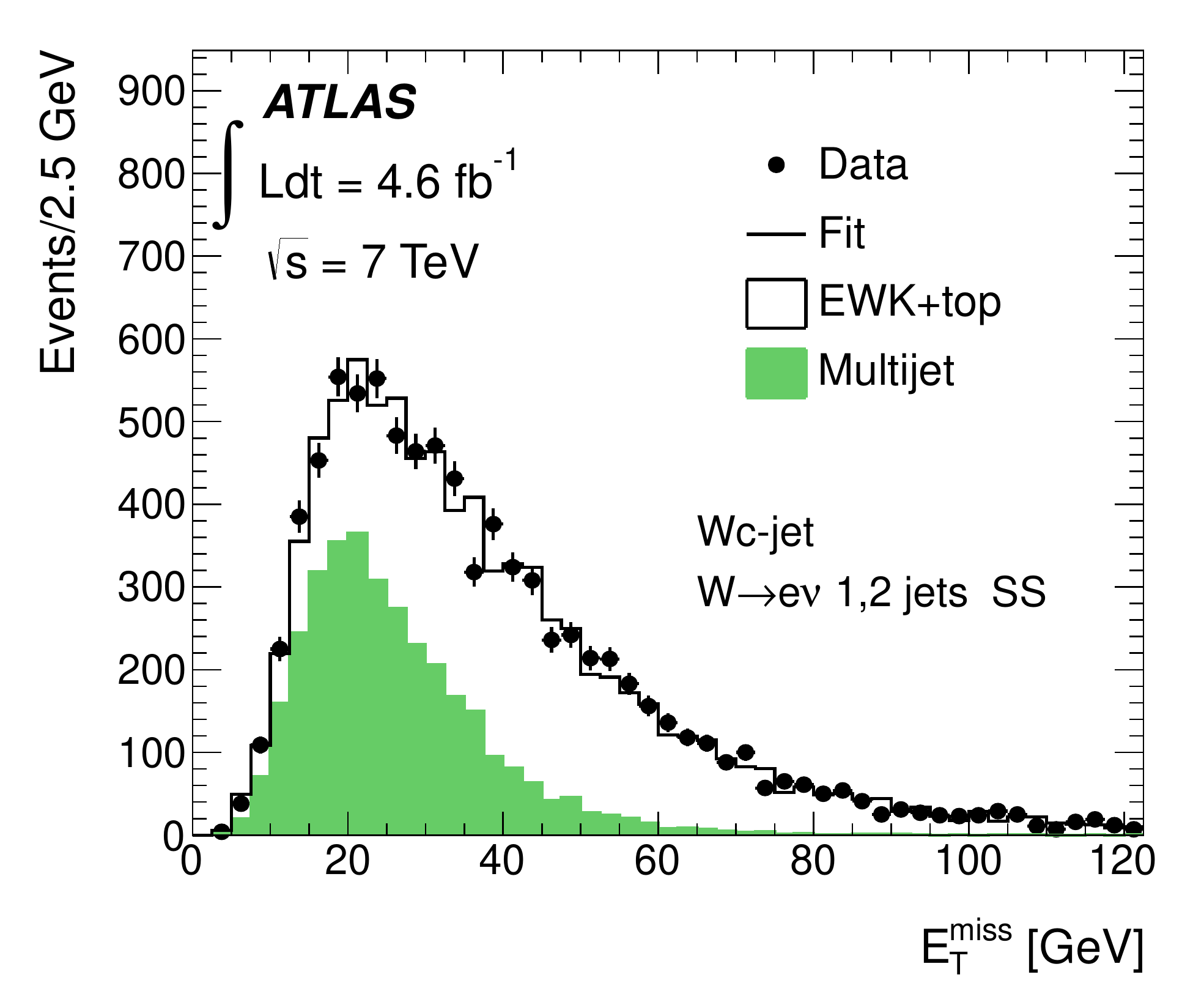} 
\caption{Results of the fits to the \met\ distribution which are used to determine the multijet background in the electron channel, in the OS (left) and SS (right) samples of $W$+1,2 jets candidate events. The data are shown by filled markers and the fit result by the solid line. The multijet template, normalised according to the fit result, is shown by the filled histogram. The shape of the distribution for the electroweak and top-quark processes is obtained from simulation. Electroweak processes include $W$, $Z$ and diboson processes.
}
\label{fig:wc_elec_multijet_fit}
\end{center}
\end{figure}

The numbers of $W$+light and multijet background events are obtained from a constrained $\chi^2$ fit to the number of events in the SS sample followed by a propagation to \OSSS\ using the equation
\begin{equation} 
N_{\rm{bkg}}^\OSSS = A_{\rm{bkg}}\cdot N_{\rm{bkg}}^{\OS+\SSS} = \frac{2\cdot A_{\rm{bkg}}}{1-A_{\rm{bkg}}}N_{\rm{bkg}}^{\SSS}.
\end{equation}

In the fit, the sum of the multijet and $W$+light backgrounds plus the remaining backgrounds and a small signal contribution is required to be equal to the total data count in the SS sample. The relative fractions of multijet and $W$+light events are allowed to vary in the fit, while all other backgrounds and the signal contribution are fixed to the values from simulation. 

The OS/SS asymmetry of the multijet background, $A_{\rm{multijet}}$, is found by performing a binned maximum-likelihood fit of templates to the \met\ distribution in data. The fit follows the procedure discussed in section~\ref{s:wdyield} and is done in an expanded sample where the \met\ selection requirement is removed. Two templates are used: one representing the multijet background and the other representing the contributions from all other sources. The template for the multijet sample is extracted from a data control sample selected by inverting the electron isolation and some of the electron identification requirements. Contamination in the control sample from $W/Z$ and top-quark events is estimated from simulation and subtracted. 
The template representing all other processes, including the signal, $W/Z$, diboson and top-quark production, is obtained from MC simulation and built separately for OS and SS samples. 
Figure~\ref{fig:wc_elec_multijet_fit} shows the results for the OS and SS $W$+1,2 jets samples. 
$A_{\rm{multijet}}$ is computed using the fit results in the signal region ($\met>25$\,\GeV) and is found to be consistent with zero within uncertainties. The uncertainties are dominated by the statistical component. The systematic uncertainties are estimated by varying the fit range and trying alternative multijet and other background templates. These uncertainties are found to be small.

The OS/SS asymmetry of the $W$+light background, $A_{W+\rm{light}}$, is obtained from MC simulation and corrected using the asymmetry measured in a data control region following the relation: 
\begin{equation}
A_{W+\rm{light}} = A^{\rm{MC}}_{W+\rm{light}}\frac{A^{\rm{data,tracks}}_{W+\rm{light}}}{A^{\rm{MC,tracks}}_{W+\rm{light}}}.
\label{e:asymwlight}
\end{equation}
$A^{\rm{MC}}_{W+\rm{light}}$ is the OS/SS asymmetry in the MC simulation for the
signal region
and $A^{\rm{MC,tracks}}_{W+\rm{light}}$ ($A^{\rm{data,tracks}}_{W+\rm{light}}$) is the OS/SS asymmetry in MC (data) events
estimated using the charges of the $W$ boson and a generic track that passes
the soft-muon kinematic requirements. $A^{\rm{MC,tracks}}_{W+\rm{light}}$ and $A^{\rm{data,tracks}}_{W+\rm{light}}$ are computed from an expanded sample selected with no soft-muon requirements (called the pretag sample). $A_{W+\rm{light}}$ is found to be approximately 10\%. The uncertainty on $A_{W+\rm{light}}$ is dominated by the statistical uncertainty on $A^{\rm{MC}}_{W+\rm{light}}$. The sub-leading systematic uncertainty contains contributions from uncertainties on the background contamination in the pretag sample and the modelling of the track properties.

The estimated numbers of background events are shown in table~\ref{tab:elec_yields}. The total number of OS(SS) events in the data SMT samples of $W$+1 jet and $W$+2 jets is 7436(3112) and 4187(2593), respectively. The corresponding number of \OSSS\ events in data is 4320$\pm$100 for the $W$+1 jet and  1590$\pm$80 for the $W$+2 jets sample.

\begin{table}[htb]
\begin{center}
\begin{tabular}{|l |c | c | c | c|}
\hline
  $N^\OSSS$ ($W\rightarrow e\nu$)   & $W+1$~jet    & $W+2$~jets &  $W+1,2$~jets \\
\hline
  $W$+light                       & 240$\pm$100  &  100$\pm$50 &   330$\pm$130 \\
  Multijet                       & 130$\pm$140  &  0$\pm$100 &  160$\pm$170 \\
  $t\bar{t}$                      &  13$\pm$5    &  79$\pm$14   &   92$\pm$16  \\
  Single top                      &   62$\pm$10   &  78$\pm$12   &  140$\pm$20 \\ 
  Diboson                         &   35$\pm$6   &  35$\pm$5   &   70$\pm$9 \\
  $Z$+jets               &    8$\pm$12   &  15$\pm$10   &  23$\pm$15 \\ 
  \hline
  Total background                & 490$\pm$160  &  300$\pm$120 & 820$\pm$200 \\ 
  \hline
  
\hline
\end{tabular}
\caption{Estimated background in \OSSS\ events in the $W$+1 jet, $W$+2 jets and $W$+1,2 jets samples for the electron channel. The uncertainties include statistical and systematic contributions. The correlations between the uncertainties for the different background estimates stemming from the constraint in the SS sample is taken into account when computing the total background uncertainties. For backgrounds estimated with data-driven methods the yields in the $W$+1 jet, $W$+2 jets, and $W$+1,2 jets sample are estimated independently.}
\label{tab:elec_yields} 
\end{center}
\end{table}

\begin{figure}[t]
\begin{center}
\includegraphics[width=0.38\textwidth]{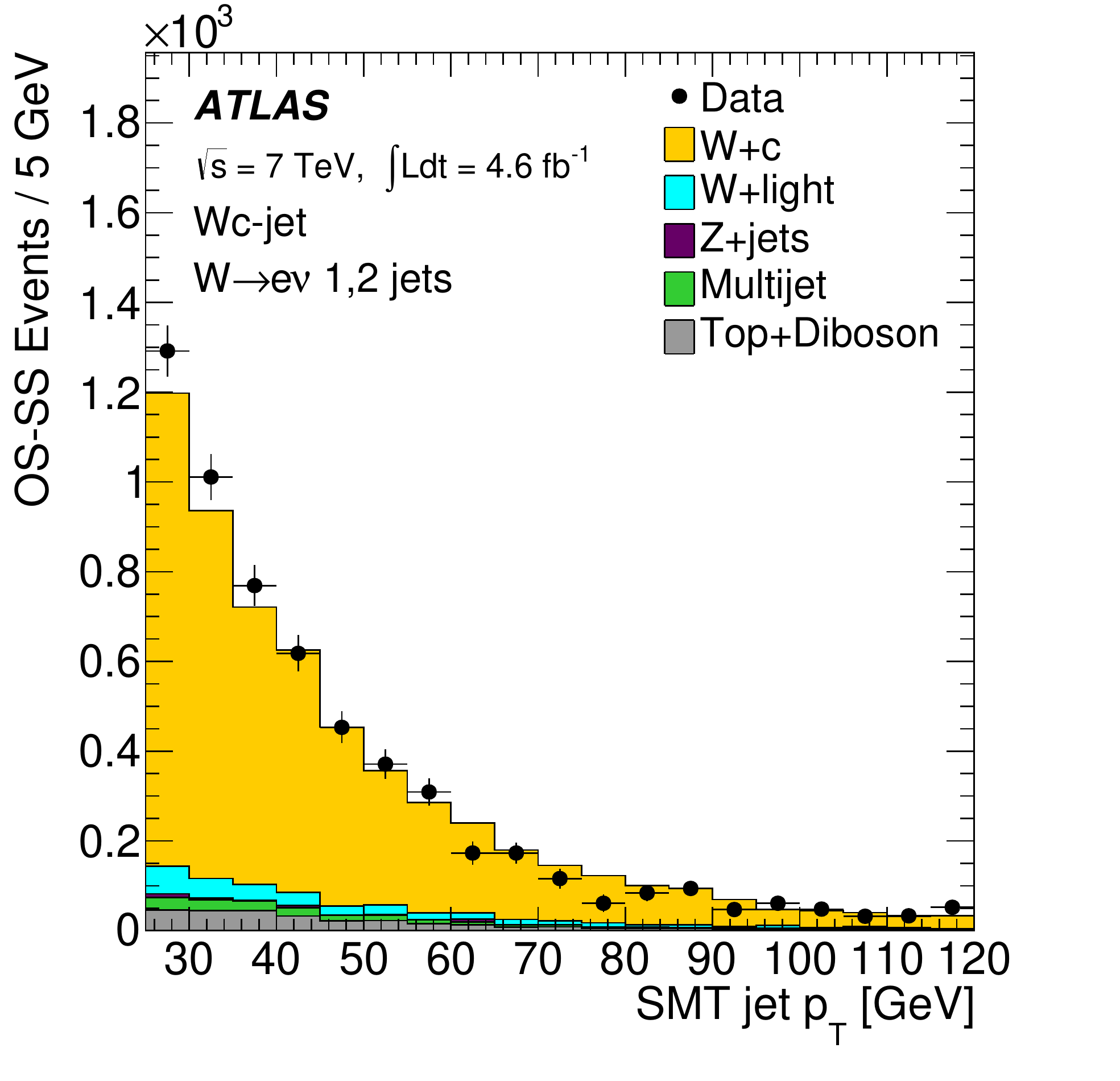}  
\includegraphics[width=0.38\linewidth]{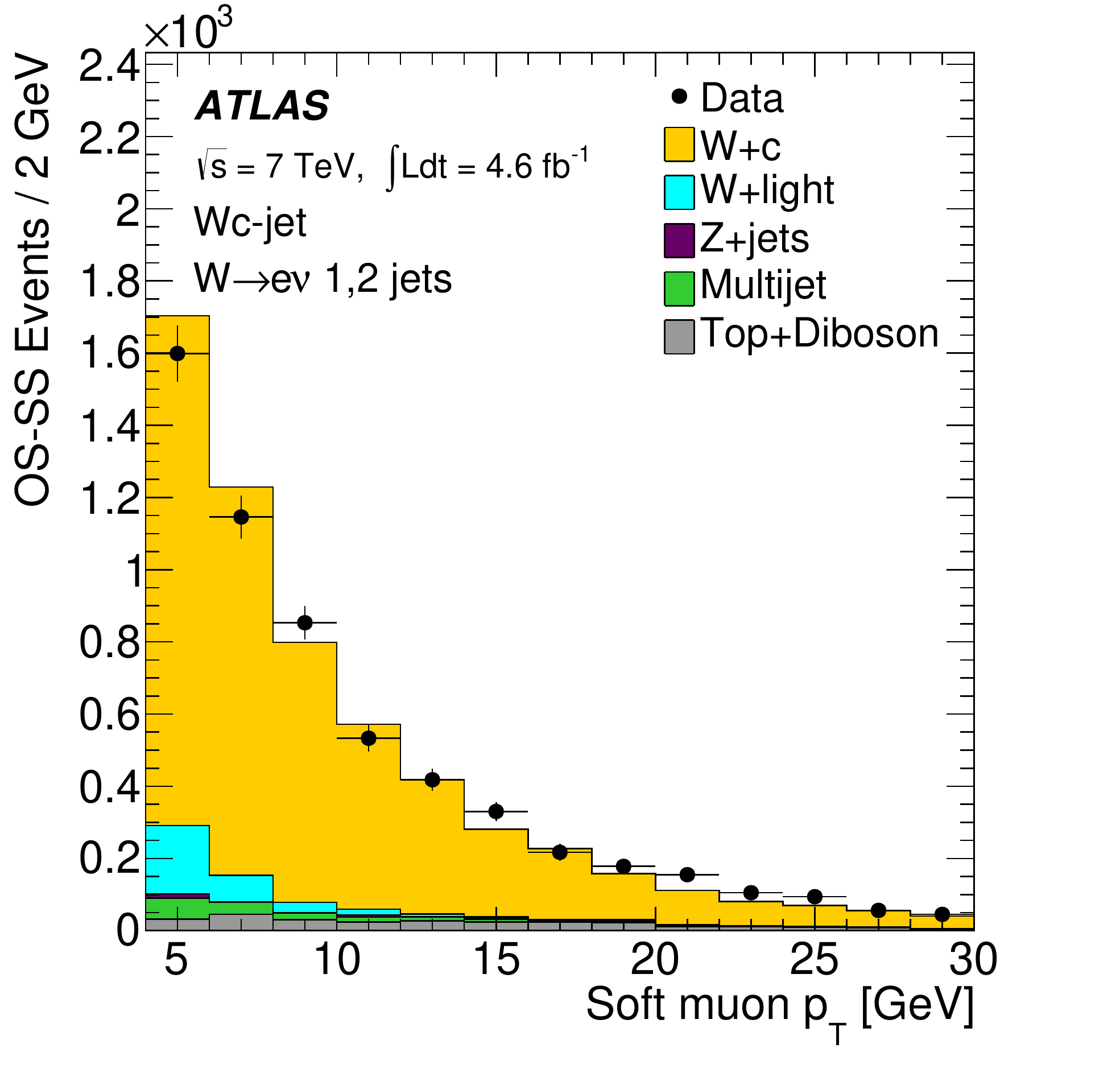} 
\caption{Distribution of the SMT jet \pt{} (left) and the soft-muon \pt{} (right) in \OSSS\ events of the $W$+1,2 jets sample for the electron channel. The normalisation of the $W$+light background and the shape and normalisation of the multijet background are obtained with data-driven methods. All other backgrounds are estimated with MC simulations and normalised to their theoretical cross sections. The signal contribution is normalised to the measured yields.}
\label{fig:wc_control_elec}
\end{center}
\end{figure}

Figure~\ref{fig:wc_control_elec} shows the \pt{} distribution of the SMT jet and the soft muon in \OSSS\ events in the $W$+1,2 jets sample for the electron channel. The signal contribution is normalised to the measured yields and the background contributions are normalised to the values listed in table~\ref{tab:elec_yields}. The MC simulation is in satisfactory agreement with data.

In addition to the inclusive samples, yields and cross sections are measured separately for $W^+$ and $W^-$ and in 11 bins of $|\eta^\ell|$.
The multijet background $|\eta^\ell|$-shape is derived from individual fits to the \met\ distribution and normalised to the inclusive total. The remaining backgrounds are taken from simulation.

\subsubsection{Backgrounds and yield in the muon channel}
\label{sec:wc_muo_yields}

The multijet background in the muon channel is substantially different from that in the electron channel 
since it is dominated by heavy-flavour semileptonic decays. 
The estimation technique is adapted to take this into account.
The multijet background in OS+SS events is determined by the equation
\begin{equation}
N^{\OS+\SSS}_{\rm{multijet}} = N^{\rm{pretag}}_{\rm{multijet}}\cdot R^{\rm{SMT}}_{\rm{multijet}},
\end{equation}
where $N^{\rm{pretag}}_{\rm{multijet}}$ is the multijet event yield in the pretag sample and $R_{\rm{multijet}}^{\rm{SMT}}$ is the soft-muon tagging rate for events in the multijet sample. 

The evaluation of $N_{\rm{multijet}}^{\rm{pretag}}$ uses a data-driven technique known as the Matrix Method~\cite{matrixMethod}. 
An expanded sample enriched in multijet events is obtained by applying all selection cuts to the data except for the muon isolation requirements. 
The efficiencies of the isolation requirements for multijet events and prompt isolated muons are needed 
to relate the expanded sample to the signal sample. The isolation efficiency for prompt muons 
is measured in an independent sample of $Z\rightarrow \mu\mu$ events. The efficiency in 
multijet events is measured both in a control sample with inverted missing transverse 
momentum and $W$-boson transverse mass requirements, and 
through a fit to the muon $d_0$ significance; it is parameterised as a function of the muon and jet kinematics.
The average of the results obtained with the two measurements is taken as the final estimate 
and half the difference is used as the systematic uncertainty.

\begin{figure}[t]
\begin{center}
\includegraphics[width=0.32\linewidth]{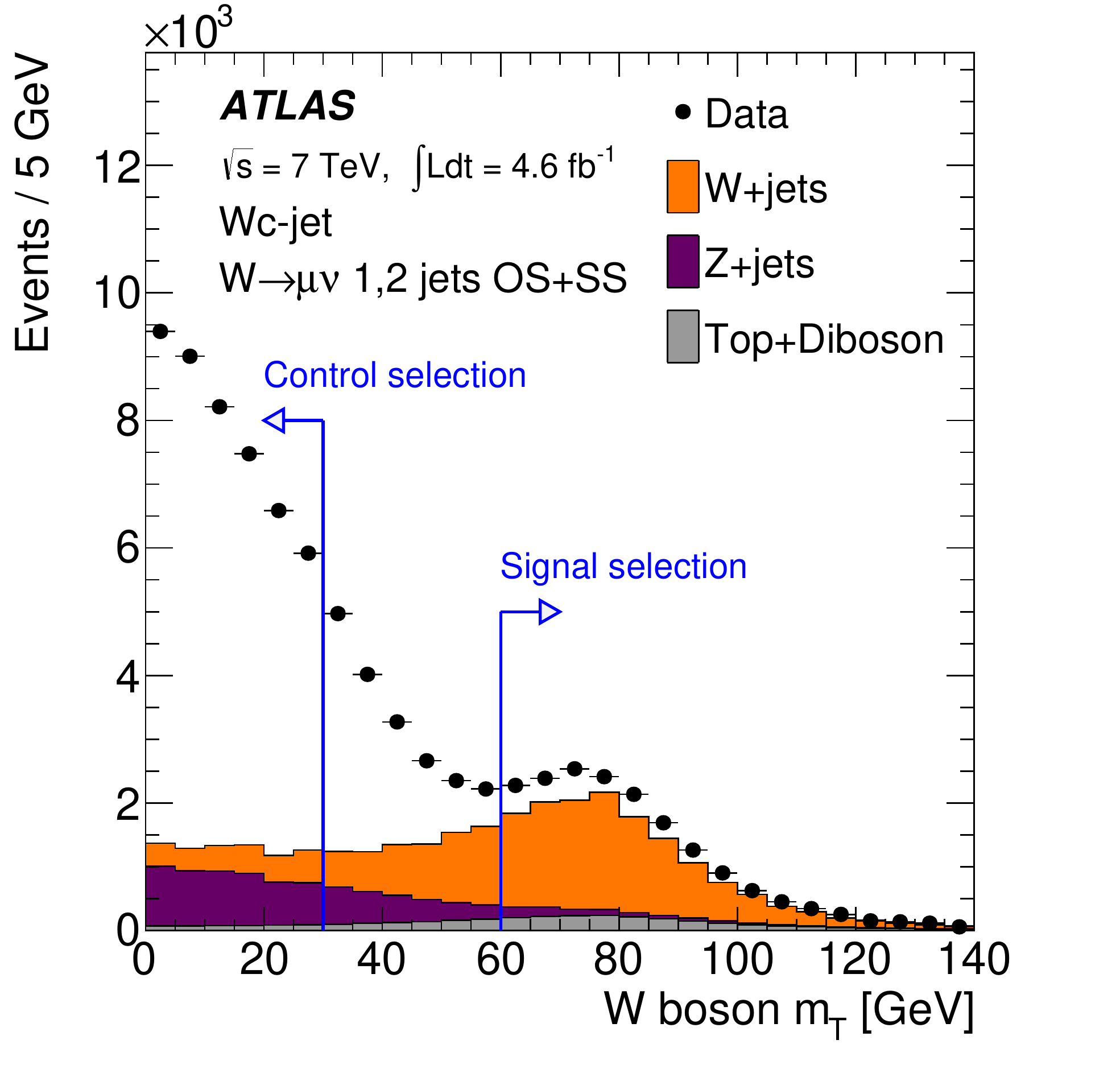} 
\includegraphics[width=0.32\linewidth]{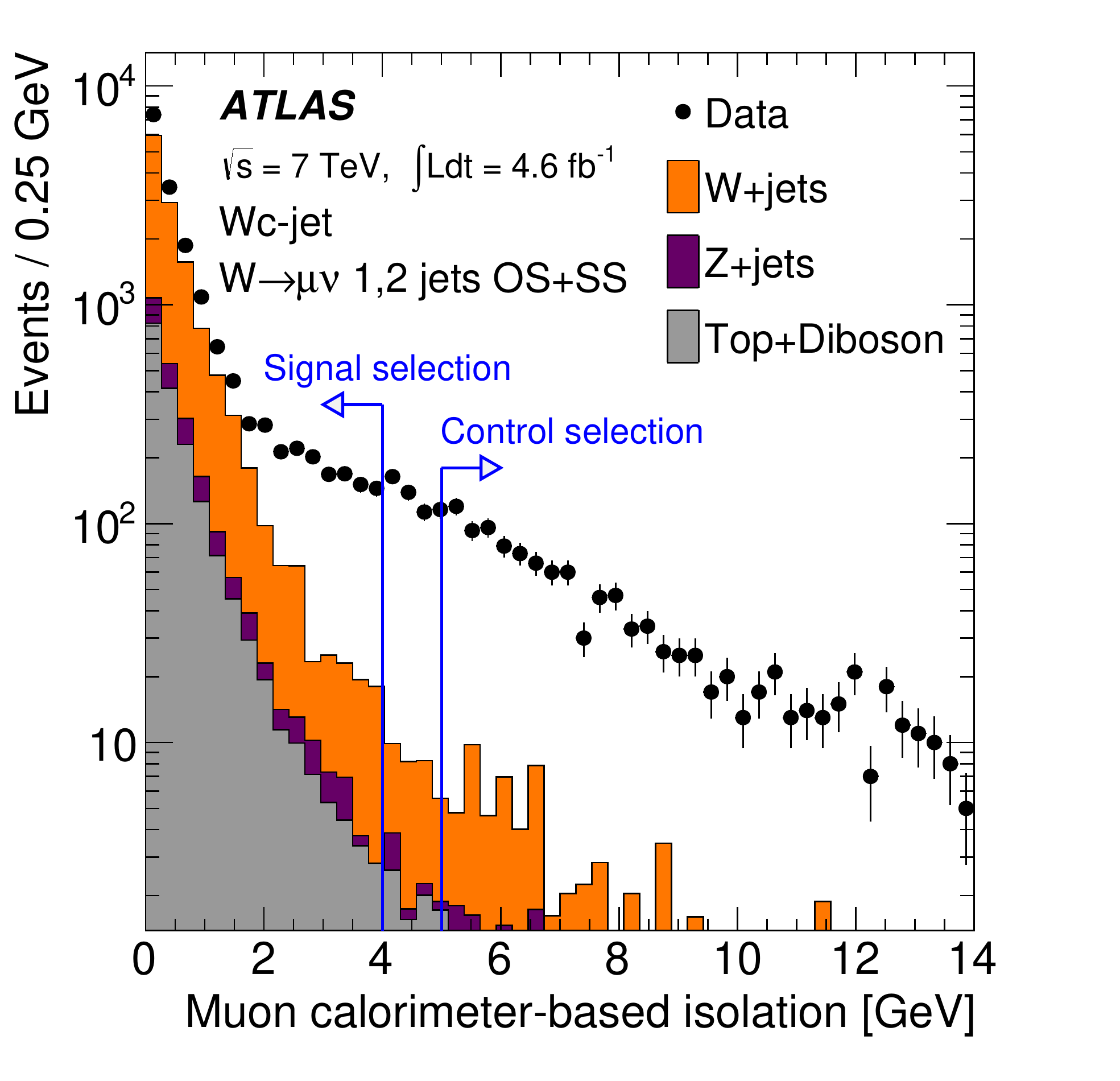}
\includegraphics[width=0.32\linewidth]{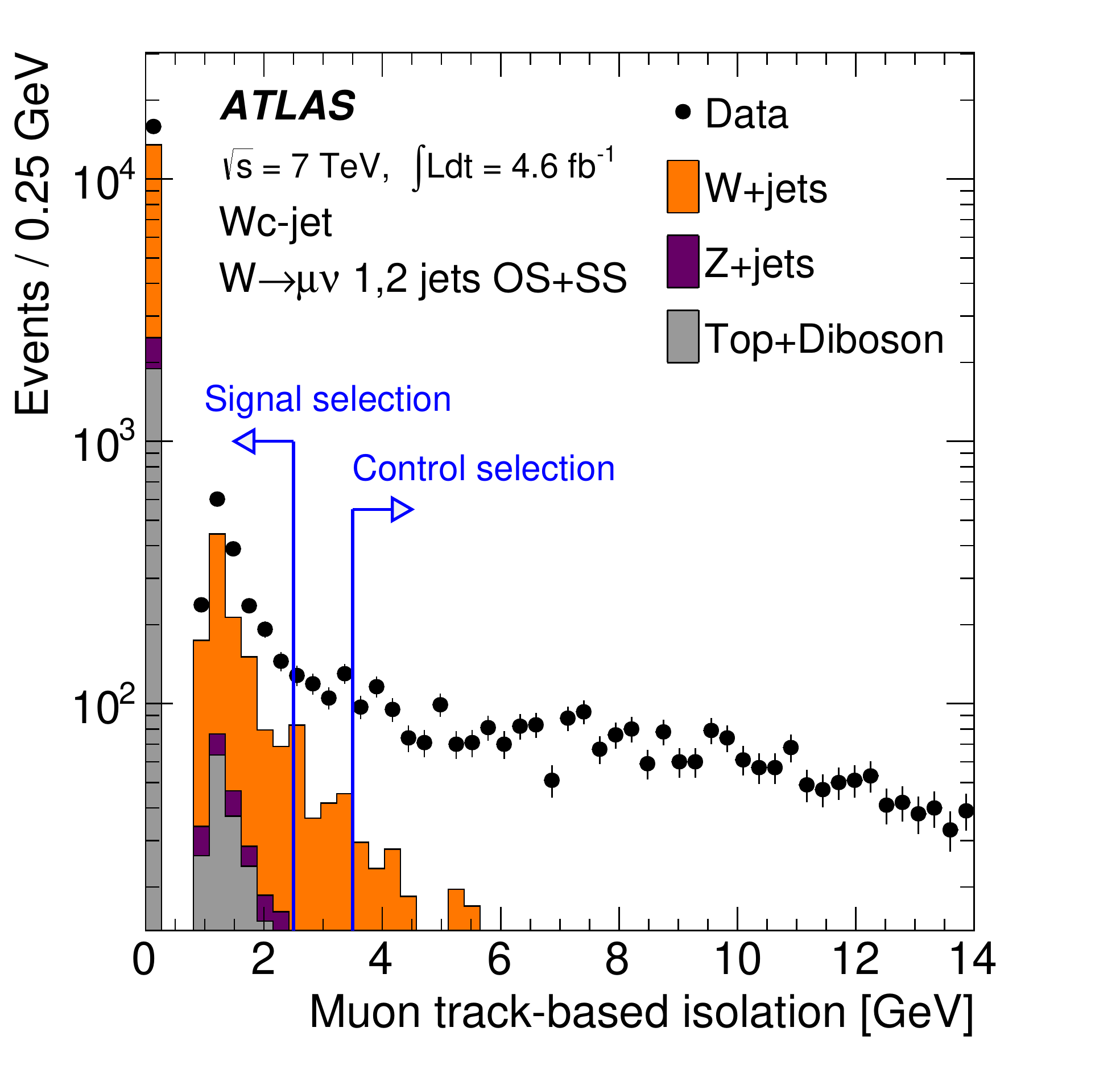}
\caption{Distribution of the $W$-boson transverse mass (left), muon calorimeter-based isolation (centre) and muon track-based isolation (right) in data and the expectation from $W/Z$+jets, top-quark and diboson events. Events with low transverse mass and large values of isolation variables are predominantly multijet events. $W/Z$+jets refers to the production of $W/Z$ bosons in association with light, $c$ or $b$ jets.}
\label{fig:wc_muon_multijet}
\end{center}
\end{figure}

$R^{\rm{SMT}}_{\rm{multijet}}$ and $A_{\rm{multijet}}$ are independently determined in two control regions enriched in multijet background. The samples are selected by inverting either the muon isolation requirements or the 
$W$-boson transverse mass requirement. The distributions of events as a function of the muon isolation variables and the $W$-boson transverse mass are shown in figure~\ref{fig:wc_muon_multijet}. The amount of contamination from $W/Z$+jets events in the multijet control regions is estimated from MC simulation. 
The contamination from top-quark and diboson production is negligible.
The value of $R^{\rm{SMT}}_{\rm{multijet}}$ is determined by measuring the soft-muon tagging rate as a function of the muon isolation in the multijet control regions and extrapolating it to the signal region assuming a linear dependence. Uncertainties from the $W/Z$+jets contamination level and from the extrapolation procedure are taken into account. The value of $A_{\rm{multijet}}$ is deduced from the average of the two control regions and is approximately 20\%.

The $W$+light background in OS+SS events is estimated according to the following equation:
\begin{equation}
N^{\OS+\SSS}_{W+\rm{light}}=N^{\rm{pretag}}_{W+\rm{jets}}\cdot f_{\rm{light}}\cdot R^{\rm{SMT}}_{W+\rm{light}},
\label{equ:wjets_much}
\end{equation}
where $N^{\rm{pretag}}_{W+\rm{jets}}$ is the yield of $W$+jets events in the pretag sample, $f_{\rm{light}}$ is the fraction 
of events in which the $W$ boson is produced in association with a light jet and $R^{\rm{SMT}}_{W+\rm{light}}$ is the soft-muon tagging rate in $W$+light events. All the terms of equation (\ref{equ:wjets_much}) are derived using data-driven methods.

$N^{\rm{pretag}}_{W+\rm{jets}}$ is calculated as the difference between the number of selected data events and the sum of all other expected background contributions, namely multijet, $Z$+jets, 
top-quark, and diboson production in the pretag sample. The multijet background is estimated using the Matrix Method as explained above while all other backgrounds are taken from simulation. The fraction $f_{\rm{light}}$ is obtained through an analysis of the tagging rate of a lifetime-based tagger in the pretag sample as done in ref.~\cite{Aad:2012hg}. $R_{W+\rm{light}}^{\rm{SMT}}$ is determined using a $W$+light MC simulation corrected by a data-derived scale factor for the soft-muon mistag rate~\cite{softmu}. The asymmetry $A_{W+\rm{light}}$ is obtained using equation (\ref{e:asymwlight}), as done in the electron channel. 

\begin{table}[htb]
\begin{center}
\begin{tabular}{| l|c| c |c |c|}
\hline
  $N^\OSSS$ ($W\rightarrow \mu\nu$)   & $W+1$~jet    & $W+2$~jets &  $W+1,2$~jets \\
\hline
  $W$+light                       & 220$\pm$80   & 40$\pm$40 &   250$\pm$90 \\
  Multijet                       &  71$\pm$27  &  52$\pm$20 &  120$\pm$40 \\
  $t\bar{t}$                      &  24$\pm$21    &  129$\pm$19   &   154$\pm$21  \\
  Single top                      &   58$\pm$18   &  82$\pm$21   &  140$\pm$23 \\ 
  Diboson                         &   37$\pm$10   &  39$\pm$13   &   76$\pm$20 \\
  $Z$+jets               &  237$\pm$22   &  207$\pm$16   &  445$\pm$34 \\ 
  \hline
  Total background                & 650$\pm$90  &  550$\pm$60 & 1190$\pm$110 \\ 
\hline

\end{tabular}
\caption{Estimated background in \OSSS\ events in the $W$+1 jet, $W$+2 jets and $W$+1,2 jets samples for the muon channel. Uncertainties include statistical and systematic contributions.}
\label{tab:muon_yields} 
\end{center}
\end{table}

\begin{figure}[t]
\begin{center}
\includegraphics[width=0.38\linewidth]{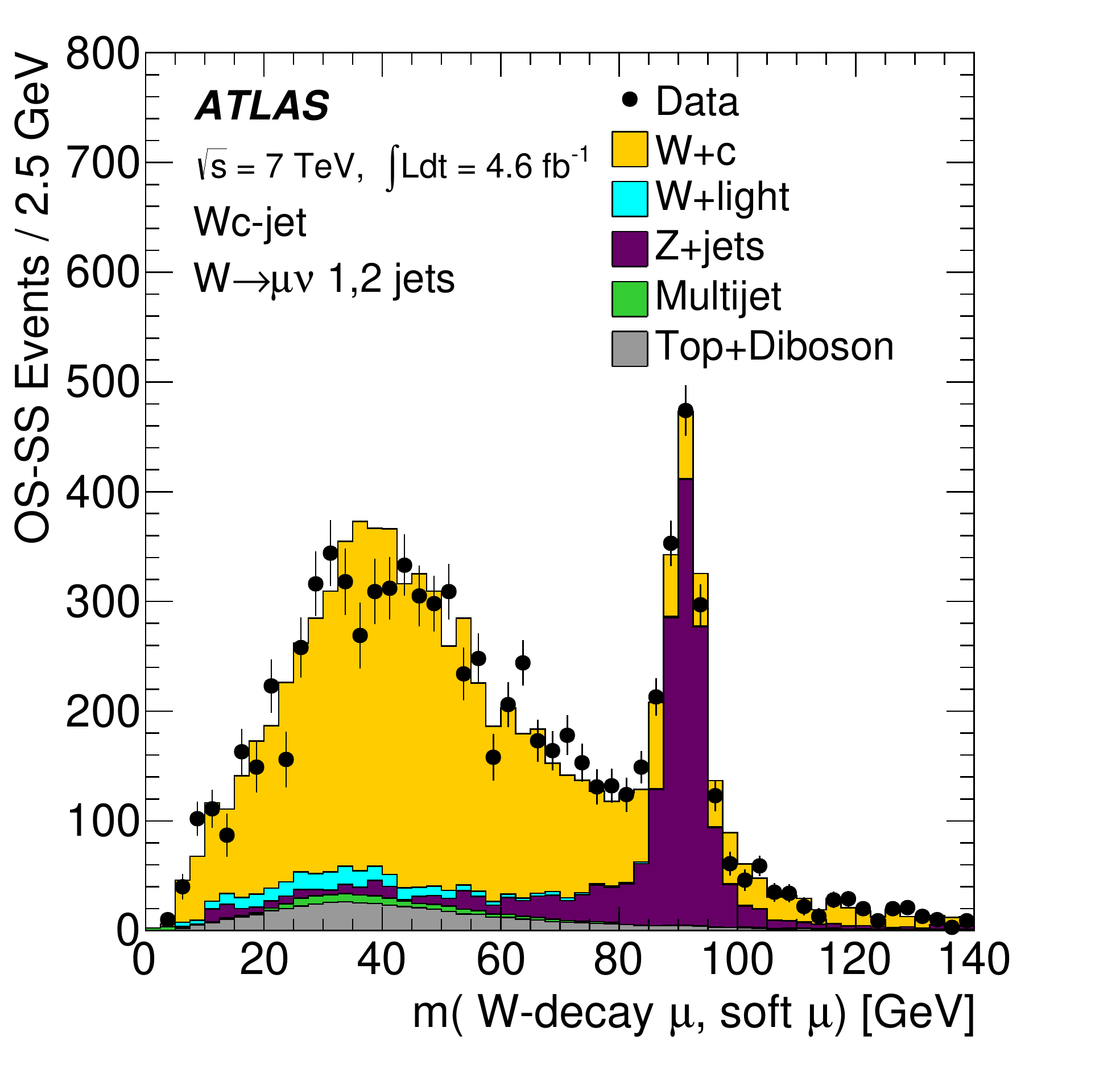}
\caption{Invariant mass constructed using the four-momenta of the soft muon and the muon from the decay of the $W$-boson candidate.}
\label{fig:wc_muon_zjet}
\end{center}
\end{figure}

The $Z$+jets background is estimated by using a data control sample to normalise the MC simulation. The control sample is defined by requiring the invariant mass of the soft muon and the muon from the decay of the $W$-boson candidate to be between 80\,\GeV\ and 100\,\GeV. 
The normalisation is carried out in \OSSS\ events, which has the advantage of minimising contributions 
from non-$Z$ events.  The $Z$+jets yield in the control region is estimated 
from data by subtracting the expected contamination of \Wca\ signal, $W$+light and diboson events 
(the latter two account for less than 1\% of the events). 
The contamination of the control sample by \Wca{} events is estimated initially through MC simulation and then refined by iteratively adjusting the \Wca{}, $Z$+jets and $W$+jets normalisations to match the data. 
A normalisation factor for $Z$+jets of $1.06\pm0.06$ is derived.
The invariant mass of the lepton from the decay of the $W$-boson candidate and the soft muon is shown in figure~\ref{fig:wc_muon_zjet} for 
$W$+1,2 jets data passing all event selection requirements except for the veto around the invariant masses of the $Z$ boson and the $\Upsilon$ meson. 
The expected contributions of all processes, normalised as described above, are also shown.
The predicted distributions provide a good description of the data.

The total number of OS(SS) events in the data samples of $W$+1 jet and  $W$+2 jets is 7736(2775) and 4376(2479), respectively. The corresponding number of \OSSS\ events in data is 4960$\pm$100 for the $W$+1 jet and 1900$\pm$80 for the $W$+2 jets sample. The expected 
backgrounds are summarised in table~\ref{tab:muon_yields}.

\begin{figure}[t]
\begin{center}
\includegraphics[width=0.38\linewidth]{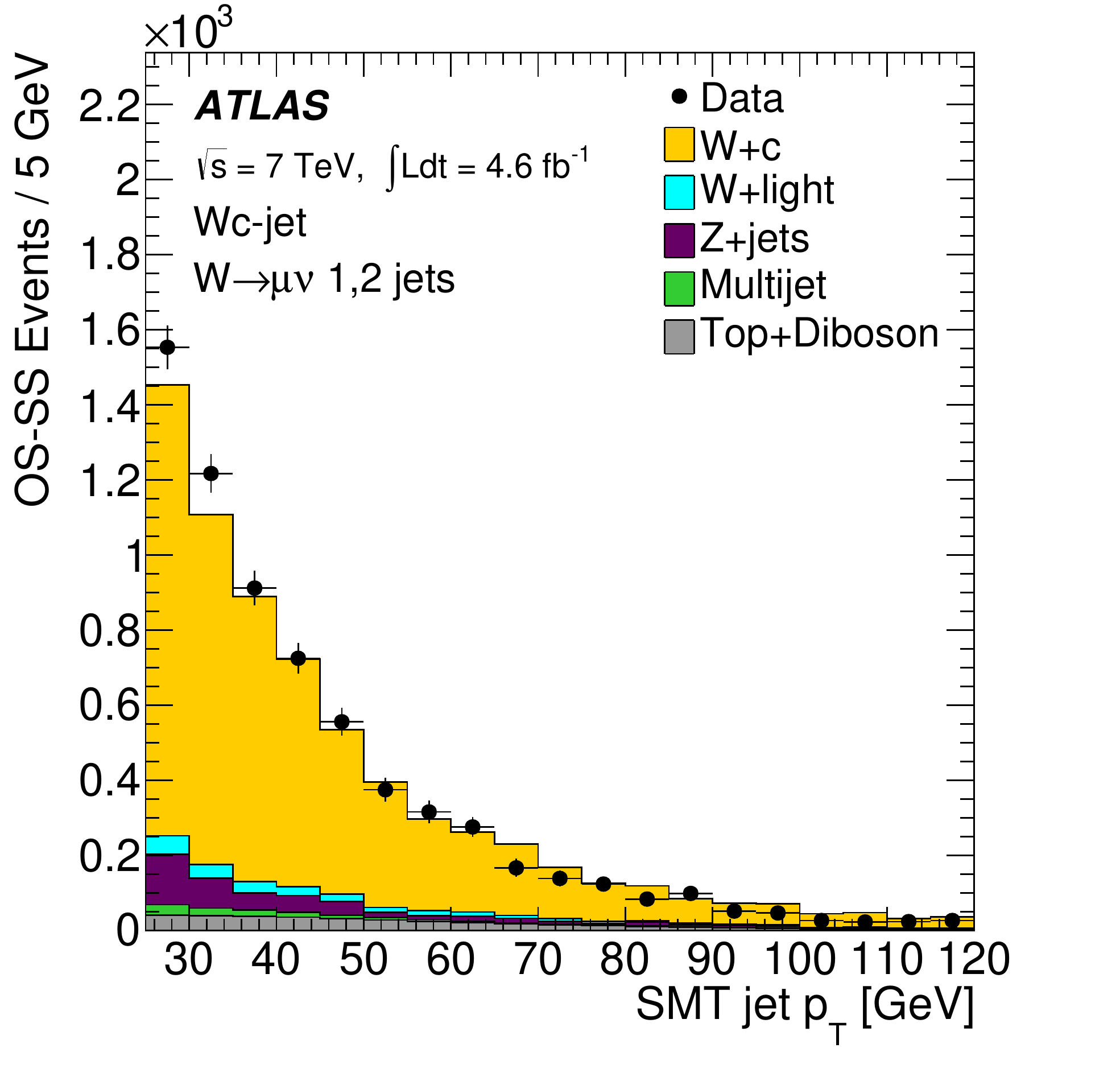} 
\includegraphics[width=0.38\textwidth]{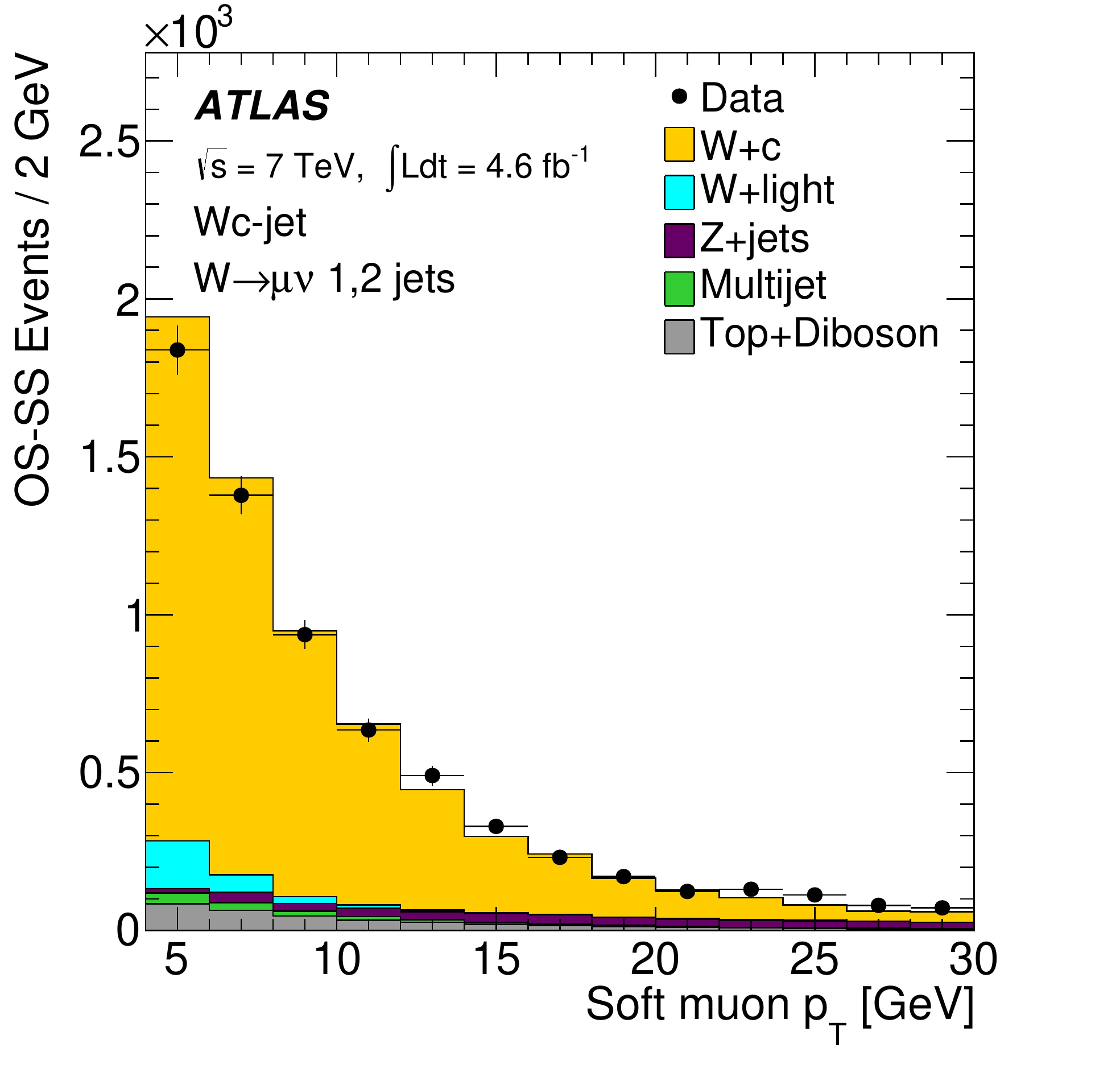}
\caption{Distribution of the SMT jet \pt{} (left) and soft-muon \pt{} (right) in \OSSS\ events of the $W$+1,2 jets sample for the muon channel. The normalisations of the $W$+light and $Z$+jets backgrounds and the shape and normalisation of the multijet background are obtained with data-driven methods. All other backgrounds are estimated with MC simulations and normalised to their theoretical cross sections. The signal contribution is normalised to the measured yields.}
\label{fig:wc_control_muon}
\end{center}
\end{figure}

Figure~\ref{fig:wc_control_muon} shows the distribution of the SMT jet \pt{} and the soft-muon \pt{} in \OSSS\ events in the $W$+1,2 jets sample for the muon channel. The signal contribution is normalised to the measured yields and the background contributions are normalised to the values listed in table~\ref{tab:muon_yields}. The MC simulation is in fair agreement with data.

In addition to the inclusive samples, yields and cross sections are measured in 11 bins of $|\eta^\ell|$, separately for $W^+$ and $W^-$, as is done for the electron channel except that the $|\eta^\ell|$ distribution of the multijet background is derived from the inverted isolation and low transverse mass control regions.

%% file: unfolding.tex
\section{Cross-section determination}
\label{s:unfolding}
\subsection{Definition of the fiducial phase space}
\label{s:phasespace}

The fiducial cross sections \xsecWD and \xsecWc measure the cross sections times the branching ratio $W\rightarrow\ell\nu$ and are determined in a common fiducial region defined in MC simulation in terms of the $W$-boson kinematics as follows:
\begin{itemize}
\item $\pt^\ell>20$\,\GeV\ and $|\eta^\ell|<2.5$,
\item $\ptnu>25$\,\GeV,
\item $\mtw>40$\,\GeV,
\end{itemize}
where $\ell$ and $\nu$ are the charged lepton and the neutrino from the decay $W\rightarrow\ell\nu$.
The leptons are defined before QED final-state radiation. As discussed in the following, the measured raw yields are corrected for detector effects to obtain the cross sections in the fiducial region of the measurement. The charm quark is identified either by a \De meson, and the corresponding cross section measures the production of events with $\ptD>8$\,\GeV\ and $|\eta^D|<2.2$, or through a muon from the semileptonic decay of a charmed hadron embedded in a particle-level jet with $\pt^{\rm{jet}}>25$\,\GeV\ and $|\eta^{\rm{jet}}|<2.5$.
In the latter case, three different cross sections depending on the jet multiplicity are measured: the production of events with exactly one $c$-jet, and either no additional jets (1-jet exclusive), exactly one additional non-$c$-jet (2-jet exclusive), or any number of additional non-$c$-jets (the 1-jet and 2-jet exclusive cross sections are discussed in section~\ref{s:addresultswc}).
Particle-level jets are constructed in simulation from stable particles, including muons and neutrinos, using the anti-$k_t$ algorithm with a radius parameter of 0.4. The lepton, all photons within a cone of $\Delta R=0.1$ around it, and the neutrino originating from the $W$ decay are not used to construct the jets. The particle-level $c$-jet is defined as the one containing a weakly decaying $c$-hadron with $\pt>5$\,\GeV, within $\Delta R<0.3$. Jets containing $c$-hadrons originating from $b$-hadron decays are not counted as $c$-jets. 

The signal yield is defined as the number of events where the $c$-hadron originates from a $c$-quark with charge sign opposite to the charge of the $W$ boson, minus the number of events where the $c$-quark and $W$ boson have the same charge sign.

\subsection{Cross-section determination}
\paragraph{Integrated cross section}
In the \WDe\ analysis the ratios of the fiducial cross sections \xsecratio are measured and the absolute cross sections \xsecWD are obtained by multiplying these cross-section ratios by the inclusive $W$ cross sections $\sigma_{\rm fid}(W)$, while in the \Wce\ analysis the cross sections are obtained directly. The production cross sections in the fiducial region, $\sigma_{\rm fid}$, are calculated using the equation:
\begin{equation}
\sigma_{\rm fid}=\frac{N-B}{C\cdot\int \rm{L} dt},
\label{equ:crosssection}
\end{equation}
where $N$ is the number of candidate events observed in data, $B$ is the number of background events and$\int\rm{L} dt$ is the integrated luminosity of the dataset. The correction factor $C$ is determined from MC simulation and accounts for detector efficiency, acceptance and resolution effects. 

\paragraph[Cross-section ratio \RC]{Cross-section ratio $\boldsymbol{\RC}$}
The ratio \RC\ is computed according to:
\begin{equation}
\RC=\frac{R^{\pm}_{\rm data}}{C^+/C^-},
\end{equation}
where $R^{\pm}_{\rm data}$ is the uncorrected ratio of signal yields in the data, and $C^{+}(C^-)$ is the correction factor defined in equation~(\ref{equ:crosssection}) and calculated separately for $W^++\overline{c}$ and $W^-+c$ events, respectively. 

\paragraph[Differential cross section as a function of lepton $|\eta|$]{Differential cross section as a function of lepton $\boldsymbol{|\eta|}$}
The differential cross sections are determined in intervals of $|\eta^\ell|$ from the same procedure used to determine the total fiducial cross section, but with yields and acceptance corrections determined separately for each $|\eta^\ell|$ bin. Since the resolution of $|\eta^\ell|$ is much higher than the bin widths chosen, simple bin-by-bin corrections are applied. 

\subsection[Determination of the \WDe{} cross sections]{Determination of the $\boldsymbol{\WDe}$ cross sections}
\paragraph[Determination of cross-section ratios \xsecratio]{Determination of cross-section ratios $\boldsymbol{\xsecratio}$}

The cross-section ratios are measured individually for $D$ and \Ds\ mesons. For the measurement of \Ds, the three decay modes are combined.  In the measurement of \xsecratioNoSp, acceptance effects in the $W$ reconstruction largely cancel and are not treated in the correction procedure. Instead, residual effects due to differences in the kinematic distributions between inclusive $W$ and \WDe\ events are estimated from simulation and included in the systematic uncertainty. The corresponding systematic uncertainty is 0.6\%.

The procedure used to correct the measured \ptD\ spectra to the true
spectra takes the reconstruction efficiency and resolution effects
into account. The Bayesian unfolding procedure~\cite{Adye:2011gm} is used and the response matrix is based on a \Pythia +\EvtGen\ simulation in which the backgrounds from reflections from other $D$-meson decay modes are also included.
After \OSSS\ subtraction, the fraction of reconstructed candidates that come from true \De mesons that decay in the mode being reconstructed
depends on the decay mode and is above 99\% for \DpDec and \DsDecNoSp, 93\% for \DsatDec and 60--91\% for \DkpppDecNoSp, depending on \ptD.

In order to obtain the response matrix, the number of reconstructed \WDe\ events per measured \ptD\ bin is determined for each true \ptD\ bin from simulation by the same fitting procedure that is applied to data (see section~\ref{s:qcdbkginWc}). Efficiency corrections are computed separately for the four $D$ decay modes and the two charges. The average reconstruction efficiencies are 32\% for the $D$ decay and 51\%, 25\% and 27\% for the three \Ds\ decay modes ($K\pi$, $K\pi\pi^0$, $K3\pi$), respectively.

After the raw \WDe\ yields are corrected for detector effects, the \ptD\ spectra are corrected for the relative branching ratios according to ref.~\cite{PDG}, resulting in one measurement of the transverse momentum spectrum of the $D$ meson and three measurements of the transverse momentum spectrum of the \Ds\ meson. The measurements of the \Ds\ transverse momentum spectrum for the three \Ds\ decay channels are consistent within uncertainties and are averaged using the procedure discussed in section~\ref{s:combination}. 

The cross-section ratios are then determined by dividing the corrected \ptD\ spectra by the inclusive $W$ yields.  The fiducial cross-section ratios \xsecratio are obtained by integrating the \ptD\ spectra. The same procedure is applied to correct the measured yields in four bins of $|\etaell|$ with the only difference being that bin-to-bin corrections rather than Bayesian unfolding are used. The measured cross-section ratios are presented in section~\ref{s:addresultswd}.

\paragraph[Determination of cross sections \xsecWD]{Determination of cross sections $\boldsymbol{\xsecWD}$}
\label{s:subsecXSecDeterm}

The cross sections \xsecWD are obtained by multiplying \xsecratio\ by the inclusive
$W$ cross section $\sigma_{\rm fid}(W)$ in the fiducial region as determined from equation~(\ref{equ:crosssection}). The efficiency corrections are determined from a simulated sample of MC
events generated with \PowHeg+\PythiaSix. The fiducial cross sections $\sigma_{\rm fid}(W)$
are calculated separately in the electron and muon channels and are afterwards combined. The uncertainties on the extrapolation from the individual lepton measurement regions to the common fiducial region are determined by the methods discussed in ref.~\cite{Aad:2011dm} and amount to 0.6\%. 

\subsection[Determination of the \Wce{} cross sections]{Determination of the $\boldsymbol{\Wce}$ cross sections}

The cross sections \xsecWc\ are determined by applying equation~(\ref{equ:crosssection}) with the efficiency factors, $C$, obtained from the \Alpgen+\PythiaSix\ MC simulations and corrected for charm fragmentation and decay as described in section~\ref{s:dataandmc}. The cross sections are evaluated separately in the exclusive 1-jet and 2-jet bins and extrapolated to the fiducial region with one $c$-jet and any number of additional jets. This constitutes a small extrapolation of the order of 5\% and the related systematic uncertainties are discussed in section~\ref{s:systematics}. The acceptance times the signal selection efficiency is about 2\% owing to the small semileptonic branching ratio.

In addition, the cross section is evaluated for a fiducial volume defined in terms of the kinematics of the muon from the $c$-hadron decay in order to minimise the extrapolation uncertainty. The definition of a particle-level $c$-jet in section~\ref{s:phasespace} is extended to require exactly one muon with $\pt>4$\,\GeV\ and $|\eta|<2.5$ within $\Delta R<0.5$ of the jet axis and with charge opposite to the charge of the $W$ boson. Muons from decays in flight are explicitly excluded.
With this definition, the acceptance times the signal selection efficiency is 35\% in the electron channel and 36\% in the muon channel. The results for this fiducial volume are given in section~\ref{s:addresultswc}.

%% file: systematics.tex
\section{Systematic uncertainties}
\label{s:systematics}

Systematic uncertainties arise from the $W$ reconstruction, the charm tagging, the yield determination and the procedures used to correct for detector effects. The uncertainties on the background and yield determinations are discussed in sections~\ref{s:wd} and~\ref{s:wc}. The other systematic uncertainties considered in this analysis are discussed below. The systematic uncertainties for the cross-section measurements are summarised in table~\ref{t:syswd} for \xsecWD{} and in table~\ref{t:syswc} for \xsecWcNoSp{}. Most of the systematic uncertainties either cancel in the measurement of the ratio \RC\ or are significantly reduced. The remaining systematic uncertainties are shown in table~\ref{t:sysratiowd} for \asymratio and table~\ref{t:sysratiowc} for \asymratiocNoSp.

\begin{table}[h]
\begin{center}
\begin{tabular}{| l | c  |c|  }
\hline
Relative systematic uncertainty in \% & $WD$ & $WD^{*}$\\
\hline
Lepton trigger and reconstruction$^\ast$ & 0.4 & 0.4 \\
Lepton momentum scale and resolution$^\ast$ & 0.2  & 0.2 \\
Lepton charge misidentification  & 0.1 & 0.1 \\
\met\ reconstruction$^\ast$ & 0.4 & 0.4\\
$W$ background estimation  &1.3 & 1.3  \\
Background in \WDe\ events & 0.7 & 0.6  \\   
$W$ efficiency correction  & 0.6 & 0.6  \\
Tracking efficiency & 2.1 & 2.2 \\
Secondary vertex reconstruction efficiency & 0.4 & 0.4\\
\Ds\ isolation efficiency & - & 2 \\
Fitting procedure & 0.8 & 0.5  \\
Signal modelling & 1.4 & 1.9  \\
Statistical uncertainty on response & 0.2 & 0.2  \\ 
Branching ratio & 2.1 & 1.5  \\  
Extrapolation to fiducial region & 0.8 & 0.8\\  
Integrated luminosity$^\ast$ & 1.8 & 1.8\\  
\hline
Total & 4.3 & 4.8\\
\hline
\end{tabular}
\end{center}
\caption{Summary of the systematic uncertainties on the \xsecWD measurement. The uncertainties are given in percent of the measured cross section. Entries marked with an asterisk are correlated between the \Wce\ and \WDe\ measurements.}
\label{t:syswd}
\end{table}
\begin{table}[h]
\begin{center}
\begin{tabular}{| l | c |c|  }
\hline
Relative systematic uncertainty in \% & $W(e\nu)c$-jet & $W(\mu\nu)c$-jet \\
\hline
Lepton trigger and reconstruction$^\ast$  & 0.7 & 0.8 \\
Lepton momentum scale and resolution$^\ast$  & 0.5 & 0.6 \\
Lepton charge misidentification  & 0.2 & - \\
Jet energy resolution$^\ast$ & 0.1 & 0.1 \\
Jet energy scale & 2.4 & 2.1 \\
\met\ reconstruction$^\ast$ & 0.8 & 0.3 \\
Background yields  & 4.0 & 1.9 \\
Soft-muon tagging  & 1.4 & 1.4 \\
$c$-quark fragmentation  & 2.0 & 1.6 \\
$c$-hadron decays  & 2.8 & 3.0 \\
Signal modelling  & 0.9 & 0.2 \\
Statistical uncertainty on response  & 1.4 & 1.4 \\
Integrated luminosity$^\ast$  & 1.8 & 1.8 \\
\hline
Total  & 6.5 & 5.3 \\

\hline
\end{tabular}
\end{center}
\caption{Summary of the systematic uncertainties on the \xsecWc measurement. The uncertainties are given in percent of the measured cross section. Entries marked with an asterisk are correlated between the \Wce\ and \WDe\ measurements.}
\label{t:syswc}
\end{table}

\begin{table}[h]
\begin{center}
\begin{tabular}{| l | c |c|  }
\hline
Relative systematic uncertainty in \% & $WD$ & $WD^{*}$\\
\hline
Lepton reconstruction and identification & $<0.1$ & $<0.1$ \\
Background in \WDe\ events & 0.6 & 0.4  \\   
Tracking efficiency & 0.2 & 0.2 \\
Statistical uncertainty on response & 0.2 & 0.2  \\
\hline
Total & 0.7 & 0.5\\
\hline
\end{tabular}
\end{center}
\caption{Summary of the significant systematic uncertainties on the measurement of the ratio \asymratioNoSp. The uncertainties are given in percent.}
\label{t:sysratiowd}
\end{table}
\begin{table}[h]
\begin{center}
\begin{tabular}{| l | c |c | }
\hline
Relative systematic uncertainty in \% & $W(e\nu)c$-jet & $W(\mu\nu)c$-jet \\
\hline

Lepton trigger and reconstruction  & $<$0.1  & $<$0.1 \\
Lepton momentum scale and resolution  & 0.2 & 0.6 \\
Lepton charge misidentification  & $<$0.1 & - \\
Jet energy resolution & 0.1 & 0.1 \\
Jet energy scale & 0.2 & 0.6 \\
\met\ reconstruction & 0.3 & 0.3 \\
Background yields  & 1.4 & 1.0 \\
Soft-muon tagging  & 0.2 & $<$0.1 \\
Signal modelling  & 1.4 & 1.4 \\
Statistical uncertainty on response  & 0.5 & 0.5 \\
\hline
Total  & 2.1 & 2.0 \\

\hline
\end{tabular}
\end{center}
\caption{Summary of the significant systematic uncertainties on the measurement of the ratio \asymratiocNoSp. The uncertainties are given in percent.}
\label{t:sysratiowc}
\end{table}

\subsection{Common systematic uncertainties}
\label{s:common_uncertainties}
Uncertainties on the basic detector response and its modelling affect both the \WDe\ and \Wce\ analyses. The trigger and reconstruction efficiencies for electrons and muons are varied in the simulation within the range of their uncertainties as determined from data, and the \WDe\ and \Wce\ cross sections are recalculated. A similar procedure is used to assess the uncertainty due to the lepton momentum scale and resolution. Lepton charge misidentification effects are also considered. The charge misidentification rates for electrons and muons are given in ref.~\cite{Aad:2011mk,Aad:2008zzm} and are significant only for the electron channel. Uncertainties related to the selection and measurement of jets affect primarily the \Wce\ analysis and to a much smaller extent the \WDe\ analysis, the latter only via the \met\ reconstruction. The main sources of uncertainty for jets are due to the jet energy scale (JES) and the jet energy resolution (JER). The impact on the cross-section measurements is evaluated by varying each of these in the simulation within their respective uncertainties as determined from data. The JES uncertainty ranges from less than 1\% to about 7\%, depending on jet $\pt$ and $\eta$ \cite{ATLAS:2013wma}, with an additional 2\% assigned to charm jets. Together, the JES and JER uncertainties contribute at the few percent level to the \Wce\ cross-section measurement. Uncertainties on the lepton and jet momentum scale and resolution are propagated to the \met\ reconstruction. Additional uncertainties on the \met\ from soft jets (those with 7\,\GeV\,$<\pt<20$\,\GeV) and calorimeter cells not associated with any reconstructed objects are accounted for separately. The uncertainty on the integrated luminosity is 1.8\%~\cite{r:lumi}.

\subsection[Systematic uncertainties on \WDe]{Systematic uncertainties on $\boldsymbol{\WDe}$}

\subsubsection{Tracking efficiency}
The primary source of the uncertainty on the tracking efficiency is the potential mismodelling of
the distribution of detector material in the simulation. The amount of material in the ID is known with a precision of 5\% in the central region and up to 15\% in the more forward region. In order to study the effect on the tracking efficiency, samples generated with distorted geometries and with increased material are used. In these samples, the efficiency is found to decrease by 2.1(6.5)\% for the \De candidates with three (five) tracks. The systematic uncertainty of the weighted average is 2.2\% for the \Ds{} channel and 2.1\% for the \Dpm{} channel. These uncertainties are symmetrised to account for the possibility that the tracking efficiency could be higher in data than in simulation. The systematic uncertainty due to the material uncertainty is significantly reduced in the measurement of the ratio \RC\ since only charge-dependent effects are important.

In addition, there can be a loss of tracking efficiency in the
core of jets at high jet $\pt$  due to hit-sharing among tracks. Studies of such hit-sharing show that the simulation and data agree well and
that the resulting systematic uncertainty is negligible for
jets with $\pt$ below 500\,\GeV~\cite{Aad:2011sc}. Hence, such effects
are not considered further for this analysis.

\subsubsection{Secondary-vertex reconstruction efficiency}
Uncertainties associated with the modelling of the efficiency of the secondary-vertex fit quality requirement are estimated by comparing the efficiency of this requirement in the MC simulation and in the semileptonic $B$-decay data control region discussed in section~\ref{s:qcdbkginWc}.

\subsubsection[\Ds\ isolation efficiency]{$\boldsymbol{\Ds}$ isolation efficiency}
The efficiency of the \Ds\ isolation requirement in simulation is sensitive to the choice of fragmentation model. The systematic uncertainty is obtained from the difference between the isolation efficiency determined from the \PythiaSix\ and \Herwig\ generators. The systematic uncertainty depends on \ptD\ and amounts to 2.2\%, 2.4\%, 1\%\ and 0.8\%\ for the four \ptD\ bins.

\subsubsection{Fitting procedure}
The systematic uncertainties from the fitting procedure are dominated by the modelling of the shape of the signal templates. The main sources are the mass resolution and the amount of background from the reflections included in the signal templates. Systematic studies are performed on simulated events, and differences in the mass resolution between the simulation and the data are observed in the control region used to derive the signal templates.
For the baseline analysis, 
the simulated mass distribution is smeared to reproduce the measured resolution in the data.
Systematic uncertainties are assessed by varying the smearing parameters within
their uncertainties. 
The amount of reflection background in the signal templates is altered by changing the normalisation of the combinatorial background component to provide the best agreement with the data in the control region. Refits to the simulated $m(D)$ or $\Delta m$ distributions which use these modified backgrounds and resolutions
result in uncertainty estimates of 0.8\% for the $D$ decay and 0.5\%, 1\% and 5\% for the three \Ds\ decay modes ($K\pi$, $K\pi\pi^0$, $K3\pi$) respectively. The uncertainty is larger in the \DzkpppDec channel due to higher background. In addition, the uncertainty due to the description of the shape of the combinatorial background in the $\Delta m$ distribution is evaluated. An alternative control region, the $D^0$ mass sideband at $m(D^0)>1.985$\,\GeV, is used and the resulting fits are compared to the baseline procedure discussed in section~\ref{combbg}. The results agree within the statistical uncertainty and no additional systematic uncertainty is introduced.

\subsubsection{Signal modelling}

A potential bias, introduced by the choice of the simulated \ptD\ spectrum assumed when calculating the correction factors, is studied by repeating the correction procedure with a reweighted \ptD\ spectrum which matches the observed spectrum.  The resulting uncertainty estimates depend on the decay mode and vary from 0.5\%\ to
1.3\%. 

Since the resolution and efficiency are $\etaD$  dependent, uncertainties in the modelling of the $\etaD$ spectrum need to be taken into account. This is done by reweighting the distribution generated by \PythiaSix\ to the NLO prediction of \aMCatNLO~\cite{Frederix:2011zi} showered with \Herwigpp. Even though differences of up to 20\% are seen in the forward regions, both predictions are compatible within uncertainties with the $\etaD$ distribution observed in data. The corresponding systematic uncertainties are 2.0\%, 1.3\% and 2.3\% for the three \Ds\ decay modes ($K\pi$, $K\pi\pi^0$, $K3\pi$) and 1.1\% for the $D$ decay. 

The systematic uncertainties due to the correction for detector effects are estimated
by repeating the correction procedure on ensembles of MC events obtained by appropriate random variations of parameters describing detector performance.

Uncertainties on the total \WDe{} yields due to the finite size of the simulated 
sample used to obtain the
correction factors vary among decay modes from 0.2\%\ to 0.4\%;
the uncertainties on the yield in the highest \ptD\ bin are 1.9\% for $D^+$ and 3.3\%, 6.2\%\ and 5\%\ for the
three \Ds\ decay modes ($K\pi$, $K\pi\pi^0$, $K3\pi$). 

\subsubsection{Branching ratio}
An additional uncertainty is introduced by the correction for the relative branching ratios, which are currently known with a precision of 2.1\%, 1.3\%, 3.6\% and 2.6\%~\cite{PDG} for the \DpDec, \DsDec, \DsatDec and \DkpppDec decay channels, respectively. 

\subsubsection{Extrapolation to fiducial region}
Due to differences in the electron and muon triggers, the minimum \pte requirement in the electron channel used in the evaluation of the cross-section ratios is higher than that of the muon channel. The systematic uncertainty on the ratios \xsecratio\ due to the extrapolation from the requirement of $\pte>25$\,\GeV\ in the electron channel to the common fiducial region with $\ptell>20$\,\GeV\ is determined from simulation.

\subsection[Systematic uncertainties on \Wce]{Systematic uncertainties on $\boldsymbol{\Wce}$}
\label{syswcsec}
\subsubsection{Soft-muon tagging}
The soft-muon tagging efficiency and mistag rates are varied in the simulation within the range allowed by the tagging efficiency ($\leq 1\%$) and mistag ($15\%$) calibrations. The soft-muon reconstruction efficiency is varied in the simulation within the calibration uncertainty ($\simeq 1\%$) and is the dominant contribution to the SMT uncertainties.

\subsubsection[$c$-quark fragmentation]{$\boldsymbol{c}$-quark fragmentation}
As in the \WDe\ analysis, the correction factor for detector effects depends on the modelling of the signal kinematics and its accuracy. In particular, the $c$-quark fragmentation and $c$-hadron decay models affect the simulated soft-muon $\pt$ spectrum and the number of $c$-hadrons decaying to muons. In this analysis, the quark fragmentation is simulated with \PythiaSix\ and then corrected for discrepancies in the type and relative population of $c$-hadrons resulting from the charm fragmentation and the fraction of the $c$-quark energy carried by the $c$-hadrons. To improve on the \PythiaSix\ modelling, the fragmentation fractions in \PythiaSix\ are reweighted to those derived from the combination of measurements in  $e^+e^-$ and $ep$ collisions \cite{r:ffrac} and the respective uncertainties are taken into account. The modelling of the momentum fraction of $c$-hadrons ($\pt^{c\rm{-hadron}}/\pt^{c\rm{-jet}}$) in \PythiaSix\ is reweighted to the fraction given by \Herwigpp. The modelling of the fragmentation function in \Herwigpp\ is validated by comparing the simulation to $e^+e^-$ data as discussed in section~\ref{subs:theory}. Based on these studies, a systematic uncertainty of 2\%\ is assigned to the mean value of the fraction of the charm-quark momentum carried by the charmed hadron.

\subsubsection[$c$-hadron decays]{$\boldsymbol{c}$-hadron decays}
Two observables are used to represent the modelling of the decay of $c$-hadrons inside jets: the branching ratios of $c$-hadron semileptonic decays to muons, and the momentum of the muon ($p^*$) in the rest frame of the $c$-hadron. The latter is important because of the minimum $\pt$ requirement of 4\,\GeV\ on the soft muon. The semileptonic branching ratios of $c$-hadrons used in \PythiaSix\ are rescaled to the world average values \cite{PDG} and the respective uncertainties are taken into account. The distribution of $p^*$ from \PythiaSix\ is reweighted to correspond to the one given by \EvtGen\ and the difference between the two is considered as a systematic uncertainty.

\subsubsection{Signal modelling}
The impact on the signal acceptance terms stemming from uncertainties on the simulated jet multiplicity is estimated by varying the amount of initial- and final-state radiation in the \PythiaSix\ parton-shower parameterisation.
Additionally, the ratio of one-jet events to two-jet events in simulation is reweighted to the ratio measured in data and the acceptance is recomputed.
The difference between the derived cross sections is less than 1\% and is taken as a systematic uncertainty in the jet multiplicity modelling. Additional uncertainties on the non-perturbative physics modelling (e.g. underlying event, parton shower, color flow) are evaluated by recomputing the acceptance based on a simulation of the \Wce\ signal by \Alpgen\ + \Herwig\, in which the \Herwig\ charm fragmentation and decay are corrected using the procedure described previously for correcting \PythiaSix. The difference between the nominal and the recomputed acceptances is less than 1\% and is used as the systematic uncertainty estimate.

The kinematics of the generated events used to calculate the acceptance is influenced by the PDF set used for the event generation. Thus the choice of PDF set affects the result. To evaluate the magnitude of the effect, the acceptance is recomputed after reweighting the simulated signal sample with four different PDF sets 
(\MSTW, \NNPDF, \HERA\ and \epWZ~\cite{Aad:2012sb})
using {\sc lhapdf} \cite{lhapdf}.
The maximum difference between the acceptances derived with a single PDF eigenvector set or the different PDF central values is taken as the systematic uncertainty.
Finally, the uncertainty on the correction factors due to the limited simulated signal sample size is 1.4\%.

%% file: results.tex
\section{Results and comparison to theoretical predictions}
\label{s:results}

\subsection{Data combination}
\label{s:combination}

The combination of the cross-section measurements of \WDse, \WDpe\ and \Wce, in the electron and muon channels, is discussed in this section. The combination procedure is applied to the integrated cross-section measurements as well as to the measurements differential in $|\eta^\ell|$. The procedure is based on the averaging method developed in ref.~\cite{Aaron:2009bp}, which takes into account statistical uncertainties as well as systematic uncertainties (bin-to-bin correlated and uncorrelated) proportional to the central values of the respective cross sections.
The combined cross sections ($m^i$) in bins $i$ are derived from the individual cross-section measurements ($\mu^i_k$) in channels $k$ by minimising the following $\chi^2$ function:

\begin{equation}
 \chi^2= \sum_{k,i} w^i_k \frac{\left[{\mu^i_k} - \left(m^i + \sum_j \gamma^i_{j,k} m^i b_j\right)\right]^2}{(\delta^i_{\mathrm{sta}, k})^2 \mu^i_k (m^i-  \sum_j \gamma^i_{j,k} m^i b_j ) + (\delta^i_{\mathrm{unc},k} m^i)^2 }+ \sum_j b_j^2, \\
\label{eq:comb_chi2_averaging}
\end{equation}

\noindent where $w^i_k=1$ if channel $k$ contributes to measurement $\mu^i_k$ in bin $i$, and $w^i_k=0$ otherwise. The parameters $b_j$ denote the shift introduced by a correlated systematic error source $j$ normalised to its respective standard deviation. The relative statistical and uncorrelated systematic uncertainties on $\mu^i_k$ are denoted by $\delta^i_{\mathrm{sta}, k}$ and $\delta^i_{\mathrm{unc},k}$ and the variable $\gamma^i_{j,k}$ quantifies the relative influence of the correlated systematic error source $j$ on the measurement $\mu^i_k$.

The sources of systematic uncertainties which are fully correlated between the different measurements and the electron and the muon channels are uncertainties due to the modelling of charm fragmentation and decay, uncertainties on the \met\ reconstruction and the luminosity uncertainty. Uncertainties on the lepton reconstruction and identification efficiencies and momentum scale and resolution are correlated among the \WDse, \WDpe\ and \Wce\ measurements, but uncorrelated between the electron and the muon channel. Uncertainties due to the track and vertex reconstruction are treated as correlated among the \WDse\ and \WDpe\ channels, while uncertainties in the $c$-jet signal reconstruction and identification are correlated for the electron and muon channels in the \Wce\ analysis. Since different methods are used to determine the backgrounds in the individual channels, the corresponding uncertainties are assumed to be uncorrelated among the different channels, but correlated bin-to-bin. 

In total there are 58 differential cross-section measurements in 38 independent bins entering the combination with 113 sources of correlated systematic uncertainties.  The measured integrated cross sections together with their statistical and systematic uncertainties resulting from the averaging procedure are reported in table~\ref{tab:results_wc_IntegralXsec_AV1xj_AVplusminus}. Tabulated values of all observed results are available in the Durham HEP database \footnote{A complete set of tables with the full results are available at the Durham HepData repository, http://hepdata.cedar.ac.uk.}.

The correlation between the total uncertainties of the integrated \Wce\ and \WDse\ measurements is found to be approximately 10\%, while it is about 5\% for \Wce\ and \WDpe\ due to the larger statistical uncertainty of the \WDpe\ sample. The correlation between \WDse\ and \WDpe\ is approximately 20\%. Furthermore, the correlations between the uncertainties in the $W^+$ and the $W^-$ channels are 76\%, 58\% and 17\% for \Wce, \WDse\ and \WDpe, respectively.
Different channels use complementary $c$-hadron decay modes and the statistical overlap between the different selected data samples is of the order of 1\%. Therefore the correlations between the statistical uncertainties are neglected. 

In addition to the cross-section measurements, the averaging procedure is also applied 
to the measurements of the cross-section ratios \asymratioc and \asymratioNoSp. 
The measurements of the ratios are dominated by statistical uncertainties, 
since most of the systematic uncertainties cancel in the ratio or are significantly reduced 
(see tables~\ref{t:sysratiowd} and~\ref{t:sysratiowc}). In particular, 
the systematic uncertainties due to the lepton reconstruction and the luminosity are 
negligible for the ratio measurements. The measurements in the \Wce\ and \WDe\ channels are therefore almost completely
uncorrelated. The measurements of the cross-section ratios \asymratioDs and \asymratioDp are combined since the measurements are performed in a similar phase space ($\ptD>8$\,\GeV, $|\etaD|<2.2$) and residual differences are predicted to be small. The measurement of \asymratioc on the other hand is sensitive to a different phase space at higher $c$-jet transverse momentum ($\pt^{\rm{jet}}>25$\,\GeV, $|\eta^{\rm{jet}}|<2.5$). 
Consequently, the \Wce\ measurement is not combined with the \WDe\ measurement, but is subject to the common averaging procedure using equation~(\ref{eq:comb_chi2_averaging}). 

\begin{table}[htbp]
\centering
\begin{tabular}{| l |c|}
\hline
 & $\sigma^{\OSSS}_{\rm fid}$ [pb] \\
 \hline
   $W^+\overline{c}$-jet  &$33.6\pm0.9\,{\rm(stat)}\pm1.8\,{\rm(syst)}$\\
   $W^-c$-jet &$37.3\pm0.8\,{\rm(stat)}\pm1.9\,{\rm(syst)}$\\
   \WpDp &$17.8\pm1.9\,{\rm(stat)}\pm0.8\,{\rm(syst)}$\\
   \WmDp &$22.4\pm1.8\,{\rm(stat)}\pm1.0\,{\rm(syst)}$\\
   \WpDs &$21.2\pm0.9\,{\rm(stat)}\pm1.0\,{\rm(syst)}$\\
    \WmDs &$22.1\pm0.8\,{\rm(stat)}\pm1.0\,{\rm(syst)}$\\

\hline

\end{tabular}
\caption{Measured integrated cross sections times the branching ratio $W\rightarrow\ell\nu$ in the fiducial regions together with the statistical and systematic uncertainties.}
\label{tab:results_wc_IntegralXsec_AV1xj_AVplusminus}
\end{table}

\subsection{Theoretical predictions}
\label{subs:theory}

The theoretical predictions for the cross sections \xsecWD and \xsecWc are obtained from the \aMCatNLO~\cite{Frederix:2011zi} MC simulation that incorporates NLO QCD matrix-element calculations into a parton-shower framework. The \aMCatNLO\ event generator is based on the \MCatNLO\ formalism~\cite{Frixione:2002ik} 
and the MadGraph5 framework~\cite{Alwall:2011uj}. 
The parton-level cross section obtained with \aMCatNLO{} was found to be in 
good agreement with the prediction obtained using MCFM~\cite{Campbell:2010ff}.
\Herwigpp~\cite{Bahr:2008pv} is used to model the
parton shower, hadronisation and underlying event
of the \aMCatNLO{} simulation. The MC predictions for the charmed-hadron production fractions
are corrected to the average of measurements obtained in $e^+e^-$ and $ep$ collisions, as compiled in ref.~\cite{r:ffrac}. The uncertainties on these
production fractions are 2.4\% for the \Ds\ meson and 3.4\% for the $D$ meson and are included in the evaluation of the systematic uncertainty on the prediction.

Events are generated in \aMCatNLO\ using the \CT\ NLO PDF set. The dependence of the results on the choice of PDF set is checked by reweighting the \aMCatNLO\ predictions using various NLO and NNLO PDF sets: the \CT, \MSTW, \HERA, \NNPDF\ and \NNPDFcoll~\cite{Ball:2012cx} NLO PDF sets are used in addition to the \epWZ\ NNLO PDF set.
Asymmetric uncertainties are calculated following the prescriptions from the PDF sets.

For \MSTW\ and \NNPDF\, the $s$-quark sea is suppressed relative to the $d$-quark sea for all values of $x$. The \epWZ\ PDF set, which is based on the analysis of ATLAS
$W$ and $Z$ cross-section measurements~\cite{Aad:2011dm} together with HERA data~\cite{Aaron:2009aa}, has an $s$-quark PDF that is not suppressed with respect to the $d$-quark sea at $x\sim0.01$. The $s$-quark sea in \CT\ is less suppressed than in \MSTW\ or \NNPDF. The \NNPDFcoll\ PDF set uses only data from 
HERA, the Tevatron and the LHC, so that the data from charm production in neutrino--nucleon scattering are excluded. The $s$-quark sea of this PDF is larger than the $d$-quark sea at most values of $x$.

Processes with charm quarks in the initial state such as $dc\rightarrow W^-uc$ and $d\bar{c}\rightarrow W^-u\bar{c}$ can contribute to the \OSSS{} $W+c$ signal if there is an asymmetry in the charm and anti-charm PDFs.
The PDF sets studied here do not include a non-perturbative (intrinsic) charm component~\cite{Brodsky:1980pb}, where significant asymmetries are possible. PDF fits that include phenomenological models of intrinsic charm~\cite{Pumplin:2007wg,Martin:2009iq,Dulat:2013hea} indicate that for the values of $x$ relevant for this analysis, these contributions are expected to be small.

The dependence of the NLO prediction on the choice of
renormalisation and factorisation scales is evaluated by
independently halving and doubling their nominal value which is chosen as the sum of the transverse mass of all final-state particles. The largest variation where the scales are varied in opposite directions is taken as the uncertainty and treated as fully correlated. 
This uncertainty is $+8/-9$\% for the \WDe\ analysis and $+8/-4$\% for the \Wce\ analysis. To study the modelling of the charm fragmentation function in \Herwigpp, $e^+e^-$ annihilation events are
generated at centre-of-mass energy $\sqrt{s}=10.6$\,\GeV\ and the distribution of $x_p \equiv p/p_{\rm{max}}$ for $D$ and \Ds\ 
is compared to the data from ref.~\cite{Seuster:2005tr}. 
The evolution of the charm fragmentation function with $Q^2$ in
\Herwigpp\ is validated by generating $e^+e^-$ annihilation events  
at $\sqrt{s} = 91.2$\,\GeV\ 
and comparing the mean value of $x_E \equiv E/E_{\rm{beam}}$ for \Ds\ to that
measured in ref.~\cite{Barate:1999bg}.
Based on these studies, a systematic uncertainty
of 2\%\ is assigned to the \aMCatNLO\ predictions of \WDe\ for the mean value of
the fraction of the charm-quark momentum carried by the charmed
hadron. The effect of the charm fragmentation uncertainty on the predicted \Wce\ cross section is negligible, while its effect on the acceptance correction is discussed in section~\ref{syswcsec}.   

The effect of the uncertainty in the parton-shower model used in the MC simulation is estimated by comparing the predictions of different MC generators. The corrections for the charm fragmentation and decay discussed in section~\ref{s:dataandmc} are applied to all MC simulations to avoid a potential double counting of the uncertainties. The comparison of the fiducial cross sections obtained with \aMCatNLO+\Herwigpp, \aMCatNLO+\Herwig, \PowHeg+\Herwig\ and \PowHeg+\Pythia{} indicates a systematic uncertainty of 3\% for \WDe\ and 1\% for \Wce\ due to the modelling of the parton shower. 

\subsection{Discussion}
\label{subs:xsec}

\begin{figure*}
\begin{center}
\includegraphics[width=0.49\textwidth]{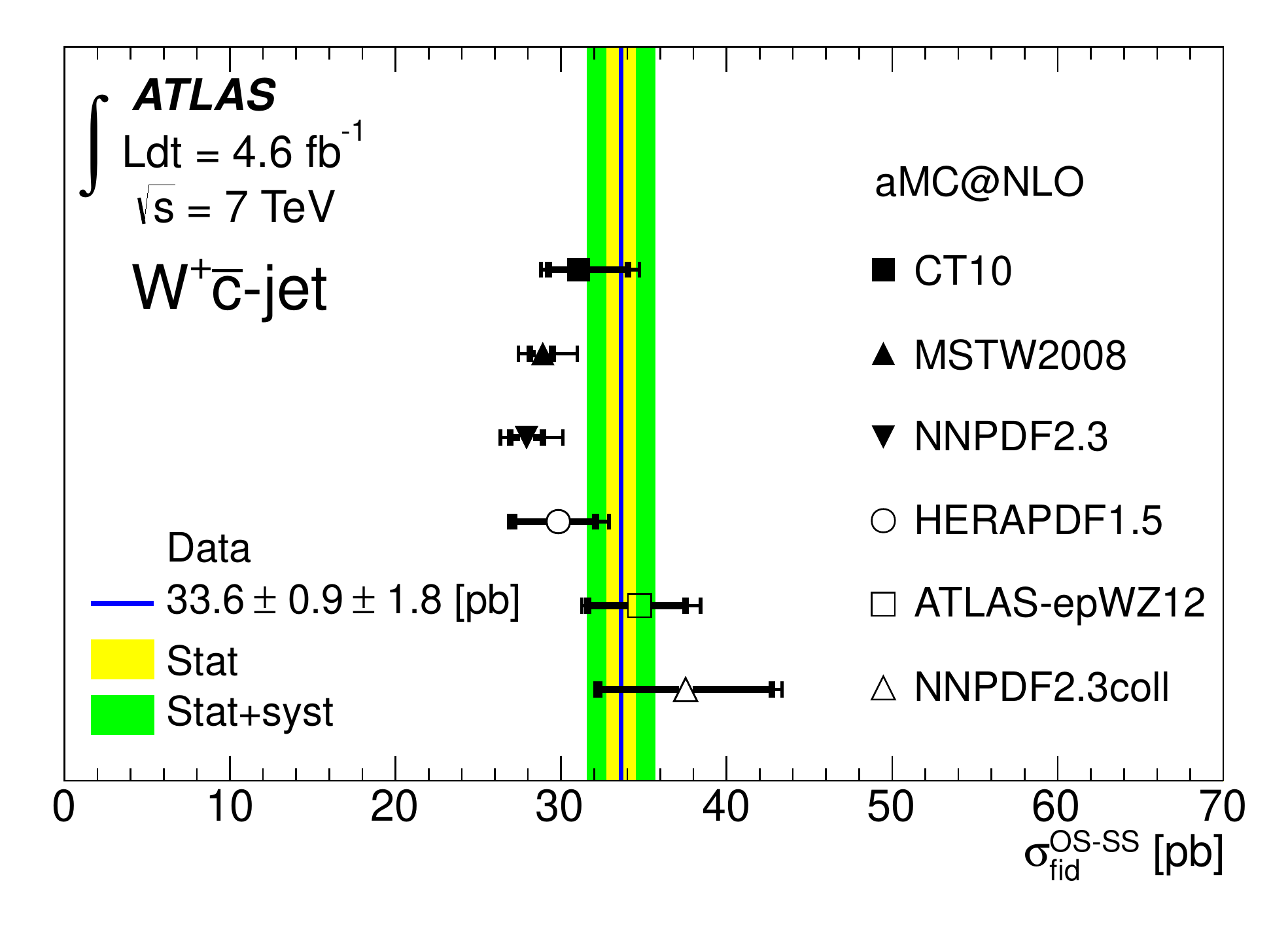}  
\includegraphics[width=0.49\textwidth]{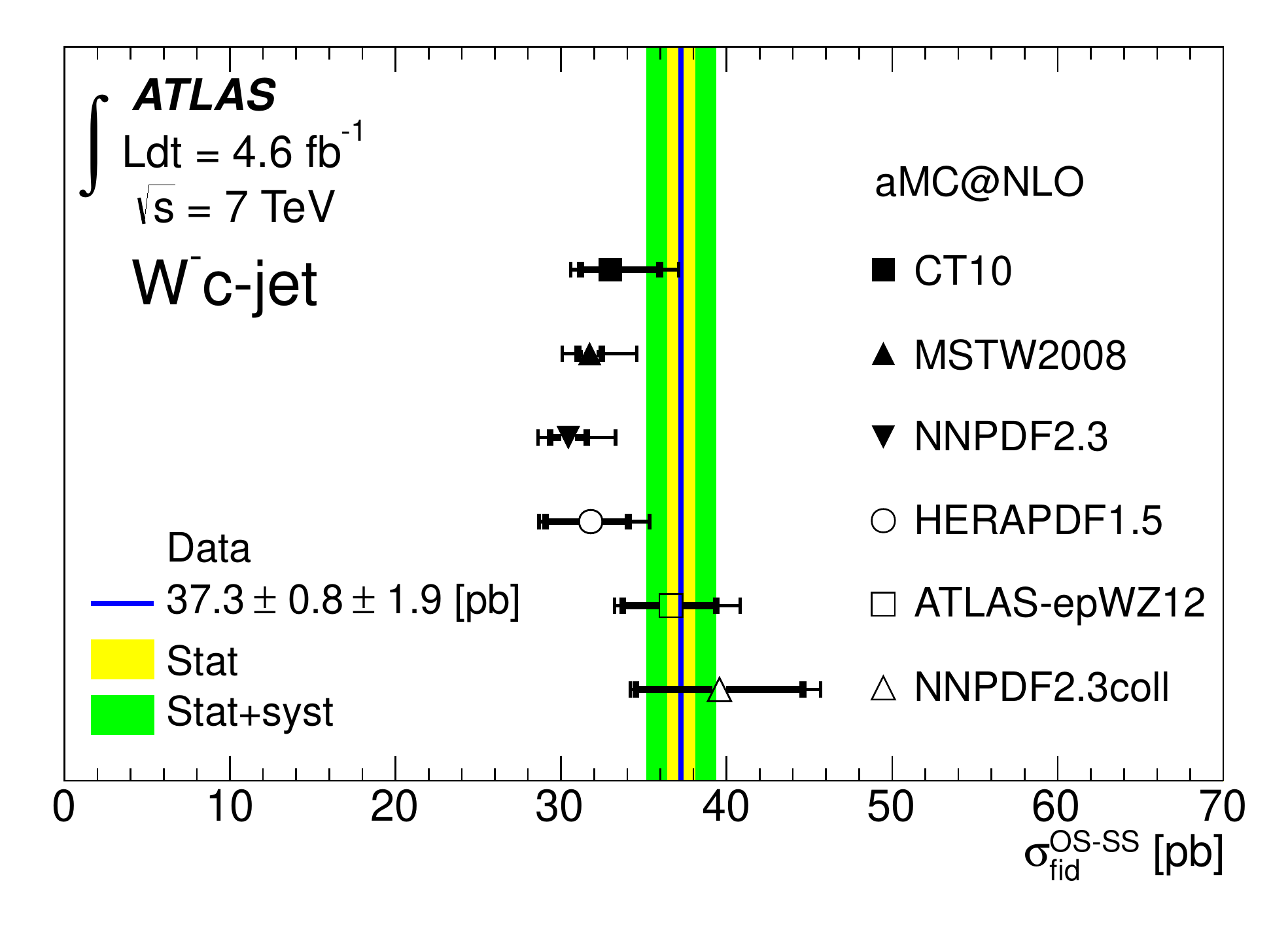}\\  
\includegraphics[width=0.49\textwidth]{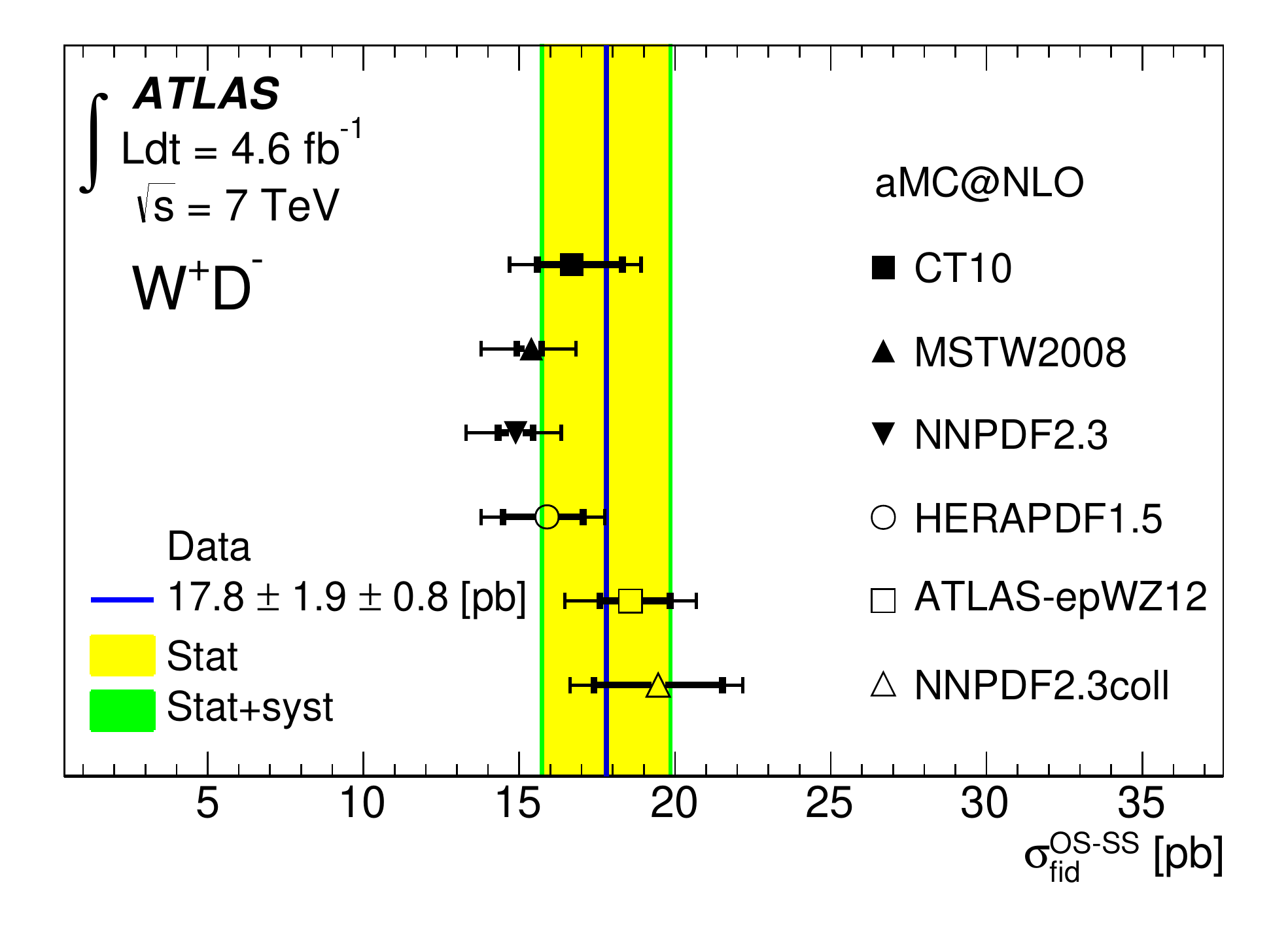}  
\includegraphics[width=0.49\textwidth]{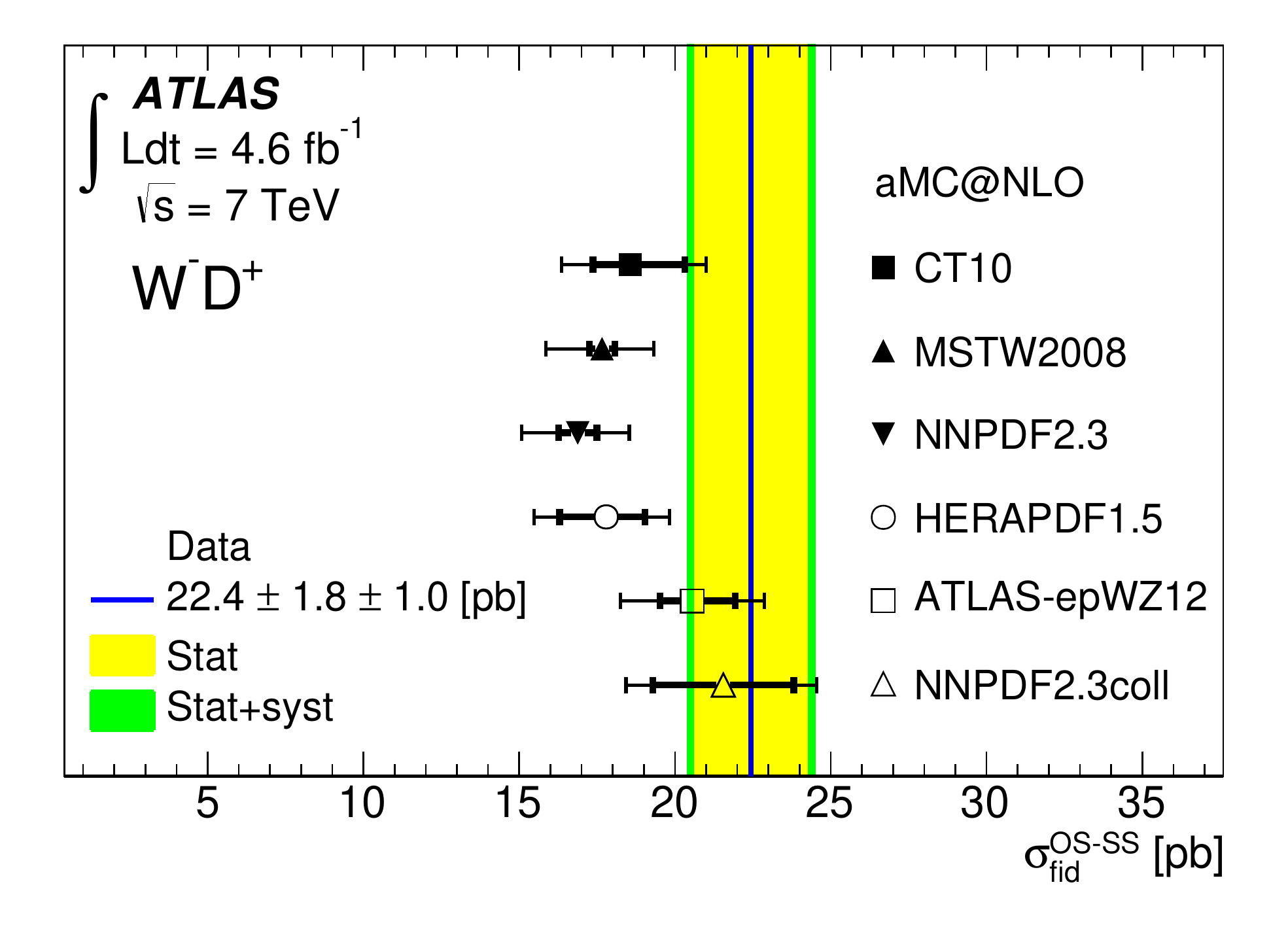} \\ 
\includegraphics[width=0.49\textwidth]{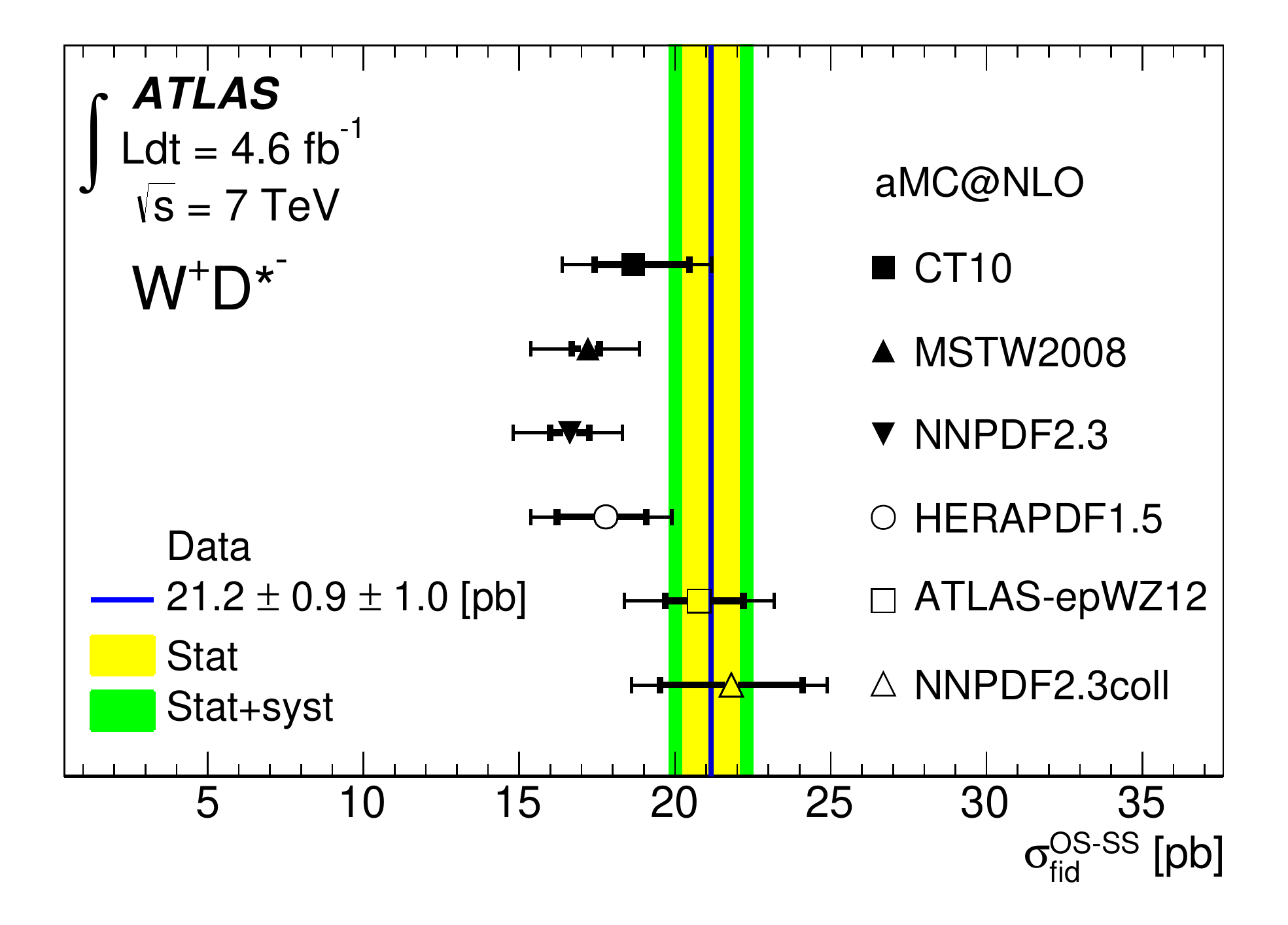} 
\includegraphics[width=0.49\textwidth]{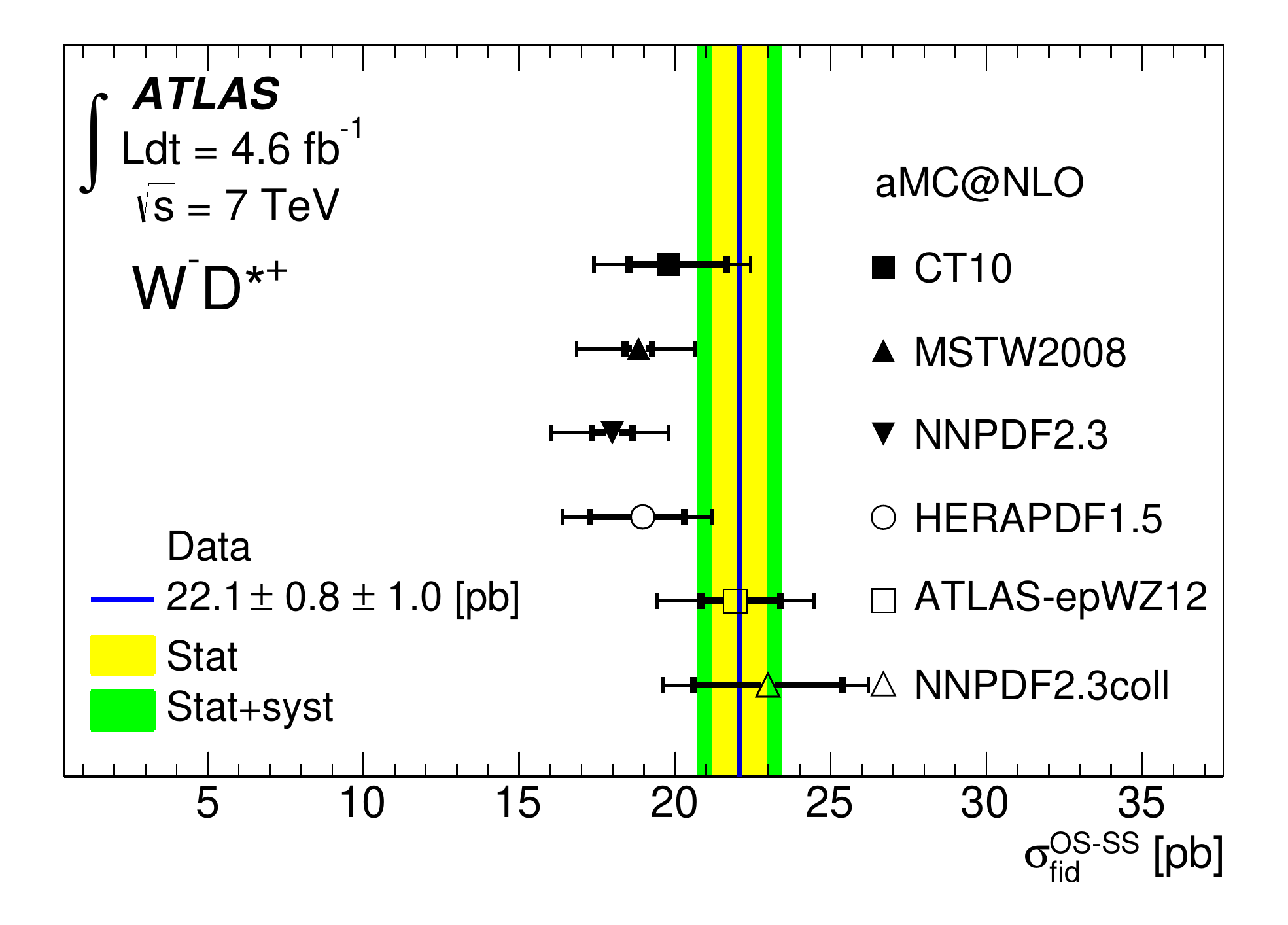}
\caption{Measured fiducial cross sections compared to various PDF predictions based on \aMCatNLO{}. The solid vertical line shows the central value of the measurement, the inner error band corresponds to the statistical uncertainty and the outer error band to the sum in quadrature of the statistical and systematic uncertainties. The PDF predictions are shown by markers. The inner error bars on the theoretical predictions show the 68\%\ confidence level uncertainties obtained from the error sets provided with each PDF set, while the outer error bar represents the total theoretical uncertainty (sum in quadrature of PDF, parton shower, fragmentation and scale uncertainties).}
\label{f:xsecamc}
\end{center}
\hspace{1mm}
\end{figure*}

The measured integrated fiducial cross sections~\xsecWD and \xsecWc are compared to the theoretical
predictions based on various PDF sets in figure~\ref{f:xsecamc}. 
The inner error bars on the theoretical predictions are the 68\%\ confidence
level (CL) uncertainties obtained from the error sets provided with each
PDF set, while the outer error bars show the sum in quadrature of these PDF
uncertainties and theoretical uncertainties due to variations in renormalisation
and factorisation scale, parton shower and charm-quark fragmentation as discussed previously. 
The predicted cross sections differ by as much as 25\%. 
The six different measurements give a consistent picture; the predictions obtained with the \epWZ\ and \NNPDFcoll\ sets are seen to overlap more with the data but simulations using \CT{}, \HERA\ and \MSTW\ 
also are in agreement with the measurements. The prediction obtained with \NNPDF\ is less favoured. A quantitative comparison of the various PDF predictions with the measured cross sections is discussed below.

\begin{figure*}
\begin{center}
\includegraphics[width=0.48\textwidth]{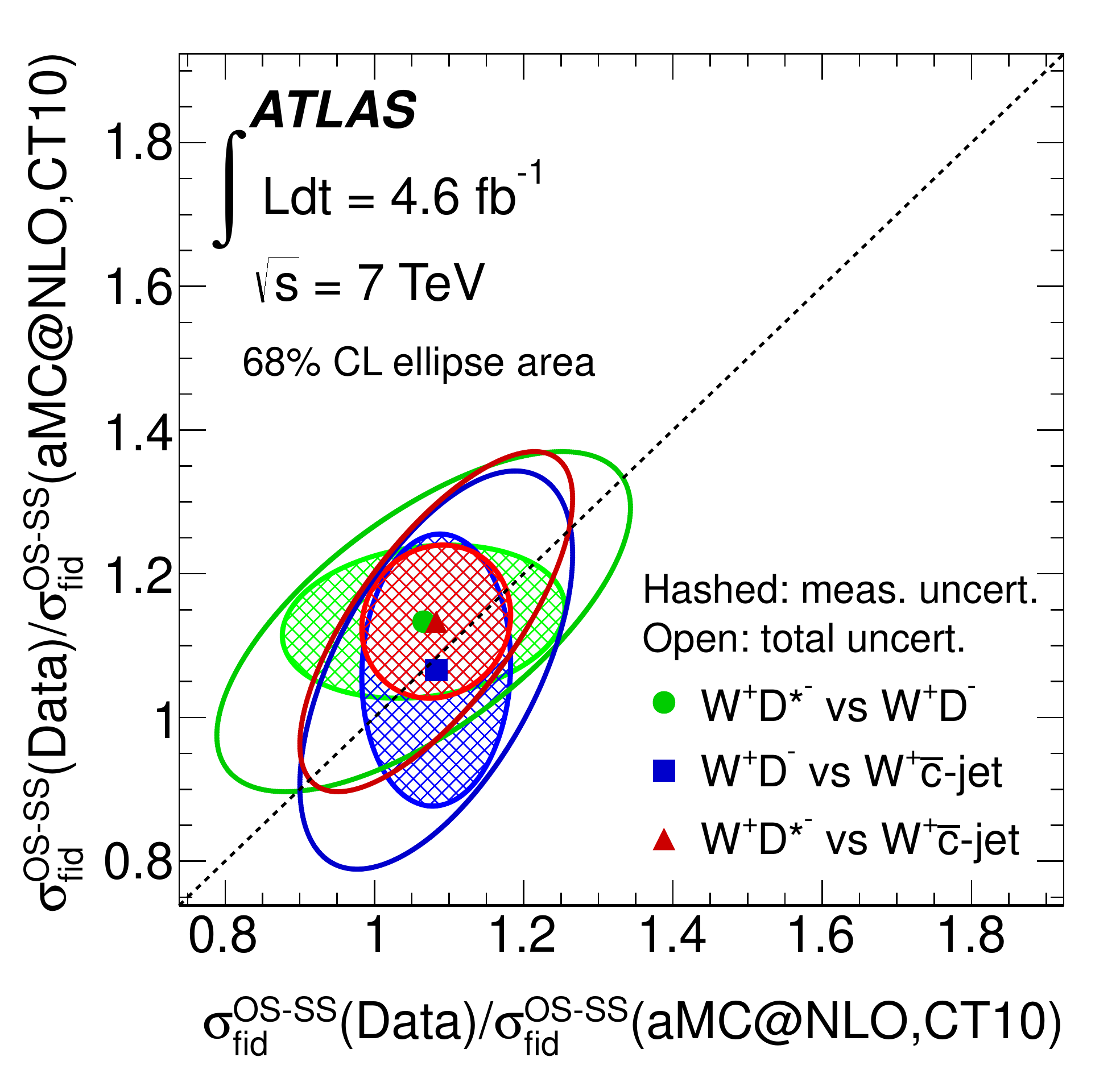}
\includegraphics[width=0.48\textwidth]{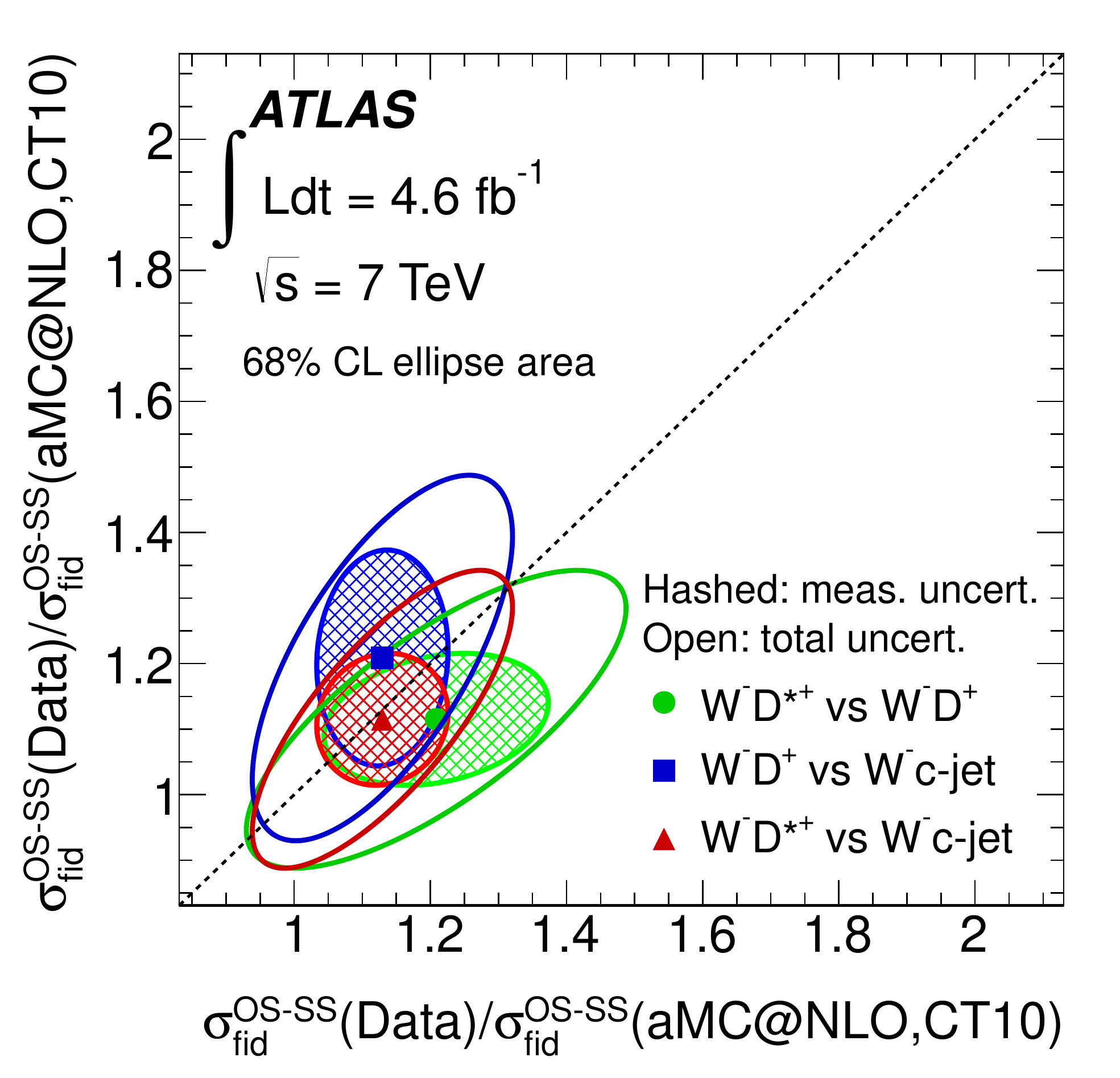}
 \caption{68\% CL contours of the measured cross sections normalised to the theoretical prediction obtained from the \aMCatNLO\ simulation using the \CT\ PDF. The filled ellipses show the experimental uncertainties, while the open ellipses show the total uncertainties, including the uncertainties on the prediction. The left figure shows the correlations among the $W^+D^{*-}$, $W^+D^{-}$ and $W^+\overline{c}$-jet cross sections, while the right figure is for  $W^-D^{*+}$, $W^-D^{+}$ and $W^-c$-jet.}
\label{fig:ellipsenorm}
\end{center}
\end{figure*}

The compatibility of the experimental
measurements from different channels is illustrated in figure~\ref{fig:ellipsenorm}, which shows the 68\% CL 
contours for the ratios of the measured cross section with respect to the theoretical prediction obtained from the \CT{} PDF. 
The large overlap of the contours with the diagonal
line reflects the good compatibility of the measurements assuming the
extrapolation among the different phase spaces as given by \aMCatNLO{}
using the \CT{} PDF.

\begin{figure*}
\begin{center}
\includegraphics[width=0.49\textwidth]{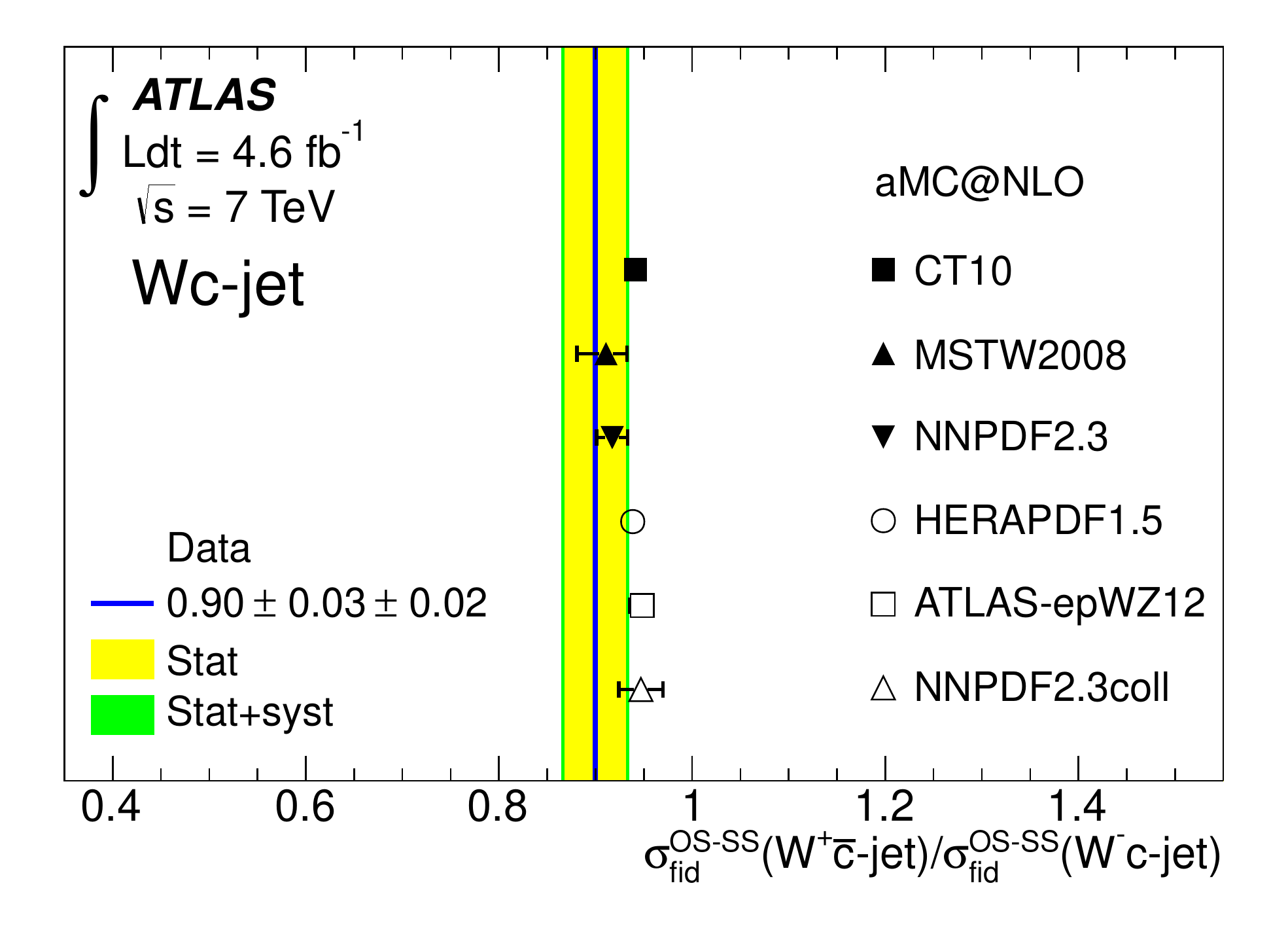}
\includegraphics[width=0.49\textwidth]{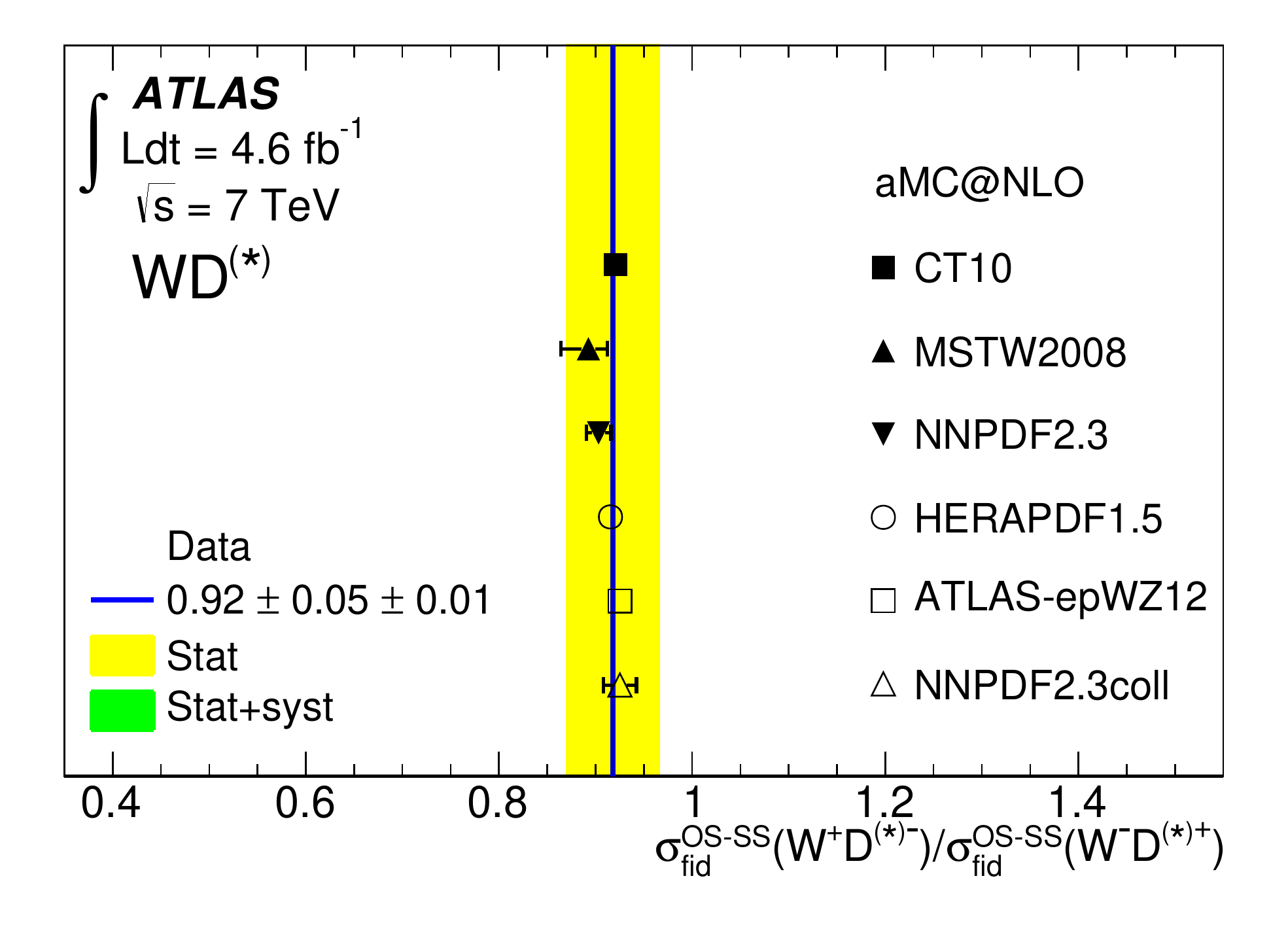}
\caption{Measured ratios \asymratioc (left) and \asymratio (right) resulting from the averaging procedure compared to various PDF predictions based on \aMCatNLO{}. The blue vertical lines show the central values of the measurements, the inner error bands show the statistical uncertainties and the outer error bands the total experimental uncertainties. The PDF predictions are shown by the black markers. The error bars on the predictions correspond to the 68\% CL PDF uncertainties.}
\label{fig:pdfasymratio}
\end{center}
\end{figure*}

Figure~\ref{fig:pdfasymratio} shows the measured ratio \RC\ compared to theoretical predictions based on various PDF sets. The predicted production ratio \RC~in $pp$ collisions can differ from unity for two reasons~\cite{Stirling:2012vh}. First, because the
proton contains valence $d$-quarks, the Cabbibo-suppressed diagrams involving
$d$-quarks enhance \Wmca\ production over \Wpca,
and thereby decrease \RC\ by about 5\%.  Second, a difference between $s$ and $\overline{s}$ PDFs, as suggested by neutrino data~\cite{nutev}, would also influence the value of the ratio: a lower population of $\overline{s}$-quarks relative to $s$-quarks in the sensitive range of the measurement would push the ratio to a lower value.
This effect is implemented in \NNPDF{} and \MSTW{}.
The contributions of the strange asymmetry in \NNPDF\ to \RC\ are small. For \MSTW, the strange asymmetry is larger and thus lowers \RC\
by about 3\%.  This pattern of predictions is consistent with those obtained
from the NLO calculation as implemented in \aMCatNLO\ and shown in figure~\ref{fig:pdfasymratio}.
The ratio measurement is consistent (within 1 $\sigma$) with all studied PDFs, and the measured uncertainty is comparable to the one obtained with \MSTW{}.

For PDFs such as \CT\ that require the $s$ and $\overline s$ distributions be equal,
the Cabibbo-suppressed diagrams are the only mechanism capable of lowering \RC. The relative size of strange asymmetry effects using NLO PDFs is studied in ref.~\cite{Stirling:2012vh}; assuming the ratio of $d$-quark to $s$-quark densities from \CT{} and that the asymmetry which is seen in the measured \RC{} is mainly due to the  $d$-quark, one can attribute the total difference $\RC(\CT)-\RC(\textrm{Data})$ to an effect of a strange asymmetry and thereby estimate the sensitivity of the current measurement. Under these assumptions the relative strange asymmetry ($A_{s\overline{s}}$) can be written as
\begin{equation}
A_{s\overline{s}} = \displaystyle\frac{\langle s(x,Q^{2})\rangle-\langle\bar{s}(x,Q^{2})\rangle}{\langle s(x,Q^{2})\rangle} \approx \RC(\CT)-\RC(\textrm{Data}),
\end{equation}
where the $s$ and $\overline{s}$ distributions are averaged over the phase space.
A value of $A_{s\overline{s}}=(2\pm3)\%$ is obtained for the combination of the \Wce{} and \WDe{} analyses.
The quoted uncertainty is dominated by statistical uncertainties.

\begin{figure*}
\begin{center}
\includegraphics[width=0.49\textwidth]{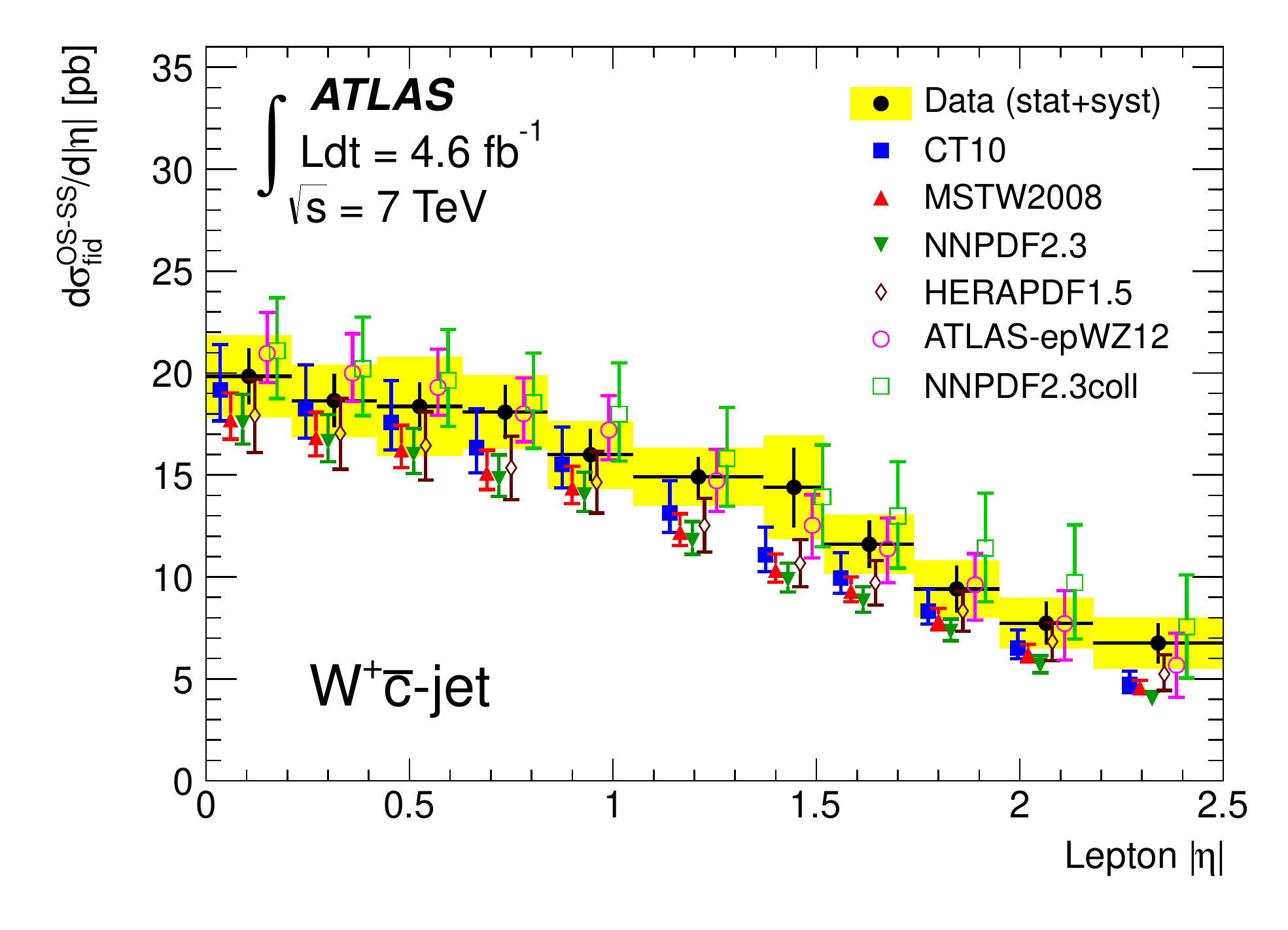} 
\includegraphics[width=0.49\textwidth]{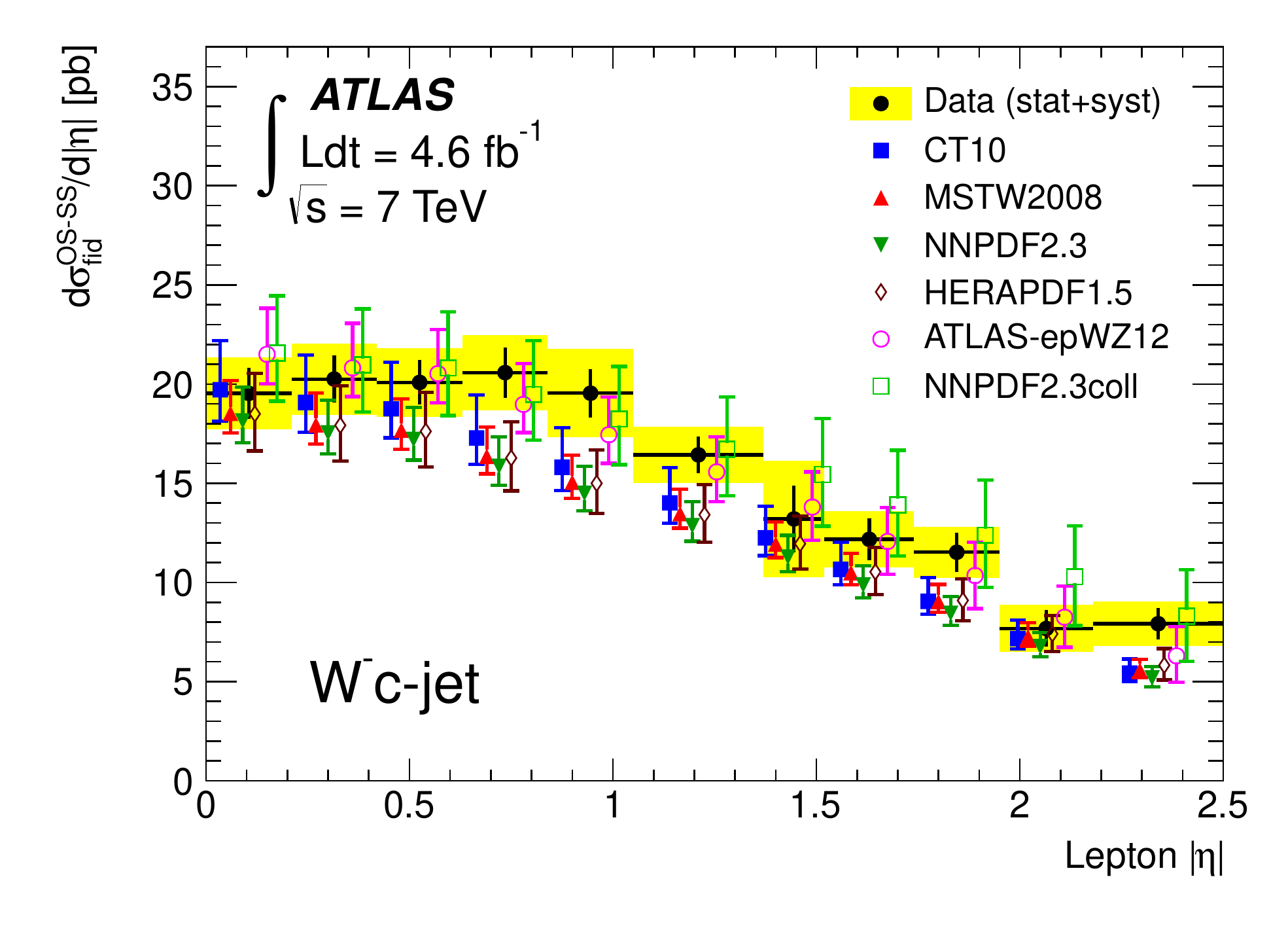} \\
\includegraphics[width=0.49\textwidth]{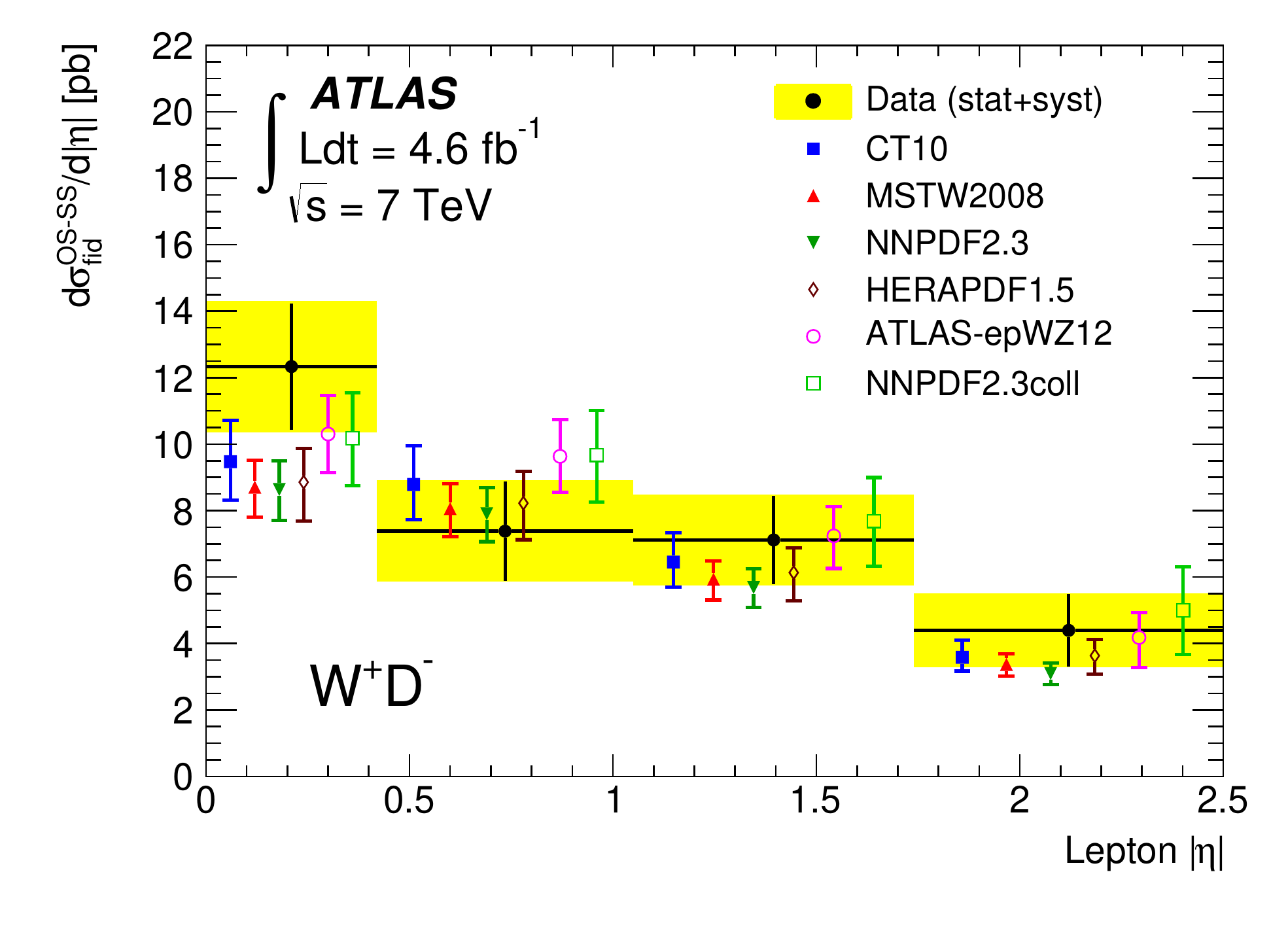}
\includegraphics[width=0.49\textwidth]{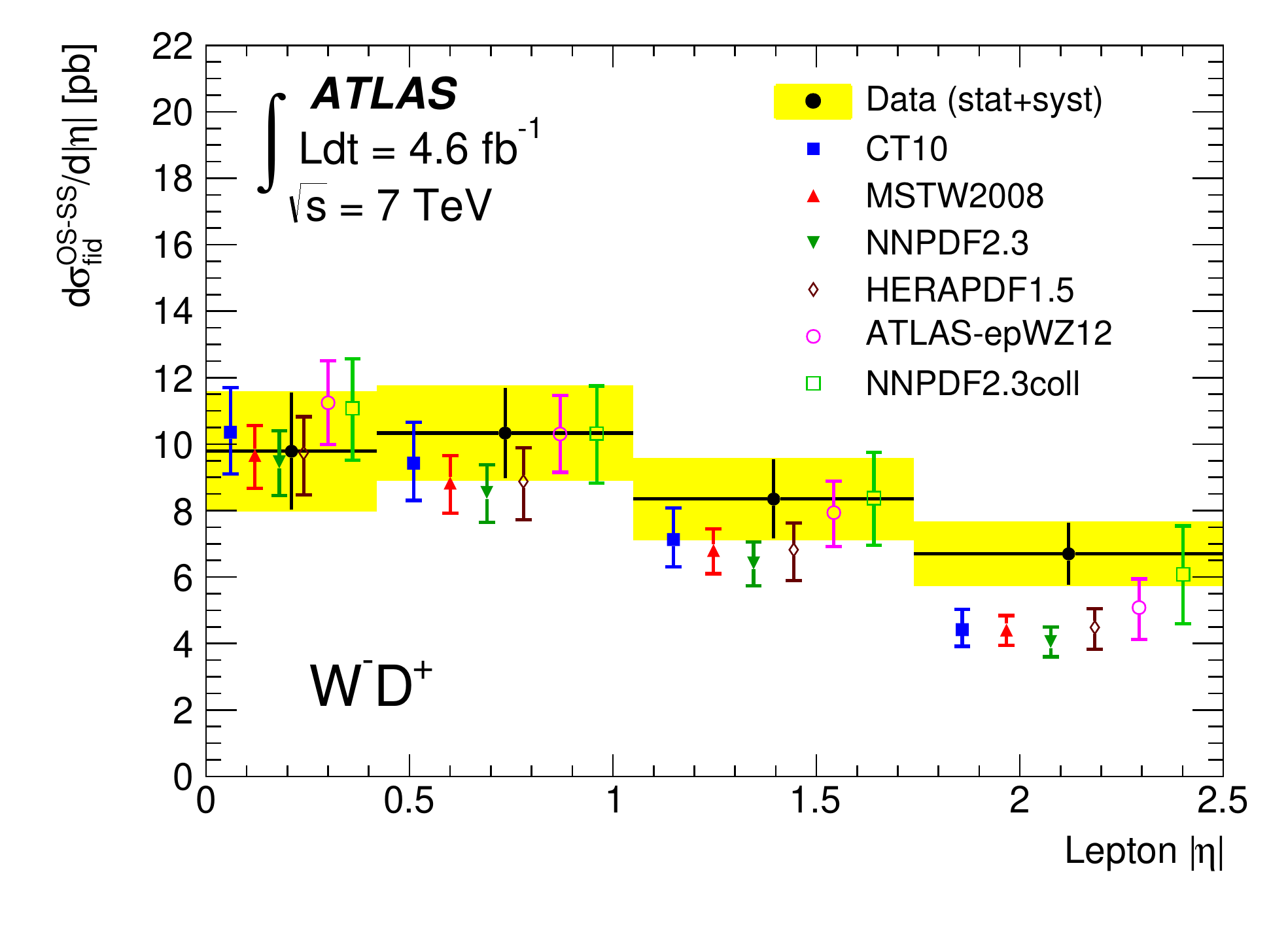}\\
\includegraphics[width=0.49\textwidth]{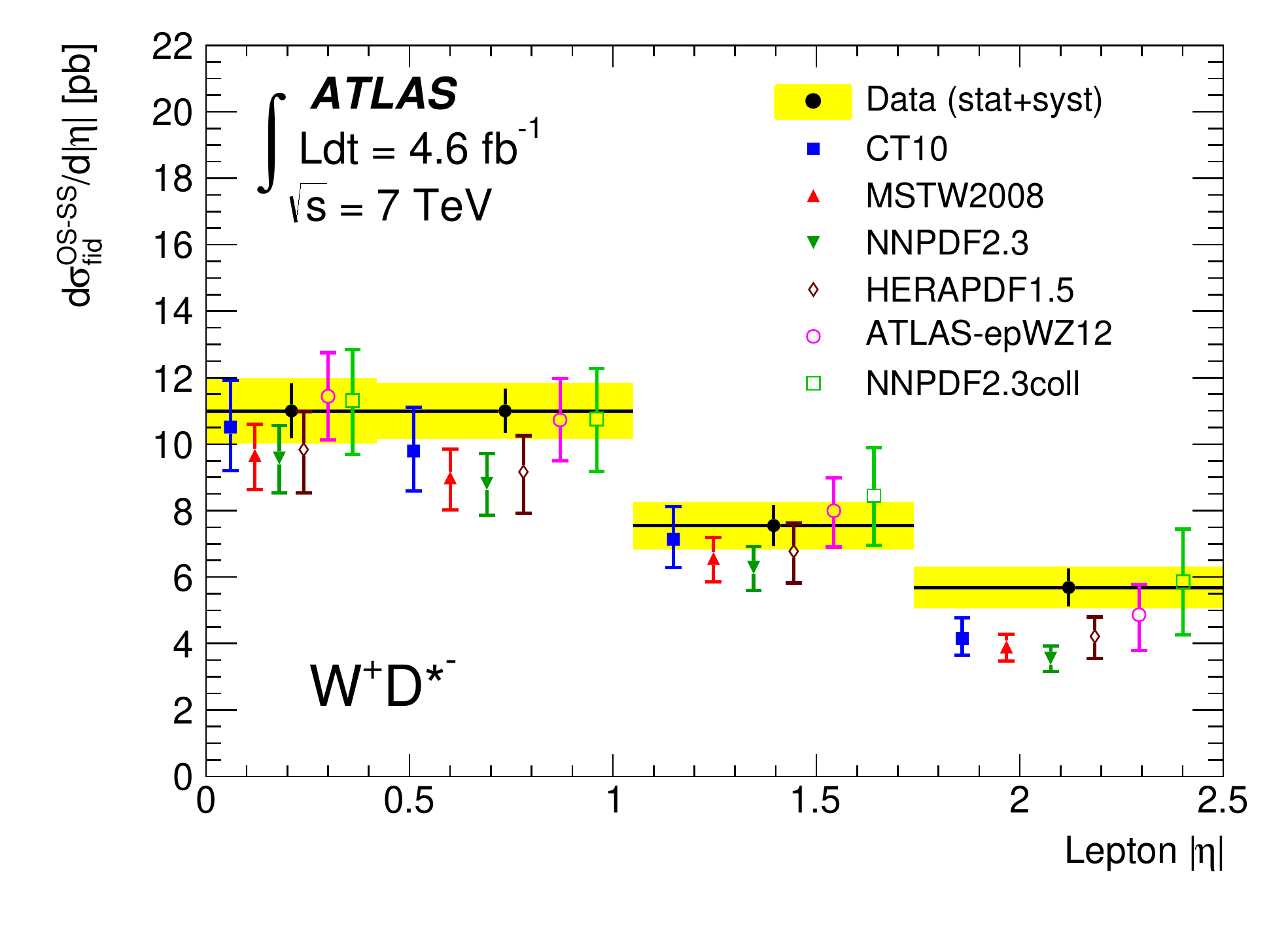} 
\includegraphics[width=0.49\textwidth]{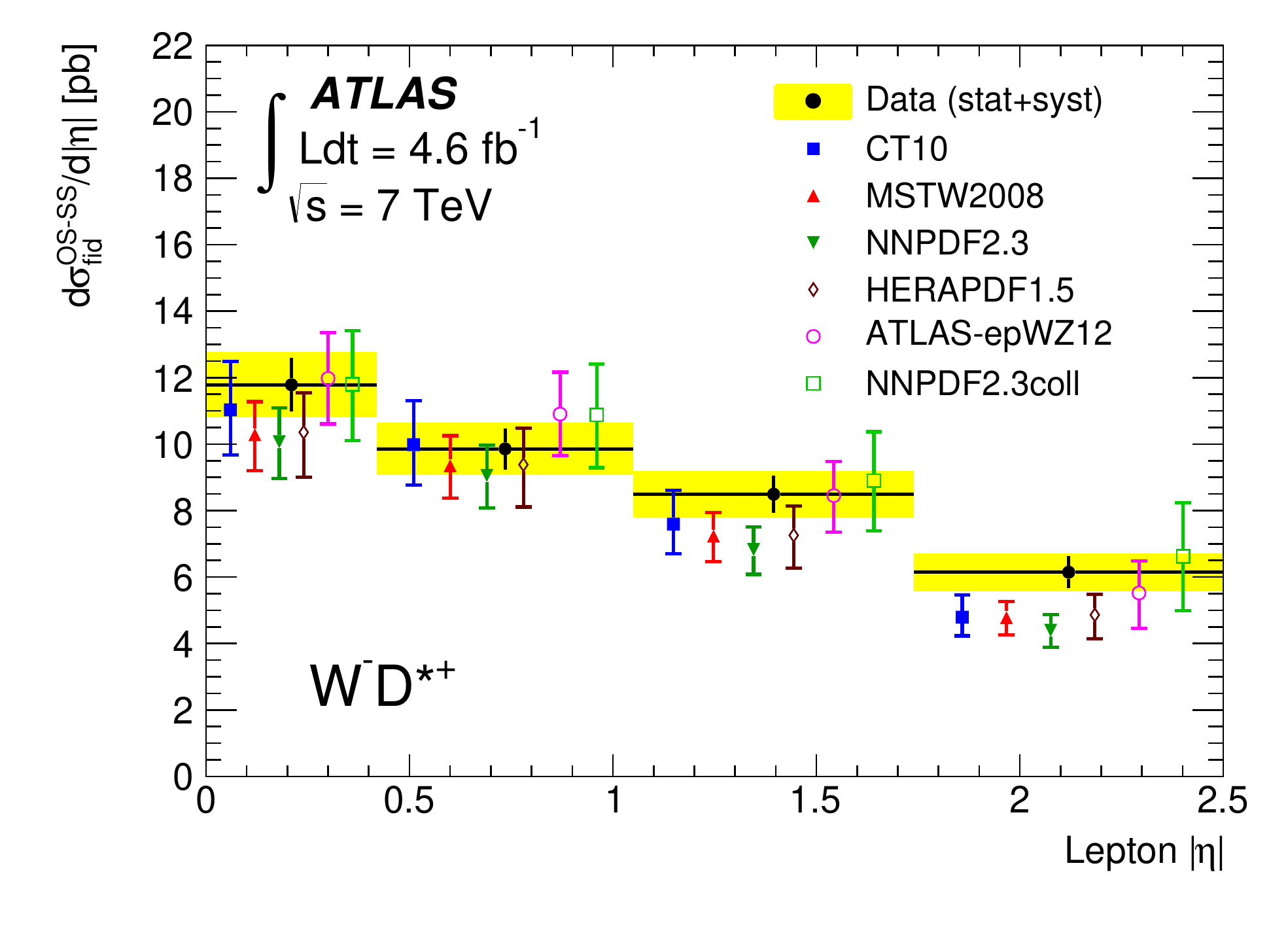}
\caption{Measured differential cross section as a function of lepton $|\eta|$ compared to predictions obtained using various PDF sets: (top left) $W^{+}\overline{c}$-jet, (top right) $W^{-}c$-jet, (middle left) $W^{+}D^{-}$, (middle right) $W^{-}D^{+}$, (bottom left) $W^{+}D^{*-}$ and (bottom right) $W^{-}D^{*+}$. The measurements are shown by the filled circles. The error bars give the statistical uncertainty, while the sum in quadrature of the statistical and systematic uncertainties is shown as an error band. The theory predictions are based on the \aMCatNLO{} simulation. The different markers correspond to the predictions obtained using various PDF sets and the corresponding error bars represent the total theoretical uncertainties (sum in quadrature of PDF, parton shower, fragmentation and scale uncertainties).}
\label{fig:pdfeta}
\end{center}
\end{figure*}

The dependence of the cross section on $|\eta^\ell|$, along with predictions of \aMCatNLO\ with various PDFs, is shown in figure~\ref{fig:pdfeta}. Similar predictions of the shapes of the $|\eta^\ell|$ distributions are obtained with the various PDF sets. The predictions differ mainly in their normalisation. The predicted shapes are in good agreement with the measured distributions.  

In order to perform a quantitative comparison of the measurements and the various PDF predictions, the $\chi^2$ function introduced in equation~(\ref{eq:comb_chi2_averaging}) is extended to include the uncertainties on the theoretical predictions:
\begin{equation}
 \chi^2 =\\
  \sum_{k,i} w^i_k
        \frac{\left[{\mu^i_k} - m^i\left(1 + \sum_j \gamma^i_{j,k} b_j + \sum_j (\gamma^{\rm theo})^i_{j,k} b^{\rm theo}_j  \right)\right]^2}
{(\delta^i_{\mathrm{sta}, k})^2 \Delta_i^k + (\delta^i_{\mathrm{unc},k} m^i)^2 }
+ \sum_j b_j^2+ \sum_j (b^{\rm theo}_j)^2,
\label{eq:comb_chi2_averagingtheory}
\end{equation}
\noindent where
\begin{equation}
 \Delta_i^k = \mu^i_k m^i\left(1- \sum_j \gamma^i_{j,k} b_j -  \sum_j (\gamma^{\rm theo})^i_{j,k} b^{\rm theo}_j\right).
\label{eq:comb_chi2_averagingtheoryadd}
\end{equation}
The notation follows the one introduced in equation~(\ref{eq:comb_chi2_averaging}). The matrix $(\gamma^{\rm theo})^i_{j,k}$ represents the relative correlated systematic uncertainties on the theory predictions and quantifies the influence of the uncertainty source $j$ on the prediction in bin $i$ and data set $k$. The parameters $b^{\rm theo}_j$ are defined analogously to the parameters $b_j$ and represent the shifts introduced by a correlated uncertainty source $j$ of the predictions. The $\chi^2$ function is minimised with respect to $b_j$ and $b^{\rm theo}_j$ with the cross-section measurements, $\mu$, fixed to the values determined in section~\ref{s:combination}.

Equation~(\ref{eq:comb_chi2_averagingtheory}) is further extended to account for asymmetric uncertainties on the predictions. The asymmetric uncertainties are described by parabolic functions
\begin{equation}
f_i(b^{\rm theo}_j) = \omega_{i,j}(b^{\rm theo}_j)^2 + \gamma_{i,j}b^{\rm theo}_j,
\end{equation}
which replace the terms $(\gamma^{\rm theo})^i_{j,k} b^{\rm theo}_j$ of equation~(\ref{eq:comb_chi2_averagingtheory}). The coefficients of $f_i(b^{\rm theo}_j)$ are determined from the values of the cross sections calculated when the parameter corresponding to source $j$ is set to its nominal value $+S^+_{i,j}$ and $-S^-_{i,j}$ where the $S^\pm_{i,j}$ are the up and down uncertainties of the respective PDF sets.\footnote{The uncertainties for the \NNPDF\ and \NNPDFcoll\ sets are obtained from the cross-section variations provided with these PDF sets by diagonalising the correlation matrix to determine the corresponding eigenvectors.} The coefficients are given by
\begin{eqnarray}
\gamma_{i,j} = \frac{1}{2}\left(S^+_{i,j}-S^-_{i,j} \right)\\
\omega_{i,j} = \frac{1}{2}\left(S^+_{i,j}+S^-_{i,j} \right).
\end{eqnarray}

The $\chi^2$-minimisation procedure implemented in the HERAFitter framework~\cite{Aaron:2009aa, r:herafitter, Aaron:2009kv, James:1975dr} is used. The cross-section measurements differential in $|\eta^\ell|$ are used to assess the quantitative agreement between the data and the PDF predictions.

\begin{table}
\scriptsize
\begin{center}
\begin{tabular}{| l |c |c| c| c |c |c|}
\hline
 & \tiny\CT & \tiny\MSTW &\tiny \HERA &\tiny \epWZ & \tiny\NNPDF & \tiny\NNPDFcoll \\
\hline
$W^+\overline{c}$-jet ($\chi^2/{\rm ndof}$)  & 3.8/11 & 6.1/11 & 3.5/11 & 3.1/11& 8.5/11  & 2.9/11\\
$W^-c$-jet ($\chi^2/{\rm ndof}$)                   & 9.0/11   & 10.3/11 & 8.3/11& 6.3/11   & 10.5/11 & 6.1/11\\
$W^+D^-$ ($\chi^2/{\rm ndof}$)              & 3.6/4 & 3.7/4 & 3.7/4 & 3.4/4  & 3.8/4 & 3.4/4\\
$W^-D^+$ ($\chi^2/{\rm ndof}$)              & 3.7/4 & 4.6/4 & 3.3/4 &  2.0/4 & 4.7/4 & 1.6/4\\
$W^+D^{*-}$ ($\chi^2/{\rm ndof}$)          & 2.9/4 & 6.0/4 & 2.2/4 &  1.7/4 & 8.1/4 & 1.6/4\\
$W^-D^{*+}$ ($\chi^2/{\rm ndof}$)          & 3.0/4 & 4.4/4 & 2.4/4 & 1.6/4  & 4.2/4 & 1.4/4\\
$N_{\rm exp}$        & 114       & 114         & 114       & 114    & 114  & 114 \\
$N_{\rm theo}$& 28        &  22         & 16         &   20    & 40    &  40   \\
Correlated $\chi^2$ (exp)                                      & 0.8    & 1.8  & 0.9 &  1.1  & 2.2  &   1.0\\
Correlated $\chi^2$ (theo)                                     & 6.2    	& 1.9 	& 2.6 	&  0.1   & 7.4   	&   0.2\\
Correlated $\chi^2$ (scale)                                  & 0.6    & 2.5          & 1.1          &  0.0   & 2.7   &   0.0\\
\hline
Total $\chi^2/{\rm ndof}$	 				& 33.6/38 & 41.3/38  & 28.0/38 &  19.2/38 & 52.1/38 & 18.2/38\\			

\hline
\end{tabular}
\end{center}
\caption{Quantitative comparison of fiducial cross sections to various PDF predictions. The table shows the partial $\chi^2/{\rm ndof}$ for the different cross-section measurements, the number of nuisance parameters for the experimental sources of systematic uncertainties ($N_{\rm exp}$), the number of nuisance parameters for the uncertainties on the predictions ($N_{\rm theo}$) as well as the correlated $\chi^2$ corresponding to the experimental uncertainties ($\chi^2$ (exp)), the uncertainties on the predictions excluding the scale uncertainties ($\chi^2$ (theo)) and the scale uncertainty ($\chi^2$ (scale)). The correlations due to the systematic uncertainties of $c$-quark fragmentation that affect both the measured cross sections and the theoretical predictions are taken into account. To avoid double-counting, these uncertainties are added to $N_{\rm exp}$ and $\chi^2$ (exp) only.  Furthermore, the total $\chi^2/{\rm ndof}$ is given.}
\label{t:pdffits}
\end{table}

The results of the $\chi^2$-minimisation procedure are shown in table~\ref{t:pdffits}. The measured cross sections are in agreement with all PDF predictions but disfavour \NNPDF{}. In addition to the total $\chi^2$, table~\ref{t:pdffits} also shows the individual contributions to the $\chi^2$ from the experimental uncertainties, the uncertainties on the predictions and the scale uncertainty. For the predictions obtained with \MSTW\ and \NNPDF\ the scale uncertainty is the dominant uncertainty. 
Improved accuracy in the theory calculation, especially reducing the scale dependence, could
enhance the sensitivity of the presented measurements to the PDF significantly.

For values $x\leq0.1$, the \HERA\ PDF is constrained mainly by the precise measurement of the proton structure function $F_2(x,Q^2)$ at HERA~\cite{Aaron:2009aa}, which fixes the quark-charge-squared weighted sum of quark and anti-quark contributions but has no sensitivity to the flavour composition of the total light-quark sea.
In the \HERA\ PDF set, the strange-quark distribution is expressed as an $x$-independent fraction, $f_s=\overline{s}/(\overline{d}+\overline{s})$. The central value $f_s = 0.31$ at $Q^2=1.9\,\GeV^2$ is chosen to be consistent with determinations of this fraction using the neutrino--nucleon scattering data with an uncertainty spanning the range from 0.23 to 0.38. This model uncertainty is parameterised as a nuisance parameter in the $\chi^2$ minimisation.

\begin{figure}
\begin{center}
\includegraphics[width=0.7\textwidth]{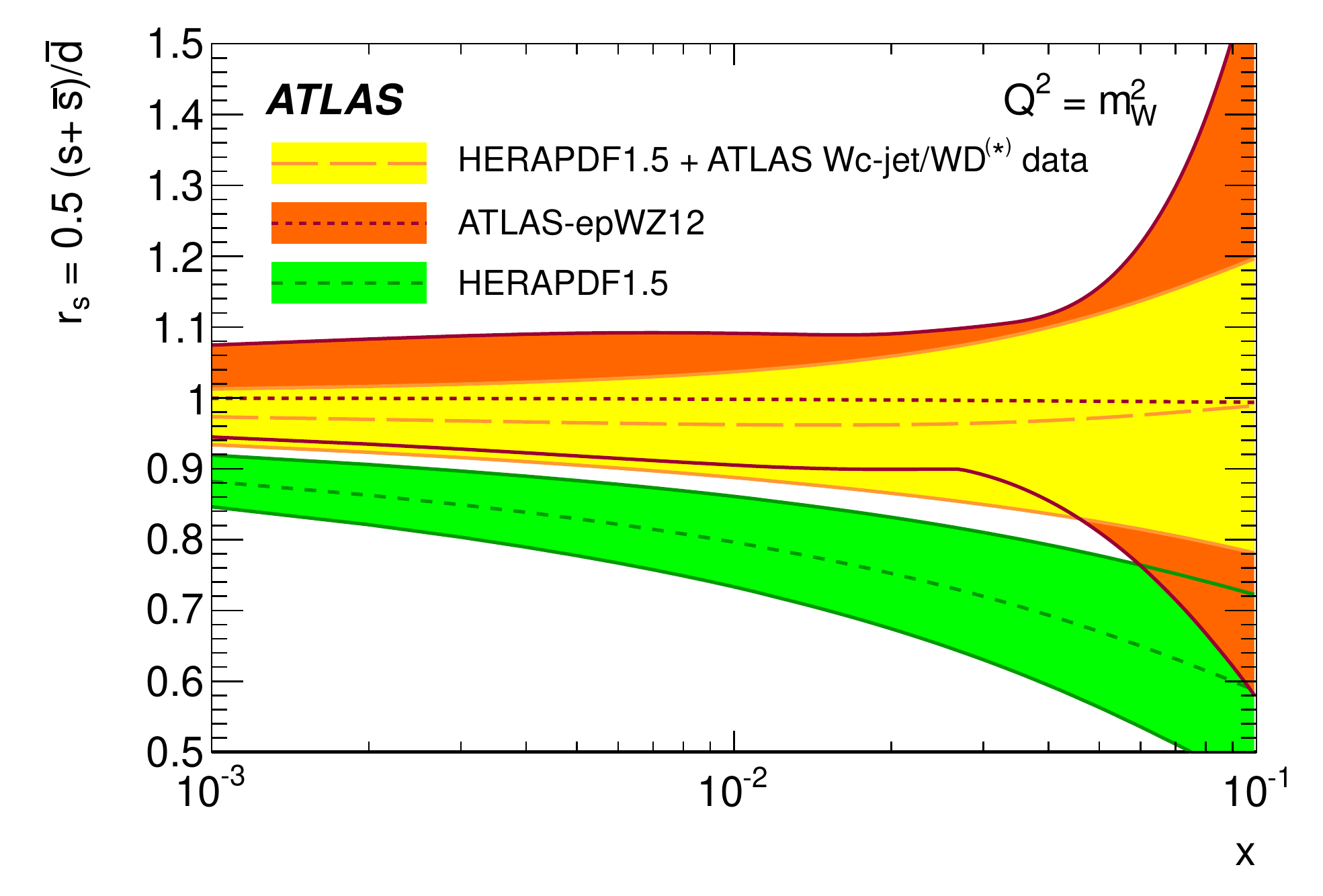}
\caption{Ratio of strange-to-down sea-quark distributions $r_s = 0.5(s+\overline{s})/\overline{d}$ as a function of $x$ as assumed in \HERA\ PDF compared to the ratio obtained from the fit including the ATLAS \Wce/\WDe\ data and the ratio obtained from \epWZ{}. The error band on the ATLAS \Wce/\WDe\ measurements represents the total uncertainty. The ratio $r_s$ is shown at $Q^2=m_W^2$.}
\label{fig:pdfconstraint}
\end{center}
\end{figure}

The $\chi^2$-minimisation procedure not only gives information about the overall compatibility of the predictions with the data, but also allows constraints on the PDF eigenvectors to be obtained. \HERA{} is the only publicly available PDF set where the effect of varying the strange-quark density is parameterised by a single parameter ($f_s$). The $\chi^2$-minimisation procedure discussed above can be used as follows to calculate a value for $f_s$ based solely on the measurements discussed here while ignoring all previous measured or assumed values of $f_s$. 
The $\chi^2$ minimisation is repeated for the \HERA\ PDF set after artificially increasing the uncertainty of the strange-quark fraction $f_s$.
This procedure corresponds to a free fit of the eigenvector representing $f_s$ while all other eigenvectors are constrained within the uncertainties determined in the \HERA\ fit.
A value of 
\begin{displaymath}
r_s \equiv 0.5(s+\overline{s})/\overline{d}=f_s/(1-f_s) = 0.96\,^{+0.16}_{-0.18}\,^{+0.21}_{-0.24}
\end{displaymath} 
is determined at $Q^2=1.9\,\GeV^2$ and is independent of $x$ as implemented in the \HERA{} PDF.  
The first uncertainty represents the experimental and theoretical uncertainties and the 
second uncertainty corresponds to the scale uncertainty of the \Wca\ calculation. 
Since the scale uncertainty is the dominant uncertainty, its effect is assessed separately by repeating the fit under the assumption of perfect knowledge of the scale. The resulting strange-quark fraction is shown in figure~\ref{fig:pdfconstraint} as a function of $x$ at $Q^2=m_W^2$. For the \HERA\ PDF the $s$-quark sea density is lower than the $d$-quark sea density at low values of $x$ and it is further suppressed at higher values of $x$. 
The ATLAS \Wce/\WDe\ data on the other hand favour a symmetric light-quark sea
over the whole $x$ range relevant to the presented measurement ($10^{-3}$ to $10^{-1}$).

The value of $r_s$ determined in this study is in good agreement with the value of $r_s=1.00^{+0.25}_{-0.28}$ 
obtained in the combined analysis of $W$ and $Z$ production at $Q^2=1.9\,\GeV^2$ and $x$ = 0.023 
by ATLAS~\cite{Aad:2012sb} and supports the hypothesis of an SU(3)-symmetric light-quark sea. 
Figure~\ref{fig:pdfconstraint} also shows that the $x$-dependence of $r_s$ obtained from the \epWZ{} PDF is
in good agreement with this study. 

%% file: additionalresults_WD.tex
\section{Additional results}
\label{s:addresults}

\subsection[Cross-section ratio \xsecratio differential in \ptD]{Cross-section ratio $\boldsymbol{\xsecratio}$ differential in $\boldsymbol{\ptD}$}
\label{s:addresultswd}

\begin{figure*}
\begin{center}
\includegraphics[width=0.4\textwidth]{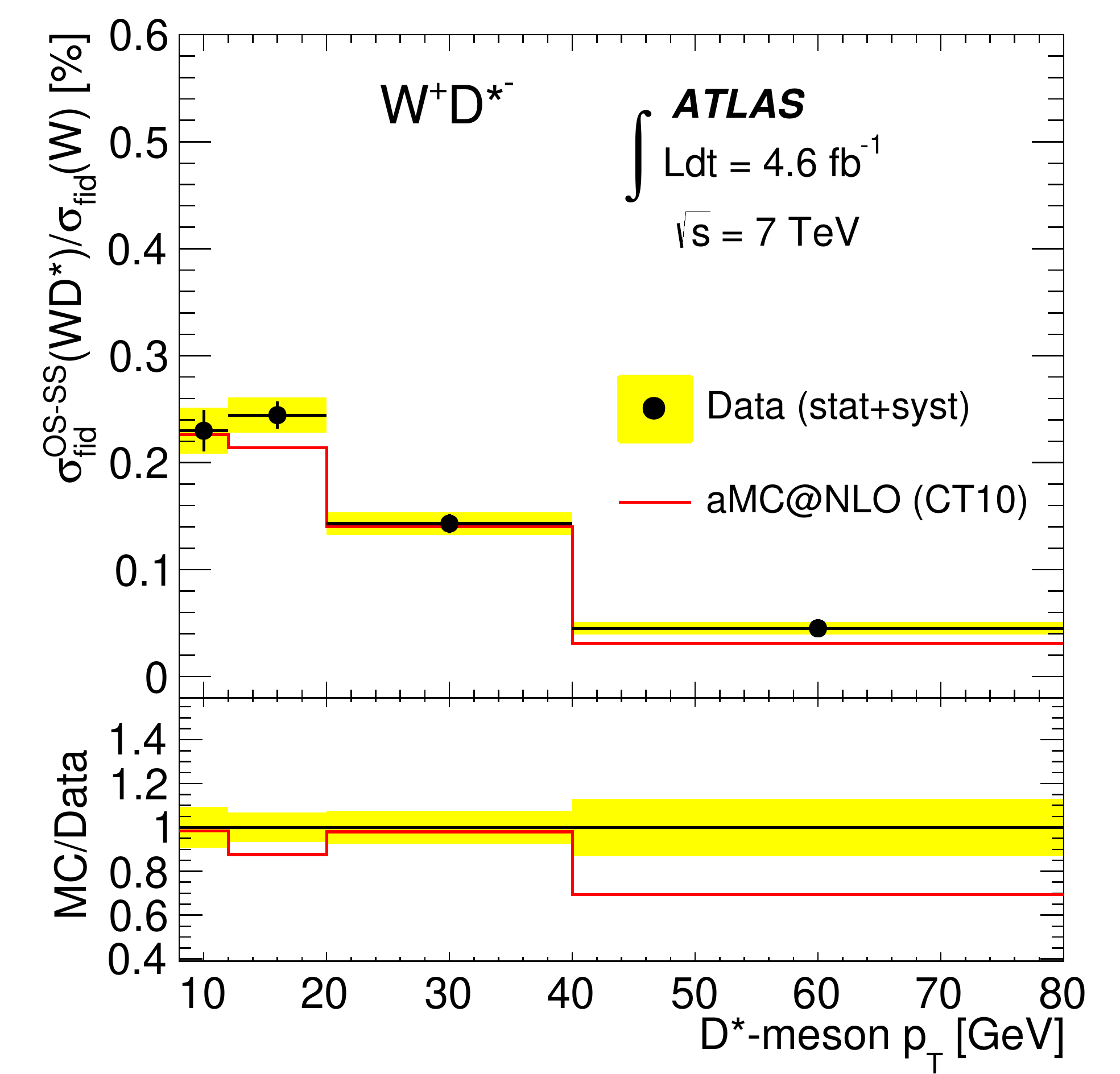}  
\includegraphics[width=0.4\textwidth]{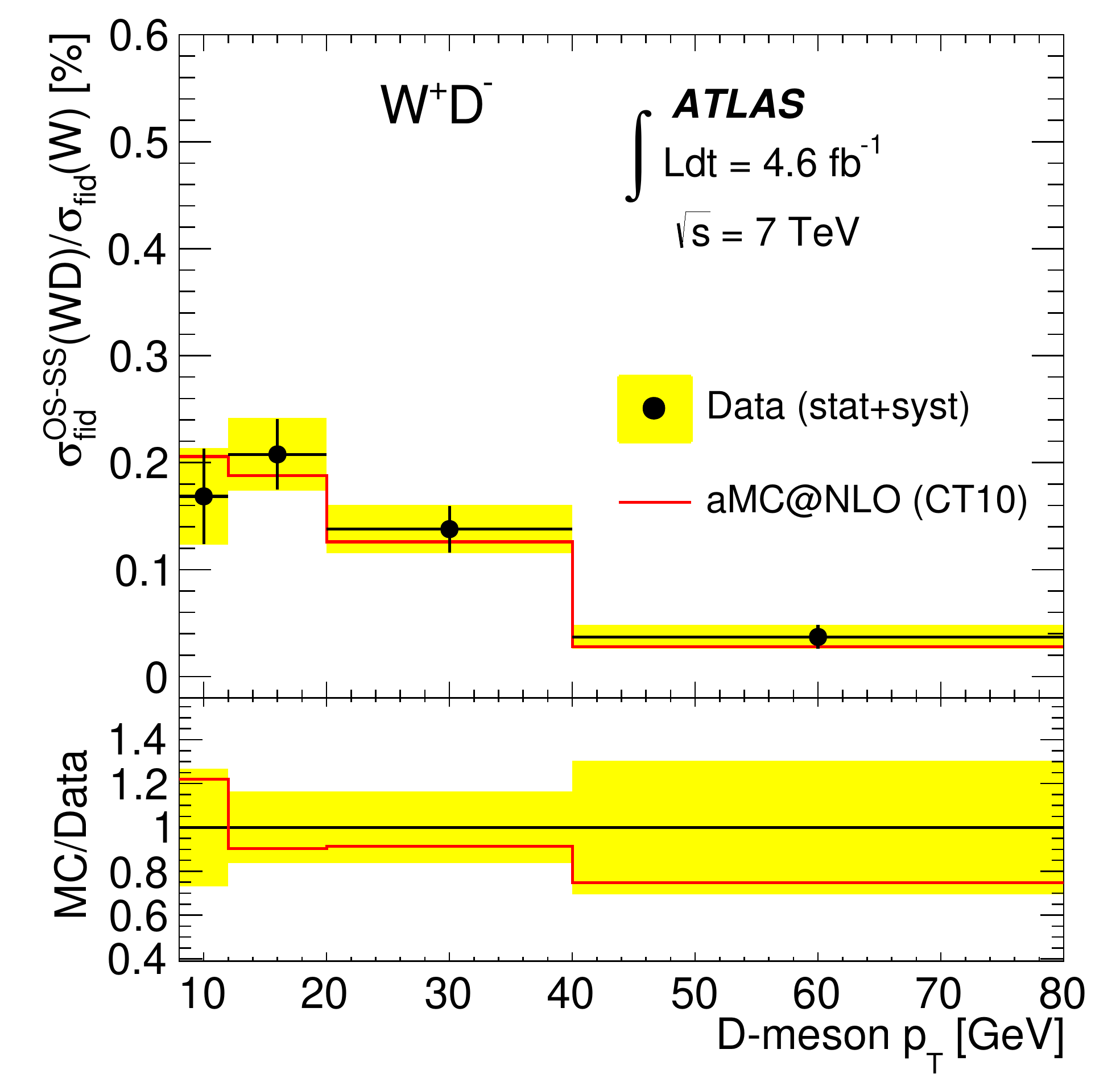} \\
\includegraphics[width=0.4\textwidth]{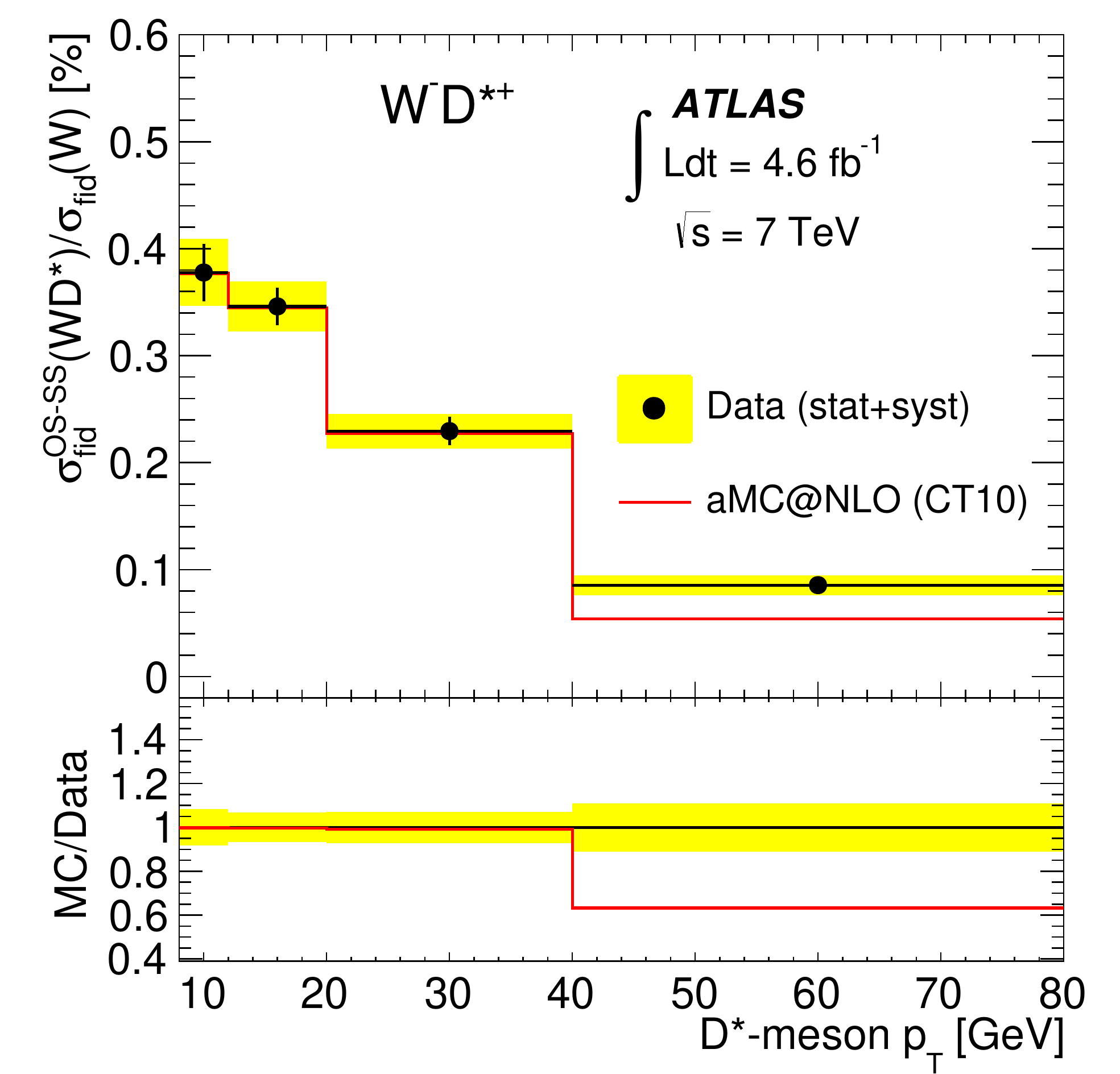}
\includegraphics[width=0.4\textwidth]{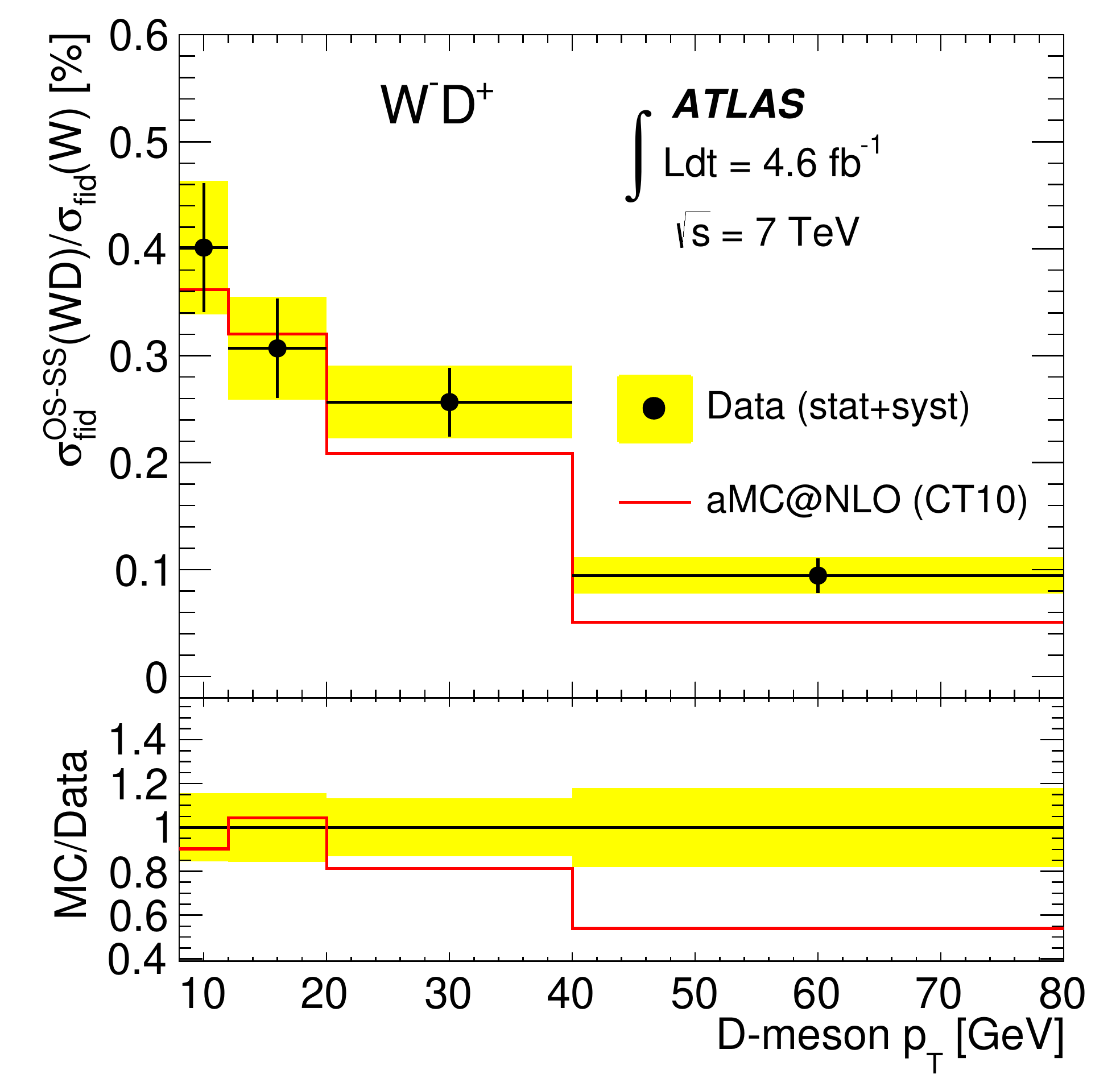}
\caption{Measured cross-section ratio $\sigma(W^{+}D^{(*)-})/\sigma(W^{+})$
  (top) and $\sigma(W^{-}D^{(*)+})/\sigma(W^{-})$ (bottom) in percent and differential in
  \ptD\ compared to the MC prediction: the left plots are for $D^{*\pm}$, while
  the right plots are for $D^\pm$. The measurement is shown by the filled
  markers. The error bars give the statistical uncertainty, while the sum in quadrature of the statistical and systematic uncertainties is shown as an error band. The solid line shows the prediction of the \aMCatNLO{} MC simulation obtained using the CT10 PDF set. The ratio of the simulated distribution to data is shown in the lower panels. Here, the error band corresponds to the sum in quadrature of the statistical and systematic uncertainties.}
\label{fig:ratioDpt}
\end{center}
\end{figure*}

In this section, the measurements of the cross-section ratio \xsecratio differential in \ptD\ are presented. The measurements are compared in figure~\ref{fig:ratioDpt} to theoretical predictions obtained from \aMCatNLO\
using the CT10 NLO PDF.  The ratio is on average $8\%$ higher in data than
in simulation.  The shape of the \ptD\ spectrum is reasonably 
well described by the MC simulation, although a slight excess in data compared to MC simulation is observed in the highest \ptD\ bin, suggesting that the \ptD\ spectrum in data might be slightly harder than the \aMCatNLO\
prediction. The measured integrated cross-section ratios in the fiducial region are shown in table~\ref{tab:wd_ratio_add}. 

\begin{table}[htbp]
\centering
\begin{tabular}{|l|r|}
\hline
 & \xsecratio [\%] \\
 \hline
   $W^+D^-$	& $0.55\pm0.06\;{\rm (stat)}\pm0.02\;{\rm (syst)}$\\
   $W^+D^{*-}$	&$0.66\pm0.03\;{\rm (stat)}\pm0.03\;{\rm (syst)}$\\
   $W^-D^+$       &$1.06\pm0.08\;{\rm (stat)}\pm0.04\;{\rm (syst)}$\\
   $W^-D^{*+}$	&$1.05\pm0.04\;{\rm (stat)}\pm0.05\;{\rm (syst)}$\\
\hline
\end{tabular}
\caption{Measured fiducial cross-section ratios \xsecratio together with the statistical and systematic uncertainty.}
\label{tab:wd_ratio_add}
\end{table}

%% file: additionalresults_Wc.tex
\subsection[Cross sections \xsecWc{} and $\sigma^{\OSSS}_{\rm fid}(Wc\mbox{-}\rm{jet}(c\rightarrow \mu))$ as function of the jet multiplicity]{Cross sections $\boldsymbol{\xsecWc}$ and $\boldsymbol{\sigma^{\OSSS}_{\rm fid}(Wc\mbox{-}\rm{jet}(c\rightarrow \mu))}$ as a function of the jet multiplicity}
\label{s:addresultswc}
\begin{figure}[tp!]
\begin{center}
\includegraphics[width=0.44\linewidth]{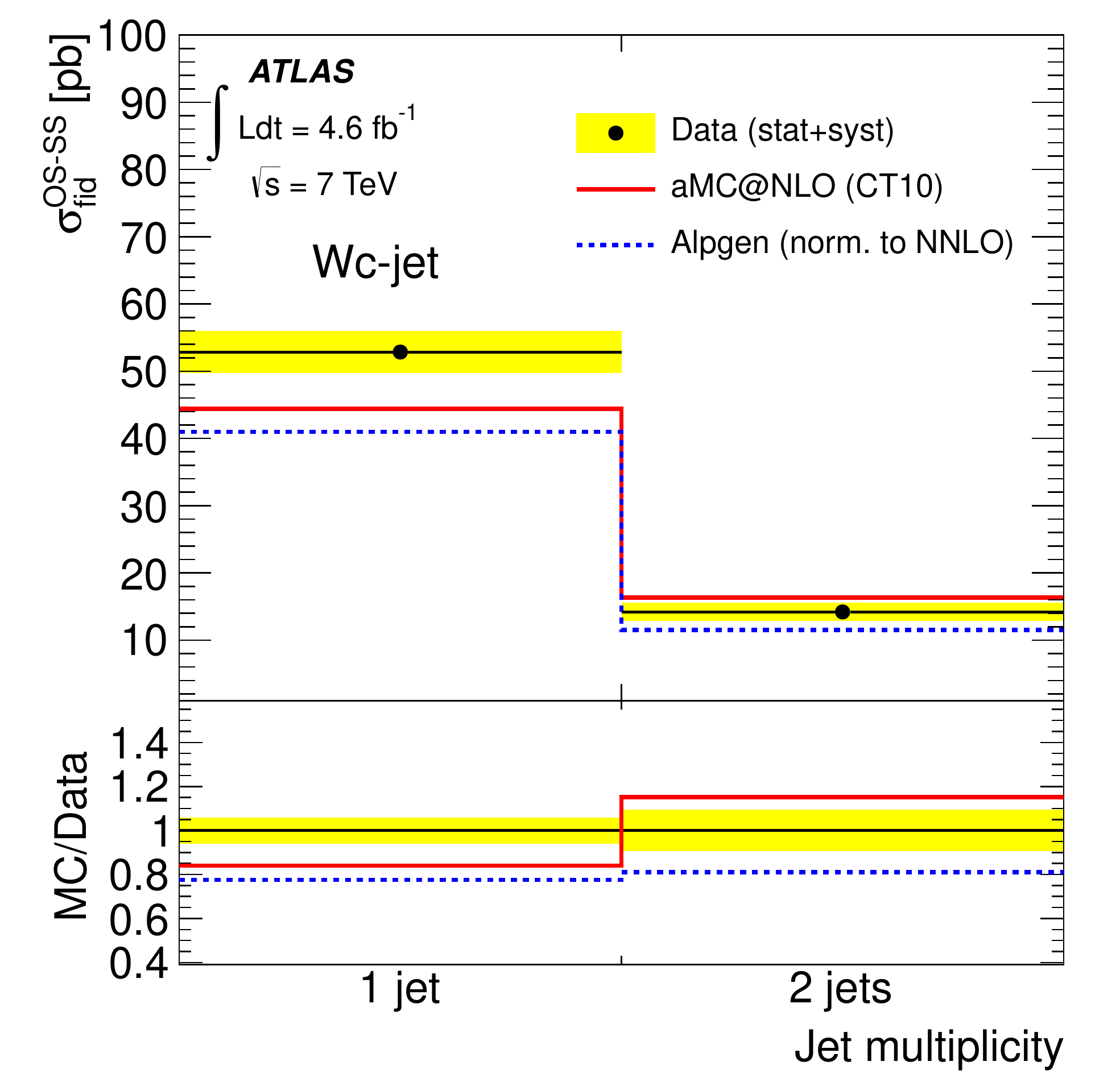} 
\caption{Measured cross sections as a function of the jet multiplicity compared to \aMCatNLO{} produced using the \CT{} NLO PDF set.
The predictions from \Alpgen{} normalised to the inclusive $W$ NNLO cross section are also shown for reference.
In the lower panels, the ratio of the simulated distribution to data is shown.}
\label{fig:wc_1jet_2jet_band}
\end{center}
\end{figure}

In addition to the \Wce\ fiducial cross section for a $W$ boson with exactly one $c$-jet and any number of additional jets, the cross section is measured with the requirements defined in section~\ref{s:phasespace}, except for requiring either exactly one or exactly two jets only one of which is identified as a $c$-jet. The results, including the ratio \RC, averaged between the electron and muon channels, are shown in table~\ref{tab:wc_exclusive_xs}.
Figure~\ref{fig:wc_1jet_2jet_band} shows the measured \Wce\ fiducial cross sections for events with exactly one or two jets compared to \aMCatNLO{} predictions with the \CT{} NLO PDF set. 
The \aMCatNLO{} central values do not describe the one-to-two-jets ratio well.
The \Alpgen{} predictions normalised to the inclusive $W$ NNLO cross section are also shown for reference.
The \Alpgen{} central values underestimate the data measurements for both the samples with one and two jets; however the one-to-two-jets ratio is well described.

Finally, in order to minimise the systematic uncertainties due to the extrapolation to the fiducial phase space, the cross sections are determined in a phase space as specified in section \ref{s:phasespace} but in which the $c$-hadron decays semileptonically to a muon with $\pt>4$\,\GeV, $|\eta|<$2.5, charge opposite to the $W$ boson and within $\Delta R<0.5$ from the $c$-jet axis. The resulting cross sections, for both the exclusive jet multiplicity and inclusive jet multiplicity definitions are also shown in table~\ref{tab:wc_exclusive_xs}, indicating a total systematic uncertainty of 4.7\% for the measurement with inclusive jet multiplicity.

\begin{table}[tp]
\centering
\begin{tabular}{|l |c|}
\hline
 & $\xsecWc$ [pb] \\
 \hline
   \Wce\      (1 jet)      &$52.9\pm0.9\,{\rm(stat)}\pm3.0\,{\rm(syst)}$\\
   \Wce\      (2 jets)   &$14.2\pm0.6\,{\rm(stat)}\pm1.2\,{\rm(syst)}$\\
   \hline
   \hline
   \RC\ (1 jet)	&$0.91\pm0.03\,{\rm(stat)}\pm0.02\,{\rm(syst)}$\\
   \RC\ (2 jets)	&$0.87\pm0.08\,{\rm(stat)}\pm0.02\,{\rm(syst)}$\\
   \hline
   \hline 
 & $\sigma^{\OSSS}_{\rm fid}(Wc\mbox{-}\rm{jet}(c\rightarrow \mu))$ [pb] \\
\hline
  \Wce\      (1 jet)     &$2.47\pm0.04\,{\rm(stat)}\pm0.13\,{\rm(syst)}$\\
   \Wce\      (2 jets)     &$0.69\pm0.03\,{\rm(stat)}\pm0.06\,{\rm(syst)}$\\
   \Wce\      (inclusive)  &$3.36\pm0.06\,{\rm(stat)}\pm0.16\,{\rm(syst)}$\\
\hline
\end{tabular}
\caption{Measured fiducial cross sections and \RC\ for exclusive jet multiplicity together with the statistical and systematic uncertainties. The lower part of the table shows the measured fiducial cross section for the production of a $W$ boson together with a soft muon from the charm-quark decay. The branching ratio $W\rightarrow\ell\nu$ is included in the fiducial cross section definition. }
\label{tab:wc_exclusive_xs}
\end{table}

\clearpage

%% file: conclusion.tex
\section{Conclusion}
\label{s:conclusion}

Integrated and differential cross sections for $W$-boson production in association with a single charm quark are measured as a function of the pseudorapidity of the lepton from the $W$-boson decay in 4.6\,$\rm{fb^{-1}}$ of $pp$ collision data at $\sqrt{s}=7$\,\TeV\ collected with the ATLAS detector at the LHC. Two methods are used to tag the charm quark: either the presence
of a muon from semileptonic charm decay within a
hadronic jet or the presence of a charged \De ($D$ or $D^*$) meson.
The integrated cross sections for the fiducial region 
$\pt^\ell>20$\,\GeV, $|\eta^\ell|<2.5$,
$\ptnu>25$\,\GeV, 
$\mtw>40$\,\GeV\ are measured for
the \Wce\ events
with jets passing the fiducial requirements of $\pt>25$\,\GeV, $|\eta|<2.5$, yielding 
\begin{eqnarray*}
  \xsecplusWc  & = & 33.6\pm0.9\,{\rm(stat)}\pm1.8\,{\rm(syst)}\,{\rm pb}\\
  \xsecminusWc & = & 37.3\pm0.8\,{\rm(stat)}\pm1.9\,{\rm(syst)}\,{\rm pb}
\end{eqnarray*}
and for \De mesons with  $\ptD>8$\,\GeV\ and $|\etaD|<2.2$, yielding
\begin{eqnarray*}
  \xsecplusDp & = & 17.8\pm1.9\,{\rm(stat)}\pm0.8\,{\rm(syst)}\,{\rm pb} \\
 \xsecminusDp  & = & 22.4\pm1.8\,{\rm(stat)}\pm1.0\,{\rm(syst)}\,{\rm pb}\\
 \xsecplusDs & = & 21.2\pm0.9\,{\rm(stat)}\pm1.0\,{\rm(syst)}\,{\rm pb}\\
 \xsecminusDs & = & 22.1\pm0.8\,{\rm(stat)}\pm1.0\,{\rm(syst)}\,{\rm pb}.
\end{eqnarray*}
Furthermore, the cross-section ratios are determined to be
\begin{eqnarray*} 
\RC(Wc\mbox{-}\rm{jet}) = & \asymratiocNoSp & =  0.90\pm0.03\,{\rm(stat)}\pm0.02\,{\rm(syst)} \\
\RC(WD^{(*)})            = &  \asymratioNoSp & =  0.92\pm0.05\,{\rm(stat)}\pm0.01\,{\rm(syst)}
\end{eqnarray*}
and are in agreement with theoretical predictions. In addition to the cross-section measurements differential in lepton pseudorapidity, measurements of the differential distributions of the $D$-meson transverse momentum and the jet multiplicity in \Wce\ events are performed. 

The predicted
cross sections depend on the choice of PDF set and have uncertainties
associated with the choice of renormalisation and factorisation scales.
With these uncertainties taken into account, the data are consistent with
a wide range of PDFs, but show a preference for PDFs with an SU(3)-symmetric light-quark sea. The ratio of the strange-to-down sea-quark distributions is determined to be $0.96^{+0.26}_{-0.30}$ at $Q^2=1.9\,\GeV^2$.

%% file: Acknowledgements.tex



\section*{Acknowledgements}

We thank CERN for the very successful operation of the LHC, as well as the
support staff from our institutions without whom ATLAS could not be
operated efficiently.

We acknowledge the support of ANPCyT, Argentina; YerPhI, Armenia; ARC,
Australia; BMWF and FWF, Austria; ANAS, Azerbaijan; SSTC, Belarus; CNPq and FAPESP,
Brazil; NSERC, NRC and CFI, Canada; CERN; CONICYT, Chile; CAS, MOST and NSFC,
China; COLCIENCIAS, Colombia; MSMT CR, MPO CR and VSC CR, Czech Republic;
DNRF, DNSRC and Lundbeck Foundation, Denmark; EPLANET, ERC and NSRF, European Union;
IN2P3-CNRS, CEA-DSM/IRFU, France; GNSF, Georgia; BMBF, DFG, HGF, MPG and AvH
Foundation, Germany; GSRT and NSRF, Greece; ISF, MINERVA, GIF, I-CORE and Benoziyo Center,
Israel; INFN, Italy; MEXT and JSPS, Japan; CNRST, Morocco; FOM and NWO,
Netherlands; BRF and RCN, Norway; MNiSW and NCN, Poland; GRICES and FCT, Portugal; MNE/IFA, Romania; MES of Russia and ROSATOM, Russian Federation; JINR; MSTD,
Serbia; MSSR, Slovakia; ARRS and MIZ\v{S}, Slovenia; DST/NRF, South Africa;
MINECO, Spain; SRC and Wallenberg Foundation, Sweden; SER, SNSF and Cantons of
Bern and Geneva, Switzerland; NSC, Taiwan; TAEK, Turkey; STFC, the Royal
Society and Leverhulme Trust, United Kingdom; DOE and NSF, United States of
America.

The crucial computing support from all WLCG partners is acknowledged
gratefully, in particular from CERN and the ATLAS Tier-1 facilities at
TRIUMF (Canada), NDGF (Denmark, Norway, Sweden), CC-IN2P3 (France),
KIT/GridKA (Germany), INFN-CNAF (Italy), NL-T1 (Netherlands), PIC (Spain),
ASGC (Taiwan), RAL (UK) and BNL (USA) and in the Tier-2 facilities
worldwide.

%% file: atlas_authlist.tex
\begin{flushleft}
{\Large The ATLAS Collaboration}

\bigskip

G.~Aad$^{\rm 84}$,
T.~Abajyan$^{\rm 21}$,
B.~Abbott$^{\rm 112}$,
J.~Abdallah$^{\rm 152}$,
S.~Abdel~Khalek$^{\rm 116}$,
O.~Abdinov$^{\rm 11}$,
R.~Aben$^{\rm 106}$,
B.~Abi$^{\rm 113}$,
M.~Abolins$^{\rm 89}$,
O.S.~AbouZeid$^{\rm 159}$
% ...
\bigskip
\\
% ...
$^{11}$ Institute of Physics, Azerbaijan Academy of Sciences, Baku, Azerbaijan\\
% ...
$^{21}$ Physikalisches Institut, University of Bonn, Bonn, Germany\\
% ...
$^{84}$ CPPM, Aix-Marseille Universit{\'e} and CNRS/IN2P3, Marseille, France\\
% ...
$^{89}$ Department of Physics and Astronomy, Michigan State University, East Lansing MI, United States of America\\
% ...
$^{106}$ Nikhef National Institute for Subatomic Physics and University of Amsterdam, Amsterdam, Netherlands\\
% ...
$^{112}$ Homer L. Dodge Department of Physics and Astronomy, University of Oklahoma, Norman OK, United States of America\\
$^{113}$ Department of Physics, Oklahoma State University, Stillwater OK, United States of America\\
% ...
$^{116}$ LAL, Universit{\'e} Paris-Sud and CNRS/IN2P3, Orsay, France\\
% ...
$^{152}$ Institute of Physics, Academia Sinica, Taipei, Taiwan\\
% ...
$^{159}$ Department of Physics, University of Toronto, Toronto ON, Canada\\
% ...
\end{flushleft}
